\documentclass[amsmath,twocolumn,prb,nofootinbib,longbibliography,superscriptaddress]{revtex4-2}

\usepackage{graphicx}
\usepackage{dcolumn}
\usepackage{bm}
\usepackage{braket}
\usepackage{xcolor} 
\usepackage{hyperref}
\usepackage{soul}
\usepackage{amssymb}

\begin{document}

\title{Charge density-waves with non-trivial orbital textures in rare earth tritellurides}

\author{Sergey Alekseev}
\affiliation{Department of Physics and Astronomy, Stony Brook University, Stony Brook, NY 11794}
\author{Sayed Ali Akbar Ghorashi}
\affiliation{Department of Physics and Astronomy, Stony Brook University, Stony Brook, NY 11794}
\author{Rafael M. Fernandes}
\affiliation{School of Physics and Astronomy, University of Minnesota, Minneapolis, MN 55455}
\author{Jennifer Cano}
\affiliation{Department of Physics and Astronomy, Stony Brook University, Stony Brook, NY 11794}
\affiliation{Center for Computational Quantum Physics, Flatiron Institute, New York, NY 10010}

\begin{abstract}
Motivated by recent experiments reporting unconventional collective modes in the charge density-wave (CDW) state of rare-earth tritellurides $R$Te$_3$, we derive from a multi-orbital microscopic model on the square net a CDW Ginzburg-Landau theory that allows for non-trivial orbital order. Our analysis reveals unconventional CDWs where order parameters with distinct orbital character coexist due to an approximate symmetry of the low-energy model, which becomes exact in the limit of nearest-neighbor-only hopping and decoupled $p_x$, $p_y$  orbitals. Because of this coexistence, the resulting CDW pattern displays an orbital texture that generally breaks additional symmetries of the lattice besides those explicitly broken by the CDW wave-vector. In particular,
we find two competing phases that differ in whether they break or preserve inversion and vertical mirror symmetries. We explain the mechanisms that favor each outcome, and
discuss experimental probes that can distinguish the different phases.
\end{abstract}

\maketitle

\section{\label{sec:introduction} Introduction}

Charge density-waves (CDWs) in rare earth tritellurides ($R$Te$_{3}$) have been extensively studied over the last three decades \cite{dimasi1995chemical,dimasi1996stability,brouet2004Fermi,komoda2004high,laverock2005Fermi,kim2006local,fang2007stm,brouet2008angle,ru2008effect,sinchenko2014unidirectional,maschek2015wave,kogar2019light,zong2019evidence,walmsley2020magnetic,sharma2020interplay,liu2020electronic,lei2021band,zhou2021nonequilibrium,gonzalez2022time,wang2022axial,straquadine2022evidence,Singh2023,Kivelson2023,chikina2023charge,raghavan2023atomic,Kim2024,singh20204ferro, akatsuka2024noncoplanar}. 
The itinerant electrons that drive the CDW transition are predominantly from two-dimensional (2D) Te ``square nets,'' whose partially filled $p_x$ and $p_y$ orbitals exhibit quasi-1D Fermi surfaces with near perfect nesting \cite{dimasi1996stability}. 
Theoretically, the Fermi surface structure can give rise to either stripe or checkerboard order; the former is favored for high transition temperatures~\cite{PhysRevB.74.245126}, consistent with the unidirectional (single-$\bm{Q}$) CDW observed in experiments.
In addition to Fermi surface nesting, strong electron-phonon coupling may also play a role in the CDW formation~\cite{johannes2008fermi,maschek2015wave,straquadine2022evidence}.

Yet, despite the long history of CDWs in this family of materials,
new facets have been revealed in recent experiments. For instance, a cascade of antiferromagnetic transitions was observed in GdTe$_3$ \cite{raghavan2023atomic} whereas emergent tetragonality and strain-driven CDW realignment were reported in ErTe$_3$ \cite{straquadine2022evidence,Singh2023,Kivelson2023}. Moreover, Ref.~\cite{wang2022axial} found a collective CDW mode displaying axial symmetry in GdTe$_3$. This is suggestive of additional symmetry-breaking beyond those naturally imposed by the CDW wave-vector $\bm{Q}$ (rotational and translational), which is inconsistent with a standard CDW order with ``$s$-wave'' orbital character, as usually obtained in microscopic models ~\cite{PhysRevB.74.245126}.
This discrepancy motivates us to revisit the theory of CDWs in the $R$Te$_3$ family by following the framework of Ref.~\cite{PhysRevB.74.245126} but considering an enlarged space of order parameters with non-trivial orbital character.

In this paper, we study CDWs with non-trivial orbital texture in Te square nets.
Starting from a microscopic weak-coupling model with weakly coupled $p_x$ and $p_y$ orbitals, combined with a group-theory analysis, we derive a Ginzburg-Landau (GL) theory and analyze its solutions assuming a single-$\bm{Q}$ phase. We show that in addition to the conventional phase described in Ref.~\cite{PhysRevB.74.245126}, the theory has solutions where two CDWs with distinct orbital characters coexist, giving rise to possible ground states with non-trivial orbital textures that break inversion or mirror-symmetry, even though the wave-vector $\bm{Q}$ preserves both.

This coexistence arises from an approximate symmetry of the microscopic model.
Specifically, in a nearest-neighbor square net tight-binding model, symmetry prevents hopping between $p_x$ and $p_y$ orbitals.
This decoupling results in an extra $U(1)$ symmetry corresponding to separate conservation of charge in each orbital sector, which is exact in the nearest-neighbor limit and approximate when realistic next-nearest-neighbor hopping is included.
Since the orbitals are nearly decoupled, 
the CDWs in each orbital sector are nearly independent, resulting in two distinct order parameters with near-degenerate coefficients in the GL theory. 
We study the competition/coexistence of these order parameters, analyzing the role of different mechanisms in selecting whether the symmetry broken by the orbital texture is inversion or a mirror. We also compare our results with recent experiments that report mirror symmetry-breaking in the CDW state of $R$Te$_3$ \cite{wang2022axial,singh20204ferro}.

The paper is organized as follows. In Sec. {\ref{sec:model}}, we introduce the square net tight-binding model and compute the CDW ordering vector from the charge density susceptibility. 
In Sec.~\ref{sec:symmetries}, we define the CDW order parameters and decompose them into irreducible representations of the crystallographic point group. 
This provides a framework to derive  and solve the GL theory in Sec.~\ref{sec:GL_theory}.
In Sec.~\ref{sec:experiment} we discuss the experimental implications of the results.
We conclude our work in Sec.~\ref{sec:discussion}.
In Appendix \ref{sec:deriving_gl}, we derive the GL theory. Appendix \ref{sec:fs} describes how we compute the Fermi surface in the presence of the CDW. Finally, Appendix \ref{sec:gl_d4h} extends the GL theory to the tetragonal lattice, as opposed to the orthorhombic lattice considered in the main text.

\section{\label{sec:model} Microscopic Model}

\begin{figure*}
    \includegraphics[width=\textwidth]{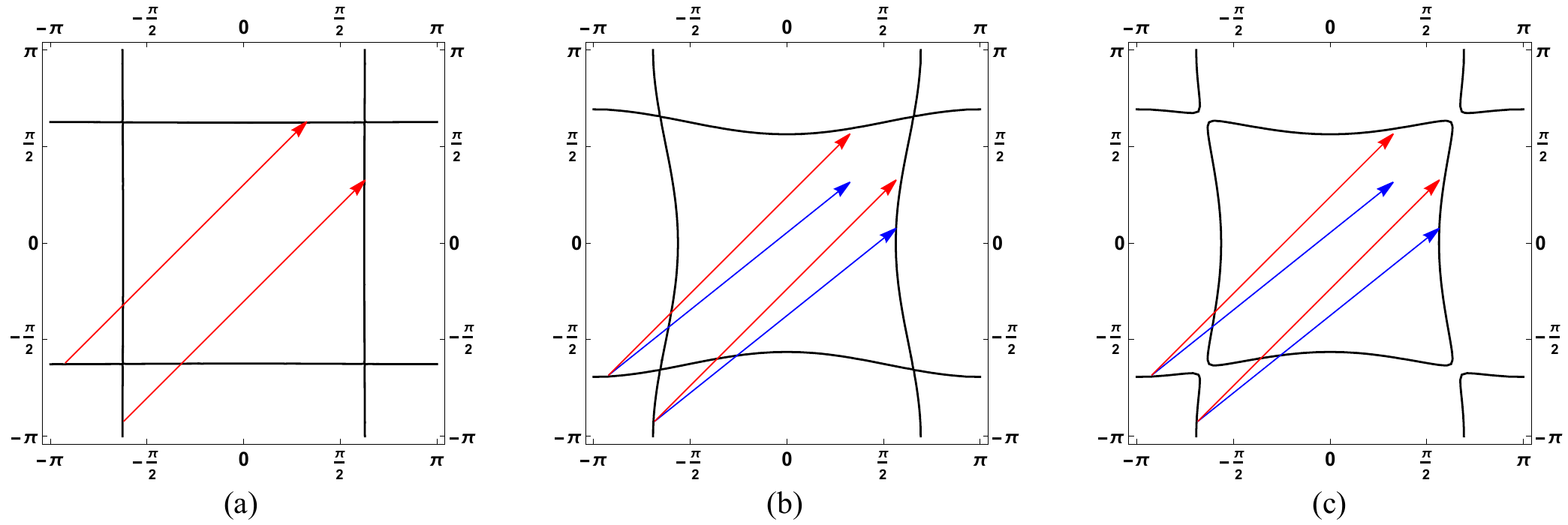}
\caption{\label{fig:fs_nesting}Fermi surfaces for $t_\sigma = 2.0$eV and (a) $t_d = 0$, $t_\pi = 0$, (b) $t_d=0$, $t_\pi= 0.37$eV, and (c) $t_d=0.16$eV, $t_\pi = 0.37$eV at chemical potential $\mu =1.53$eV. 
The vector $\bm{Q}_0 = (2k_F, 2k_F)$ (red) perfectly nests the Fermi surface in (a) and remains a good nesting vector in the presence of additional hopping terms in (b,c). The vector $(2k_F, \pi)$ (blue) nests the $p_x$-lines slightly better in (b,c) but does a poor job in nesting the $p_y$-lines.}
\end{figure*}

The $R\textrm{Te}_3$ crystal structure consists of double layers of square-planar $\textrm{Te}$ sheets separated by $R\textrm{Te}$ slabs.
The crystal is invariant under the orthorhombic space group $Cmcm$
with corresponding point group $D_{2h}$; however, since the orthorhombic distortion in $R\textrm{Te}_3$ is very weak, it is often modeled as tetragonal~\cite{wang2022axial}. 
Thus, we start by considering an idealized model of the $\textrm{Te}$ square lattice, which is invariant under the point group $D_{4h}$.

We approximate the band structure by the following tight-binding Hamiltonian written in the basis of Te $p_x$- and $p_y$-orbitals \cite{PhysRevB.74.245126,PhysRevB.101.165121}:
\begin{equation}
\label{eq:non-int_h}
    H^{(0)} = \sum_{\bm{k}}
    \begin{pmatrix}
        \psi_{\bm{k}, p_x}^\dagger & \psi_{\bm{k}, p_y}^\dagger
    \end{pmatrix}
     h_{\bm{k}}^{(0)} 
     \begin{pmatrix}
        \psi_{\bm{k}, p_x} \\
        \psi_{\bm{k}, p_y}
    \end{pmatrix}
\end{equation}
where $h_{\bm{k}}^{(0)}$ includes only nearest- and next-nearest-neighbor hopping:
\begin{equation}
h_{\bm{k}}^{(0)}=
\begin{pmatrix}
-2t_\sigma \cos k_x+2t_\pi \cos k_y & -2 t_d \sin k_x \sin k_y \\
-2 t_d \sin k_x \sin k_y & -2t_\sigma \cos k_y+2t_\pi \cos k_x
\end{pmatrix}.
\label{eq:ham}
\end{equation}
Here $t_\sigma$($t_\pi$) describes $\sigma$-bonds ($\pi$-bonds) between nearest neighbors and $t_d$ parameterizes the hopping strength diagonally across the square plaquettes. 
We estimate the hopping amplitudes $t_\sigma \approx 2.0$eV, $t_\pi \approx 0.37$eV, and $t_d \approx 0.16$eV, following Ref.~\cite{PhysRevB.74.245126}. Note that the spinor $\psi_{\bm{k}, p_i}$ contains a spin index that is implicitly summed over in $H^{(0)}$; hereafter, all our analysis is done assuming spin-rotational invariance.

An intuition for the origin and nature of the CDW arises from considering the extreme limit where $t_\pi = t_d = 0$ in Eq.~(\ref{eq:ham}). In this limit, the $p_x$ and $p_y$ orbitals are decoupled and, further, the $p_{x(y)}$ Fermi surface consists of two sets of straight lines at $k_{x(y)} = \pm k_F$, shown in Fig. \ref{fig:fs_nesting}(a), where $k_F$ is defined by:
\begin{equation}
\mu = -2 t_\sigma \cos k_F
\label{eq:chem_potential}
\end{equation}
for a fixed chemical potential $\mu$. In this simplified case, both $p_x$- and $p_y$-Fermi surfaces are perfectly nested by the vector $\bm{Q}_0 = (2k_F, 2k_F)$ and its fourfold-rotated counterparts (red arrow in Fig. \ref{fig:fs_nesting}). 
Reintroducing a small value of $t_\pi$ keeps the orbitals decoupled, but introduces curvature to the Fermi surface, shown in Fig. \ref{fig:fs_nesting}(b), so that the $p_x$- and $p_y$-lines start to slightly deviate from their average value, still given by $k_x, k_y = \pm k_F$. Finally, introducing non-zero $t_d$ couples the $p_x$ and $p_y$ orbitals and opens a gap where their Fermi surfaces cross, shown in Fig.~\ref{fig:fs_nesting}(c). As long as $t_\pi$ and $t_d$ are small compared to $t_\sigma$, the vector $\bm{Q}_0$ still does a good job of nesting the entire Fermi surface. 
One may also consider the competing nesting vector $(2k_F, \pi)$, which 
nests the $p_x$-FS even better, but does a poor job nesting the $p_y$-FS~\cite{PhysRevB.74.245126}  (blue arrow in Fig. \ref{fig:fs_nesting}. In the following, we adopt $\mu \approx 1.53$ eV, as determined from Eq. (\ref{eq:chem_potential}) with $k_F \approx 5\pi/8$ and $t_\sigma \approx 2.0$eV \cite{PhysRevB.74.245126}.
To determine which ordering vector is preferred, we numerically evaluate the charge density susceptibility matrix, which is a multi-orbital generalization of the Lindhard susceptibility, as used in, e.g., Ref.~\cite{Graser_2009}:
\begin{eqnarray}
&&\chi_{\alpha \delta}^{\beta \gamma}(\bm{Q}, T, \mu)=- \sum_{\bm{k}, nm} \frac{a_n^\alpha(\bm{k}) a_n^{\beta *}(\bm{k}) a_m^\gamma(\bm{k+Q}) a_m^{\delta *}(\bm{k+Q})}
{\xi_{\bm{k}n}-\xi_{\bm{k+Q}m}} \nonumber\\ 
&&\times\left[f\left(\xi_{\bm{k}n}\right)-f\left(\xi_{\bm{k+Q}m}\right)\right],
\label{eq:susceptibility_components}
\end{eqnarray}
where $\xi_{\bm{k} n } = \epsilon_{\bm{k} n} - \mu$ are shifted eigenenergies, $a^\alpha_m(\bm{k})$ is the matrix that diagonalizes the Hamiltonian (\ref{eq:ham}), $n$ and $m$ are the band indices, and the Greek letter superscripts indicate the orbital indices $p_x,\,p_y$. The Fermi-Dirac distribution function is denoted by $f(\xi_{\bm{k}n }, T )$. 
Eq.~(\ref{eq:susceptibility_components}) is derived from the Green’s function representation in the limit $\omega \to 0$ \cite{Graser_2009}:
\begin{eqnarray}
&&\chi_{\alpha \delta}^{\beta \gamma}(\bm{Q}, \omega, T, \mu) = \nonumber\\
&&-\beta^{-1} \sum_{\bm{k}, i \omega_n} G_{\alpha\beta}\left(\bm{k}, i \omega_n\right) G_{\gamma\delta}\left(\bm{k}+\bm{Q}, i \omega_n+i \omega\right),
\label{eq:susc_greens_function}
\end{eqnarray}
where the spectral representation of the Green’s function is given by
\begin{equation}
G_{\alpha\beta}\left(\bm{k}, i \omega_n\right)=\sum_m \frac{a_m^\alpha(\bm{k}) a_m^{\beta *}(\bm{k})}{i\omega_n-\epsilon_{\bm{k}m}}.
\label{eq:greens_function}
\end{equation}
When $t_d$ is small, the eigenstates $a^\alpha_m$ have dominantly either $p_x$ or $p_y$ character and the Green's function is nearly diagonal in the orbital indices $\alpha, \beta$. 
Consequently, the susceptibility components $\chi^{\alpha \gamma}_{\alpha \gamma}$ dominate. Thus, to determine the dominant $\bm{Q}$, we plot the fourfold-invariant combinations $\chi(\bm{Q}, T, \mu) = \chi_{11}^{11}(\bm{Q}, T, \mu) + \chi_{22}^{22}(\bm{Q},T, \mu)$ and $\tilde\chi(\bm{Q}, T, \mu) = \chi_{12}^{12}(\bm{Q}, T, \mu) + \chi_{21}^{21}(\bm{Q},T, \mu)$.

\begin{figure}
    \centering
\includegraphics[height=5cm]
{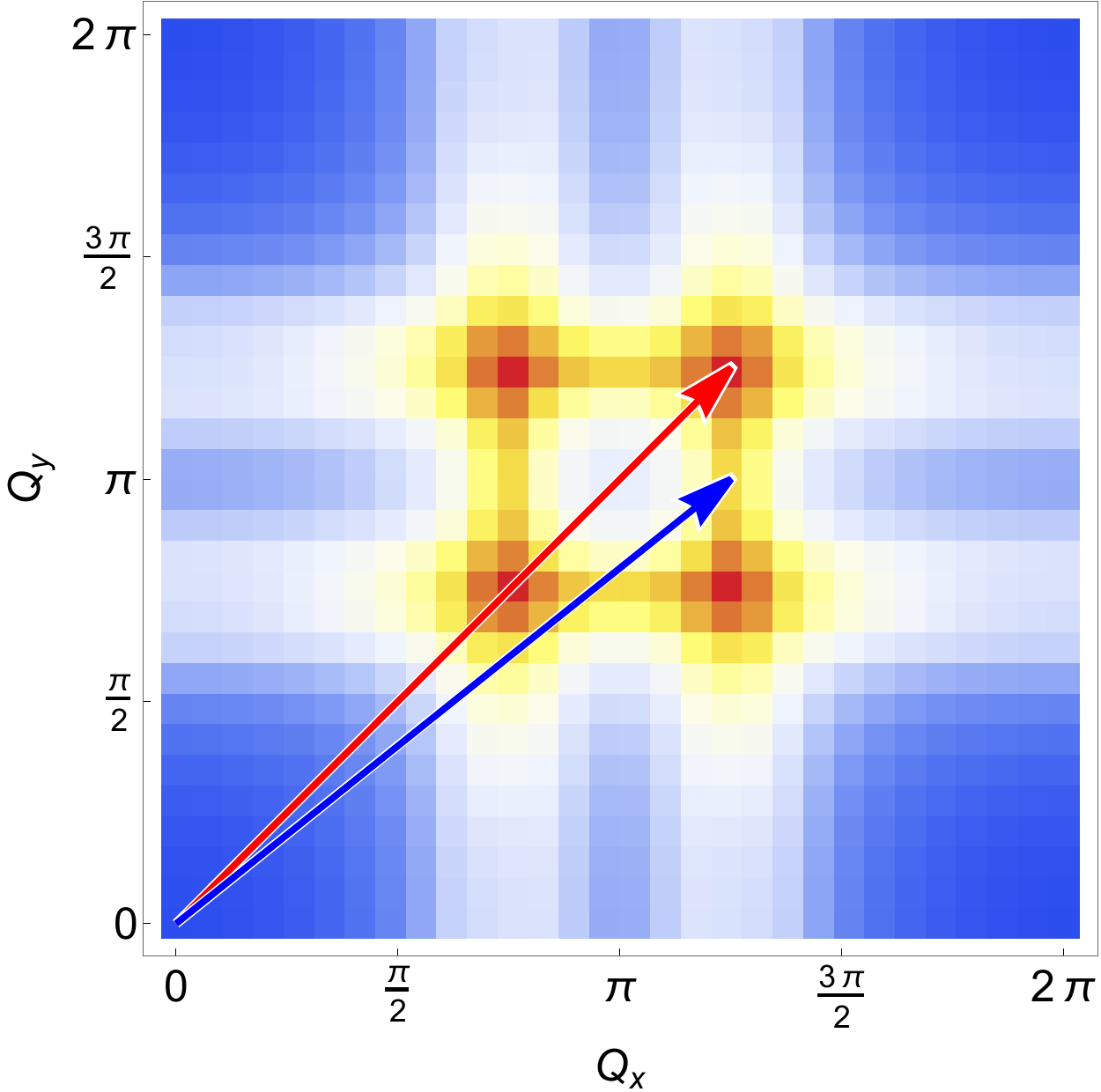}
\includegraphics[height=5cm]
{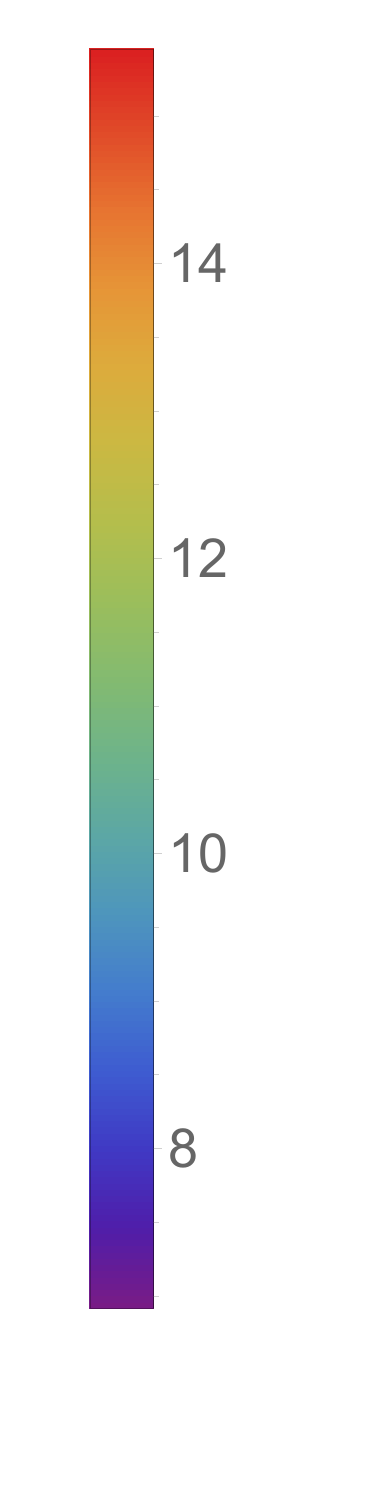}
\includegraphics[height=5cm]
{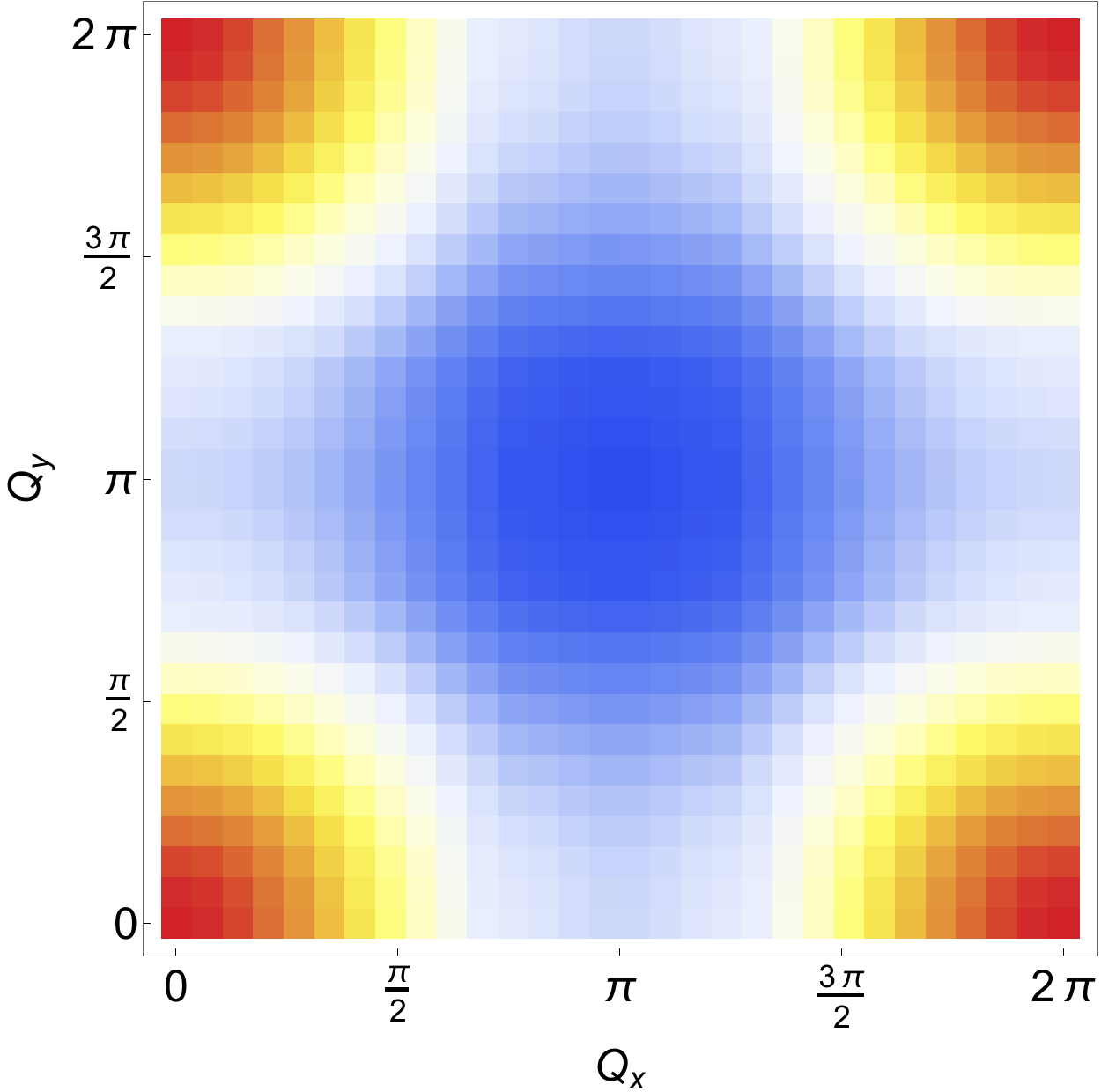}
\includegraphics[height=5cm]
{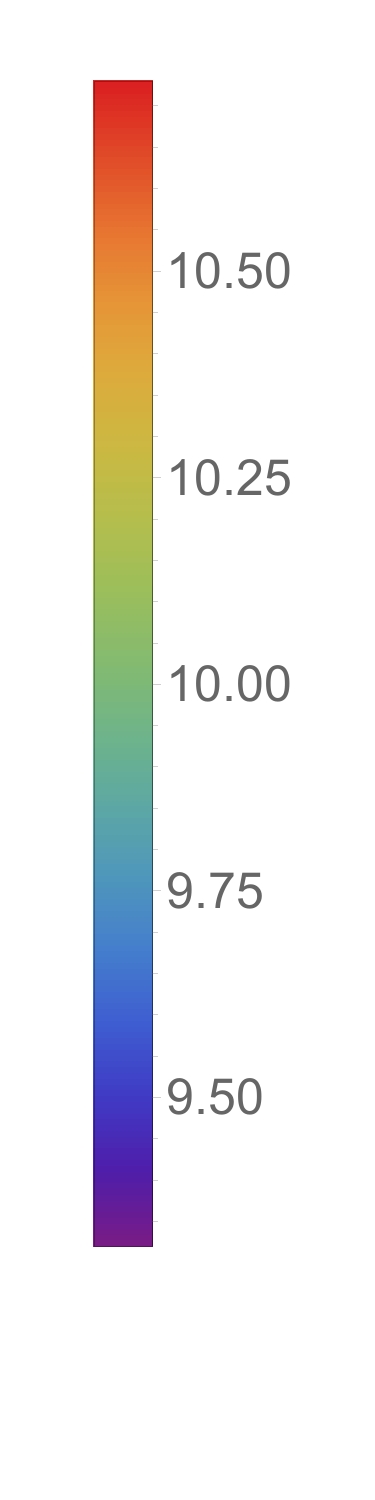}
    \caption{Susceptibilities $\chi(\bm{Q}, T, \mu)$ (upper panel)  and $\tilde\chi(\bm{Q}, T, \mu)$ (lower panel) as functions of $\bm{Q}$ at fixed temperature $T=0.15$eV and chemical potential $\mu=1.53$eV for $t_d=0.16$eV. Red and blue arrows indicate the two nesting vectors $\bm{Q}_0 = (2k_F, 2k_F)$ and $(2k_F, \pi)$ respectively for the given value of $\mu$.
    }
    \label{fig:chi_QQ}
\end{figure}

\begin{figure*}
\includegraphics[height=5.5cm]
{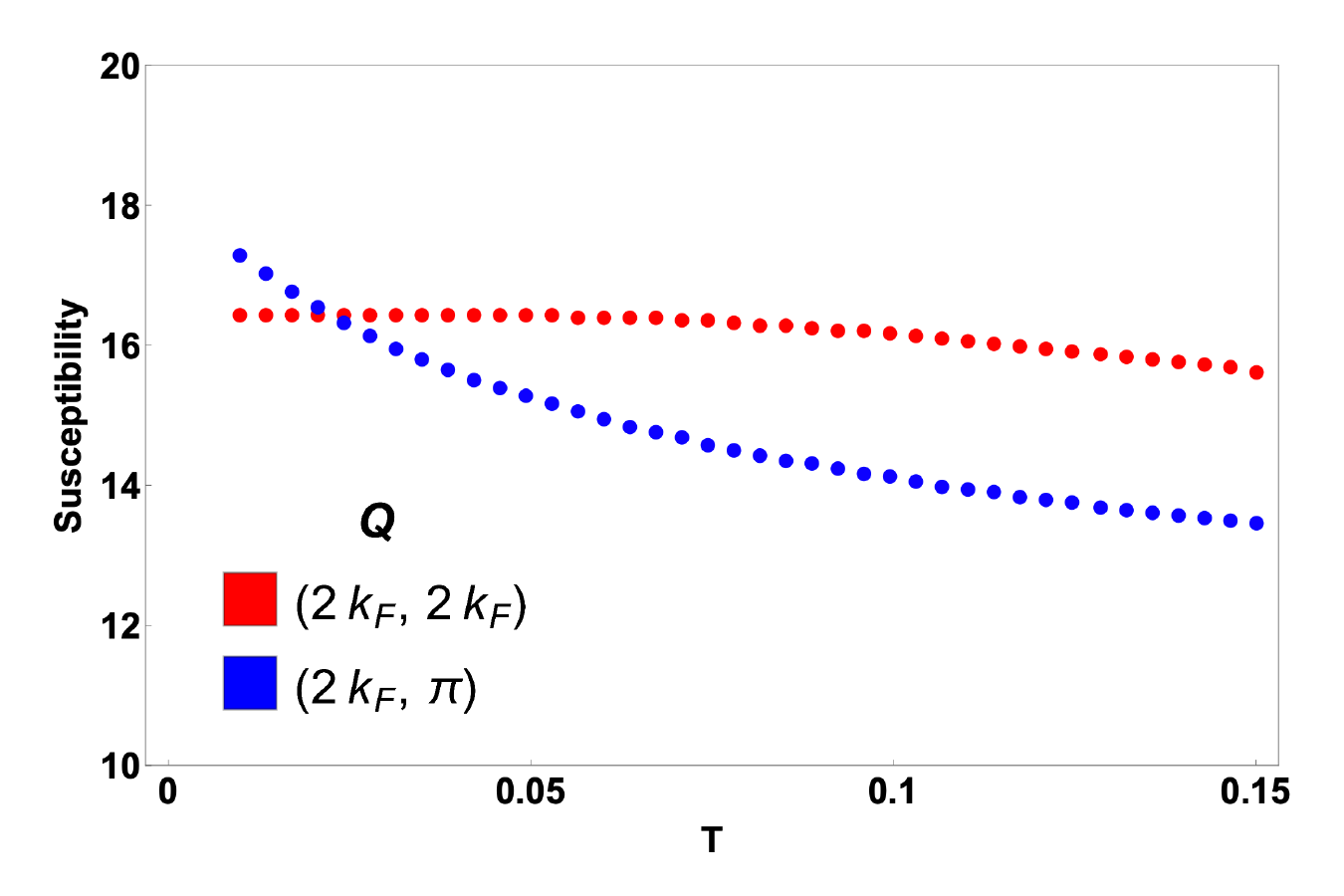}
\includegraphics[height=5.5cm]
{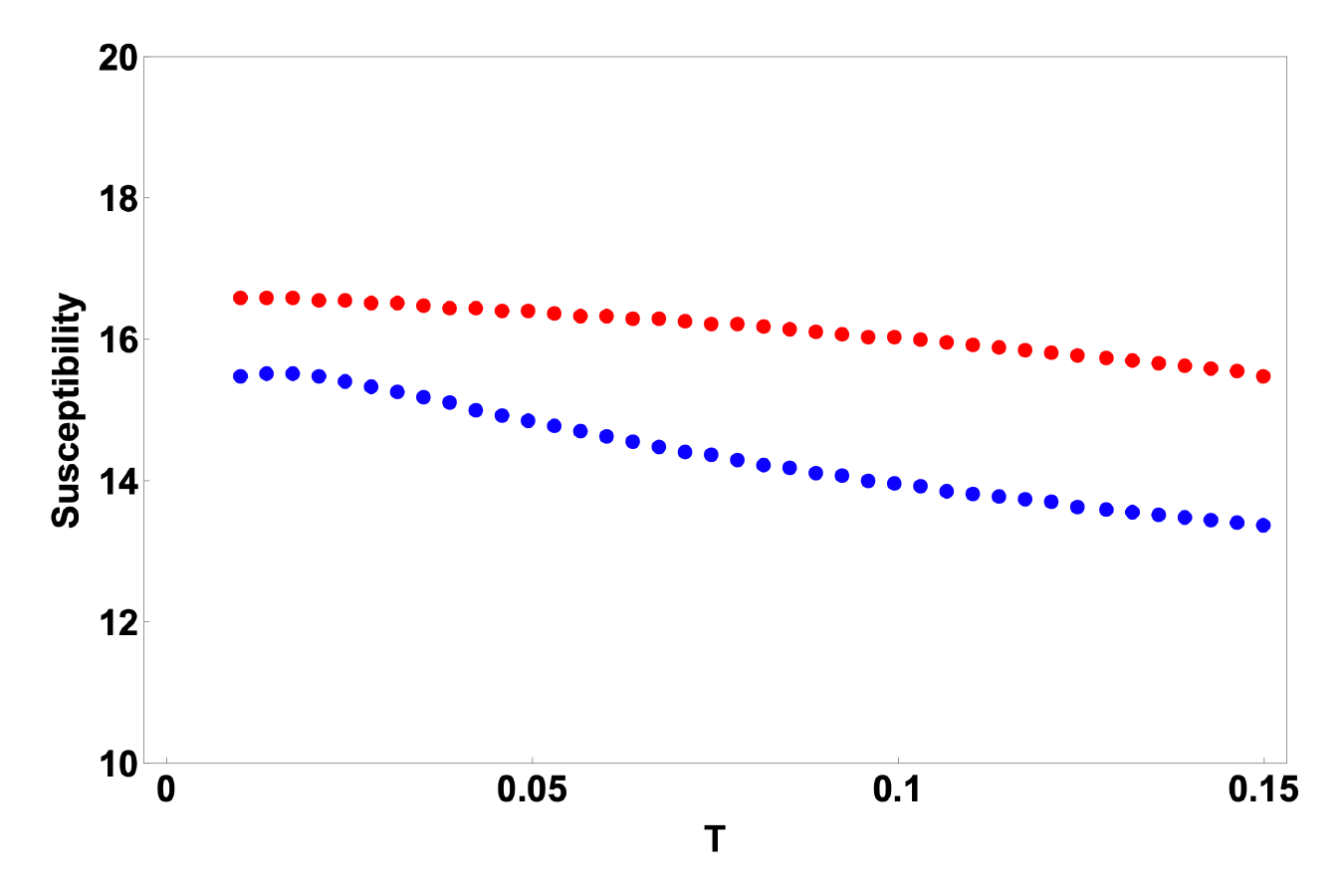}

    \caption{Susceptibility $\chi(\bm{Q}, T, \mu)$ for $\bm{Q}=(2k_F, 2k_F)$ (red) and $\bm{Q}=(2k_F, \pi)$ (blue) as a function of temperature $T$ (in eV) at fixed chemical potential $\mu=1.53$eV. Left: $t_d = 0$. Right: $t_d = 0.16$eV.}
    \label{fig:chi_Q1_vs_Q2}
\end{figure*}

In agreement with Ref.~\onlinecite{PhysRevB.74.245126}, we find that over a wide range of chemical potential and temperature, the susceptibility $\chi(\bm{Q}, T, \mu)$ peaks at approximately $\bm{Q}_0$ and its fourfold-rotated counterparts (see Fig.~\ref{fig:chi_QQ}), while $\tilde\chi(\bm{Q}, T, \mu)$ peaks at zero. In Fig.~\ref{fig:chi_Q1_vs_Q2} we plot $\chi(\bm{Q}, T, \mu)$ as a function of temperature at a fixed value of chemical potential for $\bm{Q}=\bm{Q}_0$ and $\bm{Q}=(2k_F, \pi)$. While for $t_d = 0$, the latter vector is preferred at low temperatures, for the realistic value of $t_d$, the vector $\bm{Q}_0 = (2k_F, 2k_F)$ is always preferred. 
Thus, in the following, the GL theory we derive is specific to $\bm{Q}_0$.

The peak in the non-interacting susceptibility $\chi(\bm{Q}_0)$ can result in a CDW instability in the presence of interactions. In a multi-orbital system, there are different types of local intra-orbital and inter-orbital interactions, whose combinations may favor either a spin or a charge density-wave (see, for instance, Ref. \cite{chubukov2009renormalization}). For our purposes, it is sufficient to assume a combination that favors a CDW and write an effective attractive interaction $g>0$ in the CDW channel:
\begin{equation}
V = - g
    \sum_{\substack{\bm{k}, \bm{k'}, \bm{Q} \\ \alpha\beta}} \psi^\dagger_{\bm{k}+\bm{Q}, \alpha} \psi_{\bm{k},\beta}\ \psi^\dagger_{\bm{k'}-\bm{Q},\beta}\psi_{\bm{k'},\alpha},
    \label{eq:coulomb_interaction}
\end{equation}
where $\alpha, \beta$ are the orbital indices $p_x, p_y$ and the sum over $\bm{k}, \bm{k'},\bm{Q}$ runs over the first Brillouin zone. 
The total Hamiltonian is given by
\begin{equation}
H = H^{(0)} + V.
\label{eq:full_ham}
\end{equation}
In later sections we will also consider the role of electron-phonon coupling.

The GL theory we will derive describes all possible CDW order parameters in orbital space with wave vector $\bm{Q}_0$.
The case when the charge density wave is trivial in orbital space was extensively studied in Ref.~\onlinecite{PhysRevB.74.245126}. In this situation, the symmetries broken by the CDW phase arise entirely from the direction of the wave-vector $\bm{Q}_0$, which in the case $\bm{Q}_0 = (2k_F, 2k_F)$ breaks translational symmetry and lowers the fourfold rotational symmetry to twofold. However, as will be demonstrated below, when $t_d=0$,
there are two degenerate CDW order parameters, one of which possesses a nontrivial orbital character.
For a finite and realistic value of $t_d$, the two channels remain nearly degenerate.
We study the competition between a phase with a single CDW or one where both CDWs coexist.
In the latter case, the orbital texture arising from the mixed orbital character of the CDWs causes the Fermi surface to break additional crystal symmetries beyond those broken by the wave vector $\bm{Q}_0$ itself, such as mirror or inversion. Recent Raman and second-harmonic generation (SHG) data on LaTe$_3$ are consistent with a CDW that breaks all vertical mirrors ~\cite{singh20204ferro}.

\section{\label{sec:symmetries} Symmetries and CDW Order Parameters}

Before deriving the Ginzburg-Landau theory, we define the CDW order parameters and decompose them into irreps of the crystallographic point group $D_{2h}$, the group of symmetries that either leave $\bm{Q}_0$ invariant or rotate $\bm{Q}_0 \rightarrow -\bm{Q}_0$ (i.e. the little group).
$D_{2h}$ consists of the elements:
\begin{eqnarray}
&&\{E,\, C_2,\, C_2''(y=x),\, C_2''(y=-x),\,    \nonumber \\
&&\ \ I,\, \sigma_h,\,   \sigma_d(y=x),\, \sigma_d(y=-x)\},
\end{eqnarray}
where $E$ is the identity operation, $C_2$ is a two-fold rotation about the principal ($z$) axis, $C_2''$ are two-fold rotations about the axes $y=x$ and $y=-x$, $I$ indicates inversion, $\sigma_h$ is reflection through the horizontal mirror plane,  and $\sigma_d$ denotes reflections through one of the vertical mirror planes $y=x$ or $y=-x$. For reference, we provide the character table of $D_{2h}$ in Table \ref{tab:char_tab}.

\begin{table*}
\begin{ruledtabular}
\begin{tabular}{c c c c c c c c c }
$D_{2h}$ & $E$ & $C_2$ & $C_2''(y=x)$ & $C_2''(y=-x)$ &
$I$ & $\sigma_h$ & $\sigma_d(y=x)$ & $\sigma_d(y=-x)$  \rule{0pt}{3ex}\\[1ex]
\colrule
$A_{g}$ &	$+1$ & $+1$ & $+1$ & $+1$ &	$+1$ & $+1$ & $+1$ & $+1$	\rule{0pt}{3ex}\\[1ex]
$B_{1g}$ &	$+1$ &	$+1$ &	$-1$ &	$-1$ &	$+1$ &	$+1$ &	$-1$ &	$-1$ \rule{0pt}{3ex}\\[1ex]
$B_{2g}$ &	$+1$ & 	$-1$ &	$-1$ &	$+1$ &	$+1$ &	$-1$ &	$+1$ &	$-1$ \rule{0pt}{3ex}\\[1ex]
$B_{3g}$ &	$+1$ &	$-1$ &	$+1$ &	$-1$ &	$+1$ &	$-1$ &	$-1$ &	$+1$ \rule{0pt}{3ex}\\[1ex]
$A_{u}$ &	$+1$ &	$+1$ &	$+1$ &	$+1$ &	$-1$ &	$-1$ &	$-1$ &	$-1$ \rule{0pt}{3ex}\\[1ex]
$B_{1u}$ &	$+1$ &	$+1$ &	$-1$ &	$-1$ &	$-1$ & $-1$	&	$+1$&	$+1$ \rule{0pt}{3ex}\\[1ex]
$B_{2u}$ &	$+1$ &	$-1$ &	$-1$&	$+1$ &	$-1$ &	$+1$ &	$-1$ &	$+1$ \rule{0pt}{3ex}\\[1ex]
$B_{3u}$ &	$+1$ &	$-1$  &	$+1$ &	$-1$ &	$-1$ &	$+1$ &	$+1$ &	$-1$ \rule{0pt}{3ex}\\[1ex]
\end{tabular}
\end{ruledtabular}
\caption{\label{tab:char_tab}
Character table for the point group $D_{2h}$.
}
\end{table*}

The orbital-resolved CDW order parameters are given by:
\begin{equation}
\Delta_{\bm{Q}}^{\alpha\beta} = \sum_{\bm{k}} \Braket{\psi^\dagger_{\bm{k-Q},\beta}\psi_{\bm{k},\alpha}}.
\label{eq:order_params_matrix_els}
\end{equation}
Equivalently, one may consider their linear combinations: \begin{equation}
\Delta^i_{\bm{Q}} = \sum_{\bm{k}}\Braket{\psi^\dagger_{\bm{k}-\bm{Q}}\sigma^i\,\psi_{\bm{k}}},\ i=\overline{0,3},
\label{eq:order_params_pauli}
\end{equation}
so that $ \Delta_{\bm{Q}} = \sum_i \Delta^i_{\bm{Q}} \sigma^i$. Here $\sigma^i,\ i=\overline{1,3}$, are Pauli matrices acting in orbital space, $\sigma^0$ is the identity matrix, and we have introduced an abbreviated notation where the orbital indices are not shown explicitly. Note that the order parameters satisfy $\bar{\Delta}^{i}_{\bm{Q}} = \Delta^i_{-\bm{Q}}$, where the overbar denotes complex conjugation. 

The symmetry properties of the order parameters can be directly obtained from the symmetry properties of the creation and annihilation operators.
Under the action of translations $\bm{t}$ and point group elements $g$:
\begin{eqnarray}
 &&\bm{t}:\ 
 \Delta^i_{\bm{Q}} \mapsto
 e^{-i\bm{Q}\cdot \bm{t}}\,
 \Delta^i_{\bm{Q}}, \nonumber \\
 && g:\  \Delta^i_{\bm{Q}} \mapsto \chi_{\rho_i}(g)\, 
\Delta^i_{g^{-1}\cdot\bm{Q}},
\label{eq:symmaction}
\end{eqnarray}
where $\chi_{\rho_i}(\cdot)$ is the character of the representation $\rho_i$, with $\rho_0 = \rho_x = A_g$ and $\rho_y = \rho_z = B_{1g}$ 1D irreps of $D_{2h}$.

Below, we decompose the space of order parameters into irreducible representations of the point group $D_{2h}$. We also give the irreps of their translationally invariant bilinears, which will simplify the construction of our Ginzburg-Landau theory. 
The space of order parameters is spanned by $\{\Delta^i_{\bm{Q}_0}, \Delta^i_{-\bm{Q}_0}\}$ and is decomposed into the irreps of $D_{2h}$ in Table \ref{tab:order_param_irreps_d2h}.
$\Re \Delta^i_{\bm{Q}_0}$ and $\Im \Delta^i_{\bm{Q}_0}$ denote the real and imaginary parts of the order parameter $\Delta^i_{\bm{Q}_0}$, respectively, and can be equivalently rewritten as $\Delta^i_{\bm{Q}_0}\pm \Delta^i_{-\bm{Q}_0}$, up to a coefficient.
\begin{table}[t]
\begin{ruledtabular}
\begin{tabular}{@{\hspace{5em}} c c @{\hspace{5em}}}
Basis &\ Representation  \rule{0pt}{3ex}\\[1ex]
\colrule
$\Re \Delta^i_{\bm{Q}_0}$ & $A_{g} \otimes \rho_i$ \rule{0pt}{3ex}\\[1ex]
$ \Im \Delta^i_{\bm{Q}_0} $ & $B_{3u} \otimes \rho_i$ \rule{0pt}{3ex}\\[1ex]
\end{tabular}
\end{ruledtabular}
\caption{\label{tab:order_param_irreps_d2h}
Irreducible representations of the CDW order parameters for the point group $D_{2h}$. The index $i$ takes values in the set $\{0,x,y,z\}$, with  $\rho_0 = \rho_x = A_g$, and $\rho_y=\rho_z = B_{1g}$.
}
\end{table}

\begin{table}
\begin{ruledtabular}
\begin{tabular}{@{\hspace{5em}} c c @{\hspace{5em}}}
Basis & Representation \rule{0pt}{3ex}\\[1ex]
\colrule
$\Re\Lambda_0^{ij} $ & $A_{g} \otimes \rho_i \otimes \rho_j$ \rule{0pt}{3ex}\\[1ex]
$ \Im\Lambda_0^{ij} $ & $B_{3u} \otimes \rho_i \otimes \rho_j$ \rule{0pt}{3ex}\\[1ex]
\end{tabular}
\end{ruledtabular}
\caption{\label{tab:bilinears_irreps_d2h}
Irreducible representations of the translationally invariant $\Delta$-bilinears $\Lambda^{ij}_0 = \Delta^i_{\bm{Q}_0} \Delta^j_{-\bm{Q}_0}$ for the point group $D_{2h}$. The index $i$ takes values in the set $\{0,x,y,z\}$, with $\rho_0 = \rho_x = A_g$, and $\rho_y=\rho_z = B_{1g}$.
}
\end{table}

The space of translationally invariant bilinears spanned by 
$\Lambda^{ij}_0 = \Delta^i_{\bm{Q}_0} \Delta^j_{-\bm{Q}_0}$
is analogously decomposed into irreducible representations of $D_{2h}$ in Table \ref{tab:bilinears_irreps_d2h}. In Appendix \ref{sec:deriving_gl}, we list all the $\Delta$-bilinears and their corresponding representations explicitly in Table \ref{tab:bilinears_irreps_d2h_detailed}.

It is straightforward to extend this analysis by considering the full tetragonal symmetry of the lattice, i.e. by considering the little group $D_{4h}$. In this case, one needs to also include, besides $\bm{Q}_0$ and $-\bm{Q}_0$ , the wave-vectors $\bm{Q}_1 = C_4 \cdot \bm{Q}_0$  and $-\bm{Q}_1$, where $C_4$ denotes a $\pi/2$ rotation with respect to the $z$ axis. Clearly, the case of a single-$\bm{Q}$ CDW reduces to the case studied here, i.e., the little group $D_{2h}$. Since the analysis is more transparent for the $D_{2h}$ case, we will focus on it hereafter. For completeness, we show in Appendix \ref{sec:gl_d4h} the full GL analysis for the $D_{4h}$ case. Note that in Ref. \cite{singh20204ferro}, the order parameters and bilinears are classified in terms of the irreducible representations of $D_{4h}$.

\section{\label{sec:GL_theory} Ginzburg-Landau theory}

\begin{table*}[t]
\begin{ruledtabular}
\begin{tabular}{c c c c c }

 &\multicolumn{2}{c}{Single-$\Delta$} & \multicolumn{2}{c}{Double-$\Delta$}\\
& $A_g$ & $B_{1g}$ & $B_{2u}$
& $B_{1g}$ \rule{0pt}{3ex}\\[1ex] \hline
Non-zero order parameters &	$\Re\Delta^0_{\bm{Q}_0}$ & $\Re\Delta^z_{\bm{Q}_0}$  & \shortstack{$\Re\Delta^0_{\bm{Q}_0}$, $\Im\Delta^z_{\bm{Q}_0}$ \\ $\delta \alpha = \pi/2$} & \shortstack{$\Re\Delta^0_{\bm{Q}_0}$, $\Re\Delta^z_{\bm{Q}_0}$ \\ $\delta \alpha = 0$}	\rule{0pt}{6ex}\\[1ex]
Irreps & $A_{g}$ & $B_{1g}$ & $A_g$, $B_{2u}$ & $A_g$, $B_{1g}$ \rule{0pt}{6ex}\\[1ex]
Non-zero bilinears & –– & –– & $\bar\Delta^0_{\bm{Q}_0} \Delta^z_{\bm{Q}_0} - \bar\Delta^z_{\bm{Q}_0} \Delta^0_{\bm{Q}_0}$ ($B_{2u}$) & $\bar\Delta^0_{\bm{Q}_0} \Delta^z_{\bm{Q}_0} + \bar\Delta^z_{\bm{Q}_0} \Delta^0_{\bm{Q}_0}$ ($B_{1g}$) \rule{0pt}{6ex}\\[1ex]
\shortstack{Solution in terms of \\ $\Delta^{p_x p_x}_{\bm{Q}_0}$ and $\Delta^{p_y p_y}_{\bm{Q}_0}$} & $\Delta^{p_x p_x}_{\bm{Q}_0}= \Delta^{p_y p_y}_{\bm{Q}_0}$ & $\Delta^{p_x p_x}_{\bm{Q}_0}= -\Delta^{p_y p_y}_{\bm{Q}_0}$ & \shortstack{$\Delta^{p_x p_x}_{\bm{Q}_0} = (|\Delta^0_{\bm{Q}_0}| + i|\Delta^z_{\bm{Q}_0}|) e^{i\alpha}$, \\ $\Delta^{p_y p_y}_{\bm{Q}_0}=(|\Delta^0_{\bm{Q}_0}| - i|\Delta^z_{\bm{Q}_0}|) e^{i\alpha}$} & \shortstack{$\Delta^{p_x p_x}_{\bm{Q}_0} = (|\Delta^0_{\bm{Q}_0}| + |\Delta^z_{\bm{Q}_0}|) e^{i\alpha}$, \\ $\Delta^{p_y p_y}_{\bm{Q}_0}=(|\Delta^0_{\bm{Q}_0}| - |\Delta^z_{\bm{Q}_0}|) e^{i\alpha}$}
\rule{0pt}{8ex}\\[1ex]
\end{tabular}
\end{ruledtabular}
\caption{\label{tab:solutions}
Solutions of the Ginzburg-Landau theory and their non-zero order parameters, corresponding irreps,
non-zero bilinear order parameters, and expression in terms of $p_{x,y}$ orbitals. Here $\delta \alpha = \alpha_z -\alpha_0$ is the phase difference between $\Delta^0_{\bm{Q}_0}$ and $\Delta^z_{\bm{Q}_0}$. Corresponding real-space plots illustrating the symmetry-breaking patterns are given in Fig. \ref{fig:real_space}. For simplicity, we omit the order parameter $\Delta^x_{\bm{Q}_0}$, which is non-vanishing in the $t_d \ne 0$ case whenever $\Delta^0_{\bm{Q}_0}$ is non-vanishing since the order parameters mix.
Since $\Delta^x_{\bm{Q}_0}$ transforms as the same irrep as $\Delta^0_{\bm{Q}_0}$, the symmetry is not changed by this omission. Alternatively, in the $t_d \ne 0$ case, $\Delta^0_{\bm{Q}_0}$ should be replaced by $\Delta^-_{\bm{Q}_0}$ everywhere in this table.  
}
\end{table*}

In this section, we derive the GL theory for the CDW order parameters (\ref{eq:order_params_pauli}) and classify the resulting phases, which are summarized in Table \ref{tab:solutions}. We begin by studying the $t_d=0$ case in Sec.~\ref{sec:GL_theory_td_zero}.
We show that for a nesting-driven instability of the Fermi surface, 
there is a degenerate family of solutions to the GL theory, corresponding to an arbitrary relative phase between CDWs in the $p_x$ and $p_y$ channels, i.e. to an $U(1)$ symmetry. 
This degeneracy allows, in particular, for solutions with two non-vanishing order parameters that transform as different irreps. Such solutions break inversion symmetry and one of the diagonal mirror symmetries $\sigma_d$ in addition to those required by the CDW vector $\bm{Q}_0$. We then show that introducing electron-phonon coupling may drive the system to a different phase with two coexisting order parameters which breaks both mirrors while preserving inversion. 
We then generalize the GL theory to the $t_d \ne 0$ case in Sec.~\ref{sec:GL_theory_td_non_zero} and show that the aforementioned degeneracy is resolved in favor of solutions with two coexisting channels.

\subsection{\label{sec:GL_theory_td_zero} Case $t_d = 0$}

For simplicity, we begin by analyzing the GL theory in the limit $t_d = 0$ when the $p_x$ and $p_y$ orbitals are decoupled. Due to the decoupling, the effective action takes the simple form:
\begin{eqnarray}
&& S_{\operatorname{eff}} =
2r\left(|\Delta^{p_x p_x}_{\bm{Q}_0}|^2 + |\Delta^{p_y p_y}_{\bm{Q}_0}|^2\right) \nonumber \\
&& + 
8b \left(
|\Delta^{p_x p_x}_{\bm{Q}_0}|^4 + |\Delta^{p_y p_y}_{\bm{Q}_0}|^4\right),
\label{eq:gl_simple}
\end{eqnarray}
expressed in terms of the order parameters $\Delta^{p_xp_x}_{\bm{Q}_0}$ and $\Delta^{p_yp_y}_{\bm{Q}_0}$ from Eq. (\ref{eq:order_params_matrix_els}).
While symmetry allows for terms including the remaining order parameters $\Delta^{p_xp_y}_{\bm{Q}_0}$ and $\Delta^{p_yp_x}_{\bm{Q}_0}$ in the GL theory, we argue below Eq. (\ref{eq:coeff_susceptibilities_diag_td_zero}) that they vanish.
Minimizing the effective action (\ref{eq:gl_simple}) yields $|\Delta^{p_x p_x}_{\bm{Q}_0}| = |\Delta^{p_y p_y}_{\bm{Q}_0}|$, but does not determine the phases of the order parameters, which is important for determining the symmetry of the CDW phase.

Going forward, it will be convenient to label the order parameters in terms of the point group irreps, defined in Eq. (\ref{eq:order_params_pauli}).
The GL theory to quadratic order includes all translationally invariant bilinears in Table~\ref{tab:bilinears_irreps_d2h} that transform as the trivial representation of $D_{2h}$ (the bilinears are written explicitly in Table~\ref{tab:bilinears_irreps_d2h_detailed}). 
Similarly, the quartic order consists of $A_g$-products of the bilinears from Table \ref{tab:bilinears_irreps_d2h}. 
The coefficient of each term in the GL theory can be derived from the non-interacting Green's functions defined in Eq.~(\ref{eq:greens_function}).
We refer the reader to Appendix \ref{sec:deriving_gl} for the complete derivation, where we first obtain the theory for $t_d \ne 0$ and then exploit simplifications arising in the $t_d = 0$ case. The resulting effective action to quadratic order is as follows:
\begin{equation}
S_{\operatorname{eff}}^{(2)}= a_0|\Delta_{\bm{Q}_0}^0|^2 + a_x|\Delta_{\bm{Q}_0}^x|^2 + a_y|\Delta_{\bm{Q}_0}^y|^2 + a_z|\Delta_{\bm{Q}_0}^z|^2,   
\end{equation}
where the coefficients are given in terms of the susceptibilities (\ref{eq:susceptibility_components}) and the interaction strength $g$:
\begin{eqnarray}
&&a_0 = a_z = \frac{2}{g} -
\chi_{11}^{11} - \chi_{22}^{22}, \nonumber \\
&& a_x = a_y = \frac{2}{g} -
\chi_{12}^{12} - \chi_{21}^{21}.
\label{eq:coeff_susceptibilities_diag_td_zero}
\end{eqnarray}
It is a special feature of the $t_d=0$ limit that $a_0$ and $a_z$ are exactly degenerate, as are $a_x$ and $a_y$. Importantly, this degeneracy is not enforced by the point-group symmetries: since the coefficients $a_0$ and $a_z$  (or, similarly, $a_x$ and $a_y$) correspond to order parameters that transform as different irreps, they in principle should be different. The fact that they are the same points to an emergent symmetry that arises when the orbitals are decoupled, which we will explore in further detail below. Interestingly, this same enhanced symmetry kills the bilinear coefficients that couple the pairs of CDW order parameters that transform as the same irrep according to Table \ref{tab:order_param_irreps_d2h}, namely, $\Delta_{\bm{Q}_0}^0$ and $\Delta_{\bm{Q}_0}^x$ or  $\Delta_{\bm{Q}_0}^z$ and $\Delta_{\bm{Q}_0}^y$.

Such a degeneracy implies that the phase boundaries coincide for $\Delta_{\bm{Q}_0}^0$ and $\Delta_{\bm{Q}_0}^z$, as well as for $\Delta_{\bm{Q}_0}^x$ and $\Delta_{\bm{Q}_0}^y$. Numerical evaluation of the coefficients shows that, as temperature decreases, the coefficient $a_0=a_z$ becomes negative first, in a region where $a_x=a_y$ remains positive (see Section \ref{sec:GL_theory_td_non_zero} for details). Consequently, $\Delta^x_{\bm{Q}_0}$ and $\Delta^y_{\bm{Q}_0}$ do not condense from the disordered state. From Eq. (\ref{eq:order_params_pauli}), these order parameters are expressed solely in terms of $\Delta^{p_xp_y}_{\bm{Q}_0}$ and $\Delta^{p_yp_x}_{\bm{Q}_0}$, justifying their omission in Eq.~(\ref{eq:gl_simple}).

While the analysis of the quadratic coefficients implies that both CDW order parameters $\Delta_{\bm{Q}_0}^0$ and $\Delta_{\bm{Q}_0}^z$ can acquire a non-vanishing expectation value, this is not sufficient to establish whether the energy is minimized by forming domains in which only one of the order parameters condense at a time (macroscopic phase separation) or a single phase where both are simultaneously condensed (microscopic coexistence). This distinction is equivalent to asking whether the multi-critical point corresponding to the coincidence of the two phase boundaries is bicritical or tetracritical, respectively (see, e.g., Ref. \cite{Fernandes2010}).
To determine whether $\Delta^0_{\bm{Q}_0}$ and $\Delta^z_{\bm{Q}_0}$ can simultaneously become non-zero everywhere in the system below the critical temperature -- which we refer to as the ``double-$\Delta$''  or coexistence phase -- one must consider the Ginzburg-Landau theory to fourth order in $\Delta^0_{\bm{Q}_0}$ and $\Delta^z_{\bm{Q}_0}$  (for the derivation, see Appendix \ref{sec:deriving_gl}):
\begin{eqnarray}
&& S_{\operatorname{eff}} =
r\left(|\Delta^0_{\bm{Q}_0}|^2 + |\Delta^z_{\bm{Q}_0}|^2\right) \nonumber \\
&& + 
b \left(
|\Delta^0_{\bm{Q}_0}|^2 + |\Delta^z_{\bm{Q}_0}|^2\right)^2 + b \left(
 \bar\Delta^0_{\bm{Q}_0} \Delta^z_{\bm{Q}_0} + 
 \bar\Delta^z_{\bm{Q}_0} \Delta^0_{\bm{Q}_0}
\right)^2. \nonumber \\
&&
\label{eq:d2h_gl_td_zero}
\end{eqnarray}
This is identical to the action (\ref{eq:gl_simple}), as follows from $\Delta^{0,z}_{\bm{Q}_0} = \Delta^{p_x p_x}_{\bm{Q}_0} \pm \Delta^{p_y p_y}_{\bm{Q}_0}$. 
This action is valid in the region of the phase diagram in temperature $(T)$ and interaction strength $(g)$ where both $a_x(T, g) = a_y(T, g) > 0$, and thus $\Delta^x_{\bm{Q}_0} = \Delta^y_{\bm{Q}_0} = 0$, while $a_0(T, g) = a_z(T, g) = r < 0$. It has the following set of \textit{global} minima:
\begin{equation}
     |\Delta^0_{\bm{Q}_0}|^2 + |\Delta^z_{\bm{Q}_0}|^2 = -\frac{r}{2b},\ 
 \bar\Delta^0_{\bm{Q}_0} \Delta^z_{\bm{Q}_0} + 
 \bar\Delta^z_{\bm{Q}_0} \Delta^0_{\bm{Q}_0} = 0.
\label{eq:d2h_gl_td_zero_global_minima}
\end{equation}
The second equation is satisfied when either $\Delta^0_{\bm{Q}_0}$ or $\Delta^z_{\bm{Q}_0}$ is zero, resulting in a single-$\Delta$ solution, or when both $\Delta^0_{\bm{Q}_0}$ and $\Delta^z_{\bm{Q}_0}$ are non-zero with a relative phase $\pm \pi/2$, resulting in a double-$\Delta$ solution. The single-$\Delta$ phases are characterized by either $\Re \Delta^0_{\bm{Q}_0}$ or $\Re \Delta^z_{\bm{Q}_0}$, which transform as $A_g$ and $B_{1g}$, respectively, following Table~\ref{tab:order_param_irreps_d2h}. As we explain below, the global phase can be eliminated by a lattice translation.

Note that the double-$\Delta$ solution has exactly the same free energy as the single-$\Delta$ solutions despite the fact that we extended the analysis of the GL theory to quartic order. This is a consequence of the emergent symmetry of the $t_d = 0$ case. 
We anticipate -- and show in the next section -- that this degeneracy is lifted when $t_d \ne 0$, favoring the double-$\Delta$ solution.

Thus, we now turn to discuss the symmetry of the double-$\Delta$ phase in the $t_d = 0$ case. As temperature is lowered, $\Delta^0_{\bm{Q}_0}$ and $\Delta^z_{\bm{Q}_0}$ may simultaneously acquire expectation values with a relative phase of $\pm\pi/2$. Suppose $\Delta^0_{\bm{Q}_0} = e^{i\alpha}|\Delta^0_{\bm{Q}_0}|$ and $\Delta^z_{\bm{Q}_0} = \pm ie^{i\alpha}|\Delta^z_{\bm{Q}_0}|$. When $\alpha = 0$, the resulting phase is characterized by the two $D_{2h}$-irreps $\Re \Delta^0_{\bm{Q}_0}$ and $\Im \Delta^z_{\bm{Q}_0}$, which transform as $A_{g}$ and $B_{2u}$, respectively. 
\begin{figure*}
    \includegraphics[height=6.0cm]{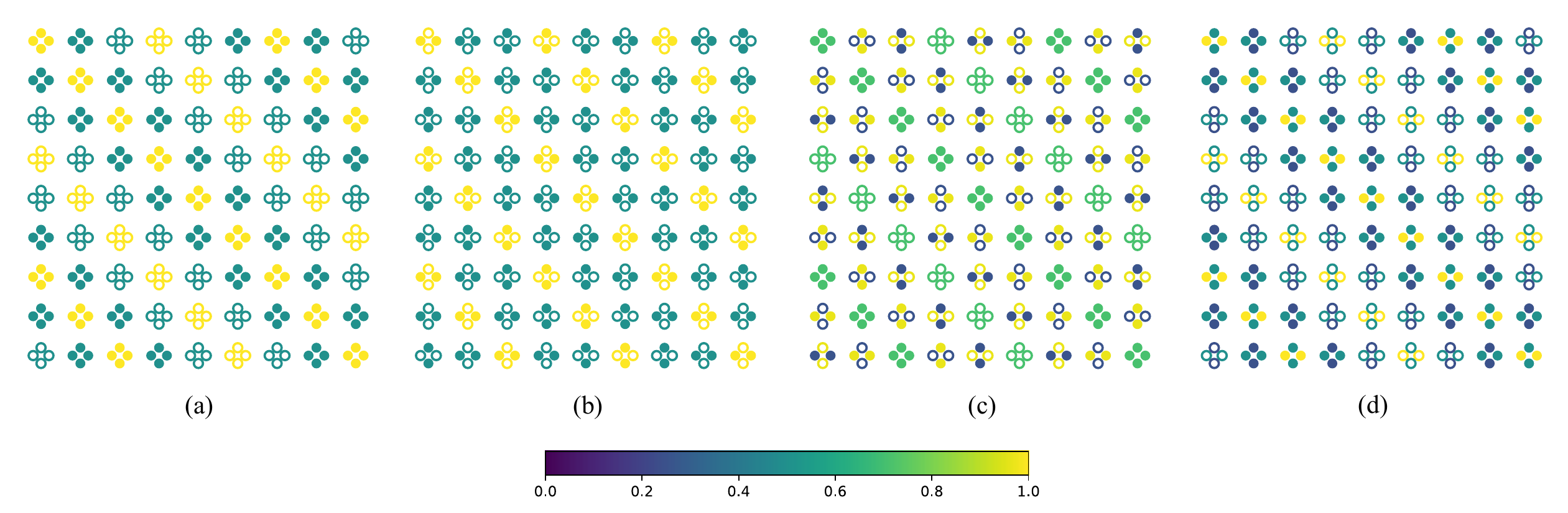}
    \caption{\label{fig:real_space}Real-space plots of the orbital occupations $\rho_{p_x}(\bm{r})$ and $\rho_{p_y}(\bm{r})$ for $\bm{Q} = (\pi/3, \pi/3)$ illustrating different symmetry breaking patterns: (a) Single-$\Delta$ $A_g$: $\rho_{p_y}=\rho_{p_x}=\cos(\bm{Q}\cdot\bm{r})$, no additional symmetries are broken besides the translational and rotational symmetries broken explicitly by $\bm{Q}$.
    (b) Single-$\Delta$ $B_{1g}$: $\rho_{p_y}=-\rho_{p_x}=\cos(\bm{Q}\cdot\bm{r})$, no additional symmetries are broken; 
    (c) Double-$\Delta$ $B_{2u}$: $\rho_{p_y}= \cos(\bm{Q}\cdot\bm{r} - \pi/4)$, $\rho_{p_x}=\cos(\bm{Q}\cdot\bm{r} + \pi/4)$, one of the diagonal mirrors is broken, inversion is broken; (d) Double-$\Delta$ $B_{1g}$: $\rho_{p_y}= 0.5\cos(\bm{Q}\cdot\bm{r})$, $\rho_{p_x}=\cos(\bm{Q}\cdot\bm{r})$, two diagonal mirrors are broken, inversion is preserved. 
    Color indicates the charge density magnitude, while positive and negative density values are represented by filled and empty circles respectively.
    }
\end{figure*}

However, for a generic $\alpha$, we obtain a different set of irreps with respect to $D_{2h}$: $\Re \Delta^0_{\bm{Q}_0}$, $\Im \Delta^0_{\bm{Q}_0}$, $\Re \Delta^z_{\bm{Q}_0} $, and $\Im \Delta^z_{\bm{Q}_0}$ transform as $A_{g}$, $B_{3u}$, $B_{1g}$, and $B_{2u}$, respectively. 
One might worry, then, that the choice of $\alpha$ -- which does not enter the free energy -- changes the symmetry of the resulting phase. As we now show, this is not the case. The irreps obtained for a generic $\alpha$ transform precisely as $A_g$ and $B_{2u}$ for the
``shifted'' point group $D_{2h}^{\bm{t}} = \{\bm{t}^{-1} g \bm{t}\, |\, g \in D_{2h}\}$, 
where $\bm{t}$ is the lattice translation for which $\Delta^i_{\bm{Q}_0}$ acquires the phase shift $\alpha$ according to Eq.~(\ref{eq:symmaction}).
(The point group $D_{2h}^{\bm{t}}$ leaves the lattice point $-\bm{t}$ invariant.)
The space of order parameters $\{\Delta_{\bm{Q}_0}^i, \Delta_{-\bm{Q}_0}^i=\bar \Delta_{\bm{Q}_0}^i\}$ decomposes into irreps of $D_{2h}^{\bm{t}}$ as follows: 
\begin{equation}
\{e^{-i\alpha} \Delta_{\bm{Q}_0}^i + e^{i\alpha} \Delta_{-\bm{Q}_0}^i,\ 
e^{-i\alpha} \Delta_{\bm{Q}_0}^i - e^{i\alpha} \Delta_{-\bm{Q}_0}^i \},
\end{equation}
where the first linear combination belongs to $A_g \otimes \rho_i$, while the second one is in $B_{3u} \otimes \rho_i$. Consequently, for a non-zero phase $\alpha$, we obtain the same irreps, $A_{g}$ and $B_{2u}$, with respect to the shifted point group $D_{2h}^{\bm{t}}$. Specifically, $e^{-i\alpha} \Delta_{\bm{Q}_0}^0 + e^{i\alpha} \Delta_{-\bm{Q}_0}^0$ and $e^{-i\alpha} \Delta_{\bm{Q}_0}^z - e^{i\alpha} \Delta_{-\bm{Q}_0}^z$ transform as $A_g$ and $B_{2u}$ respectively.
Thus, each value of $\alpha$ represents a CDW that transforms as the same irrep with respect to a shifted point group.
Equivalently, we can state that different values of the absolute phase $\alpha$ yield the same representation of the space group. Note that there is a subtlety depending on whether $\bm{Q}_0$ is commensurate or incommensurate: we assume the latter, so that $\alpha$ varies continuously.
In the commensurate case, additional terms in the GL theory will restrict the possible values of $\alpha$ to a finite set.

These results suggest that the resulting double-$\Delta$ phase is characterized by the translationally invariant bilinear order parameter $\bar\Delta^0_{\bm{Q}_0} \Delta^z_{\bm{Q}_0} -
\bar\Delta^z_{\bm{Q}_0} \Delta^0_{\bm{Q}_0} \ne 0$, which transforms as $B_{2u}$ (see Tables \ref{tab:bilinears_irreps_d2h} and \ref{tab:bilinears_irreps_d2h_detailed}). This is the same transformation properties as an in-plane electric field pointing along one of the diagonals, which therefore breaks inversion $I$ and one of the two vertical mirrors, $\sigma_d(y=\pm x)$, see Table \ref{tab:char_tab}.

In our weak-coupling analysis, the quartic coefficients of the GL theory are determined entirely by the non-interacting electronic dispersion. However, coupling to other degrees of freedom can give additional contributions to the quartic terms, which may in fact alter the orbital texture of the CDW phase. We consider specifically the coupling between the CDW and a $B_{1g}$ phonon mode. According to first-principles calculations and experimental data in GdTe$_3$ \cite{chen2019raman}, there are relatively soft $B_{1g}$  phonon modes in the spectrum, with energies ranging from about $5$ meV to about $20$ meV. As we show below, soft modes can give large contributions to the GL coefficients. Denoting the corresponding generalized phonon coordinate as $\Phi_{B_{1g}}$, a new term arises in the GL theory of the form  $\gamma \Phi_{B_{1g}} \left(\bar\Delta^0_{\bm{Q}_0} \Delta^z_{\bm{Q}_0} + \bar\Delta^z_{\bm{Q}_0} \Delta^0_{\bm{Q}_0} \right)$.
Integrating out the phonon effectively reduces the coefficient of $\left(
 \bar\Delta^0_{\bm{Q}_0} \Delta^z_{\bm{Q}_0} + 
 \bar\Delta^z_{\bm{Q}_0} \Delta^0_{\bm{Q}_0}
\right)^2$ in the action (\ref{eq:d2h_gl_td_zero}), yielding:
\begin{eqnarray}
&& S_{\operatorname{eff}} =
r\left(|\Delta^0_{\bm{Q}_0}|^2 + |\Delta^z_{\bm{Q}_0}|^2\right) \nonumber \\
&& + 
b \left(
|\Delta^0_{\bm{Q}_0}|^2 + |\Delta^z_{\bm{Q}_0}|^2\right)^2 + b' \left(
 \bar\Delta^0_{\bm{Q}_0} \Delta^z_{\bm{Q}_0} + 
 \bar\Delta^z_{\bm{Q}_0} \Delta^0_{\bm{Q}_0}
\right)^2 \nonumber \\
&&
\label{eq:d2h_gl_td_zero_phonon}
\end{eqnarray}
While when $t_d = 0$, 
$b'=b$, we will show in the next section that $b'\neq b$ generically when $t_d \neq 0$. In our current case, the phonon contribution gives $b' - b \propto -\gamma^2/\Omega_{\mathrm{ph}}$, where $\Omega_{\mathrm{ph}}$ is the $B_{1g}$ phonon frequency.
The key point is that Eq.~(\ref{eq:d2h_gl_td_zero_phonon}) has the following \textit{local} extremum:
\begin{equation}
|\Delta^0_{\bm{Q}_0}|^2 =  -\frac{r}{4(b+b')},\ 
 \Delta^0_{\bm{Q}_0} = \pm \Delta^z_{\bm{Q}_0},
 \label{eq:d2h_gl_td_zero_phonon_minima}
\end{equation}
at which $S_{\operatorname{eff}}= -r^2/4(b+b')$. This is in addition to the local extremum already obtained from Eq. (\ref{eq:d2h_gl_td_zero_global_minima}), at which $S_{\operatorname{eff}}= -r^2/4b$. 
The global minimum is determined by the magnitudes of $b$ and $b'$; specifically, the new solution in (\ref{eq:d2h_gl_td_zero_phonon_minima}) is  favored when $-b<b'<0$.

In this case, the resulting phase is characterized by the non-zero bilinear $\bar\Delta^0_{\bm{Q}_0} \Delta^z_{\bm{Q}_0} + \bar\Delta^z_{\bm{Q}_0} \Delta^0_{\bm{Q}_0}$, which transforms as $B_{1g}$, i.e., similar to in-plane shear strain. Thus, this phase breaks both mirrors $\sigma_d(y=x)$ and $\sigma_d(y=-x)$, while preserving inversion symmetry (see Table \ref{tab:char_tab}).
The absence of vertical mirrors implies that there is in-plane ``handedness'', which is a direct consequence of the fact that the bilinear $\bar\Delta^0_{\bm{Q}_0} \Delta^z_{\bm{Q}_0} + \bar\Delta^z_{\bm{Q}_0} \Delta^0_{\bm{Q}_0}$ can also be interpreted as an out-of-plane electric toroidal moment (ferroaxial moment) \cite{Cheong2018,Hayami2022}, since the latter also transforms as $B_{1g}$ in the $D_{2h}$ point group. We emphasize that, in the $D_{4h}$ point group, the corresponding bilinear transforms as $A_{2g}$, see Appendix \ref{sec:gl_d4h}, which is also the notation used in Ref. \cite{singh20204ferro}. This phase is consistent with the Raman experimental results of Refs.~\onlinecite{wang2022axial,singh20204ferro}, which identified an axial collective mode for the CDW phase, as well as the SHG results of Ref. \onlinecite{singh20204ferro}, which directly identified mirror symmetry breaking.

We will see in the next section that the quartic term $\left(
 \bar\Delta^0_{\bm{Q}_0} \Delta^z_{\bm{Q}_0} -
 \bar\Delta^z_{\bm{Q}_0} \Delta^0_{\bm{Q}_0}
\right)^2$, which can arise from coupling to a $B_{2u}$ phonon, is also allowed in the GL theory when $t_d\ne0$ and provides another route to a symmetry-broken phase.

We now describe the phases we have found in terms of the orbital-resolved order parameters $\Delta^{p_xp_x}_{\bm{Q}_0} = (\Delta^0_{\bm{Q}_0} + \Delta^z_{\bm{Q}_0})/2$ and $\Delta^{p_yp_y}_{\bm{Q}_0} = (\Delta^0_{\bm{Q}_0} - \Delta^z_{\bm{Q}_0})/2$, defined in Eq.~(\ref{eq:order_params_matrix_els}).
The single-$\Delta$ solution with $\Delta^0_{\bm{Q}_0} \ne 0$ and $\Delta^z = 0$ implies that $\Delta^{p_x p_x}_{\bm{Q}_0}=\Delta^{p_y p_y}_{\bm{Q}_0}$, while the single-$\Delta$ solution with $\Delta^0_{\bm{Q}_0}=0$ and $\Delta^z_{\bm{Q}_0}\neq 0$ implies $\Delta^{p_x p_x}_{\bm{Q}_0}=-\Delta^{p_y p_y}_{\bm{Q}_0}$. 
In the $B_{2u}$ double-$\Delta$ phase that also follows from Eq. (\ref{eq:gl_simple}), where both $\Delta^0_{\bm{Q}_0}$ and $\Delta^z_{\bm{Q}_0}$ are non-zero and their relative phase is $\pm \pi/2$, $\Delta^{p_xp_x}_{\bm{Q}_0}$ and $\Delta^{p_yp_y}_{\bm{Q}_0}$ have equal magnitude, but the phase between them is arbitrary.
On the other hand, in the $B_{1g}$ double-$\Delta$ phase obtained from Eq. (\ref{eq:d2h_gl_td_zero_phonon}), where the relative phase between $\Delta^0_{\bm{Q}_0}$ and $\Delta^z_{\bm{Q}_0}$ is trivial, $\Delta^{p_x p_x}_{\bm{Q}_0}$ and $\Delta^{p_y p_y}_{\bm{Q}_0}$ have the same phase but different magnitudes. Therefore, the double-$\Delta$ phases display a non-trivial orbital texture determined by the relative phase and relative magnitude between the two CDW order parameters. As we will discuss
in Sec. \ref{sec:experiment}, this orbital texture is the reason why, despite sharing the same irreducible representation, the $B_{1g}$ single-$\Delta$ phase and the $B_{1g}$ double-$\Delta$ phase exhibit distinct experimental signatures.

Given that $\Delta^{p_x p_x}_{\bm{Q}} = |\Delta^{p_x p_x}_{\bm{Q}}|e^{i\alpha_{p_x}}$ and $\Delta^{p_y p_y}_{\bm{Q}} = |\Delta^{p_y p_y}_{\bm{Q}}|e^{i\alpha_{p_y}}$, the corresponding onsite electronic occupations of the $p_x$- and $p_y$-orbital are given by:
\begin{eqnarray}
    && \rho_{p_x}(\bm{r}) = |\Delta^{p_x p_x}_{\bm{Q}}| \cos \left(
    \bm{Q} \cdot \bm{r} + \alpha_{p_x}
    \right), \nonumber \\
    && \rho_{p_y}(\bm{r}) = |\Delta^{p_y p_y}_{\bm{Q}}| \cos \left(
    \bm{Q} \cdot \bm{r} + \alpha_{p_y}
    \right).
    \label{eq:charge_densities}
\end{eqnarray}
We present real-space plots of the orbital occupations $\rho_{p_x}(\bm{r})$ and $\rho_{p_y} (\bm{r})$ for each of the four phases in Fig. {\ref{fig:real_space}}, highlighting the non-trivial orbital texture, where we take $\bm{Q} = (\pi/3, \pi/3)$ for clarity.

To summarize, the $t_d = 0$ case illustrates our main point: CDWs with non-trivial orbital character must be considered on the same footing as the CDW that is trivial in orbital space.
Although for $t_d=0$ the single-$\Delta$ and double-$\Delta$ phases are energetically equivalent, if we assume that the double-$\Delta$ phase is favored by additional degrees of freedom, we obtain a CDW phase that breaks additional symmetries beyond those broken explicitly by the wave vector $\bm{Q}_0$.

\subsection{\label{sec:GL_theory_td_non_zero} Case $t_d \ne 0$}
\begin{figure*}
    \centering
    \includegraphics[height=5.5cm]{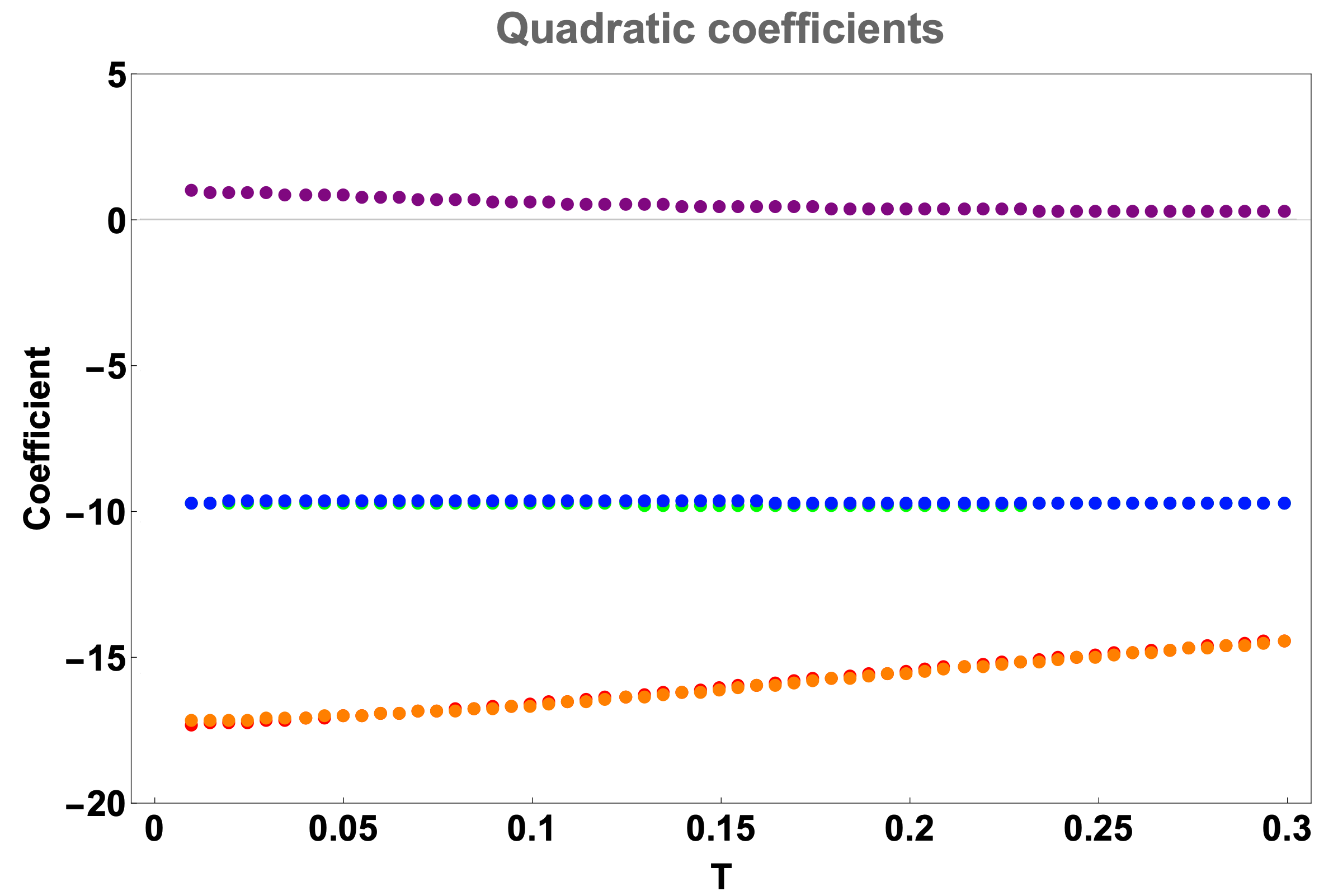}
    \includegraphics[height=5.5cm]{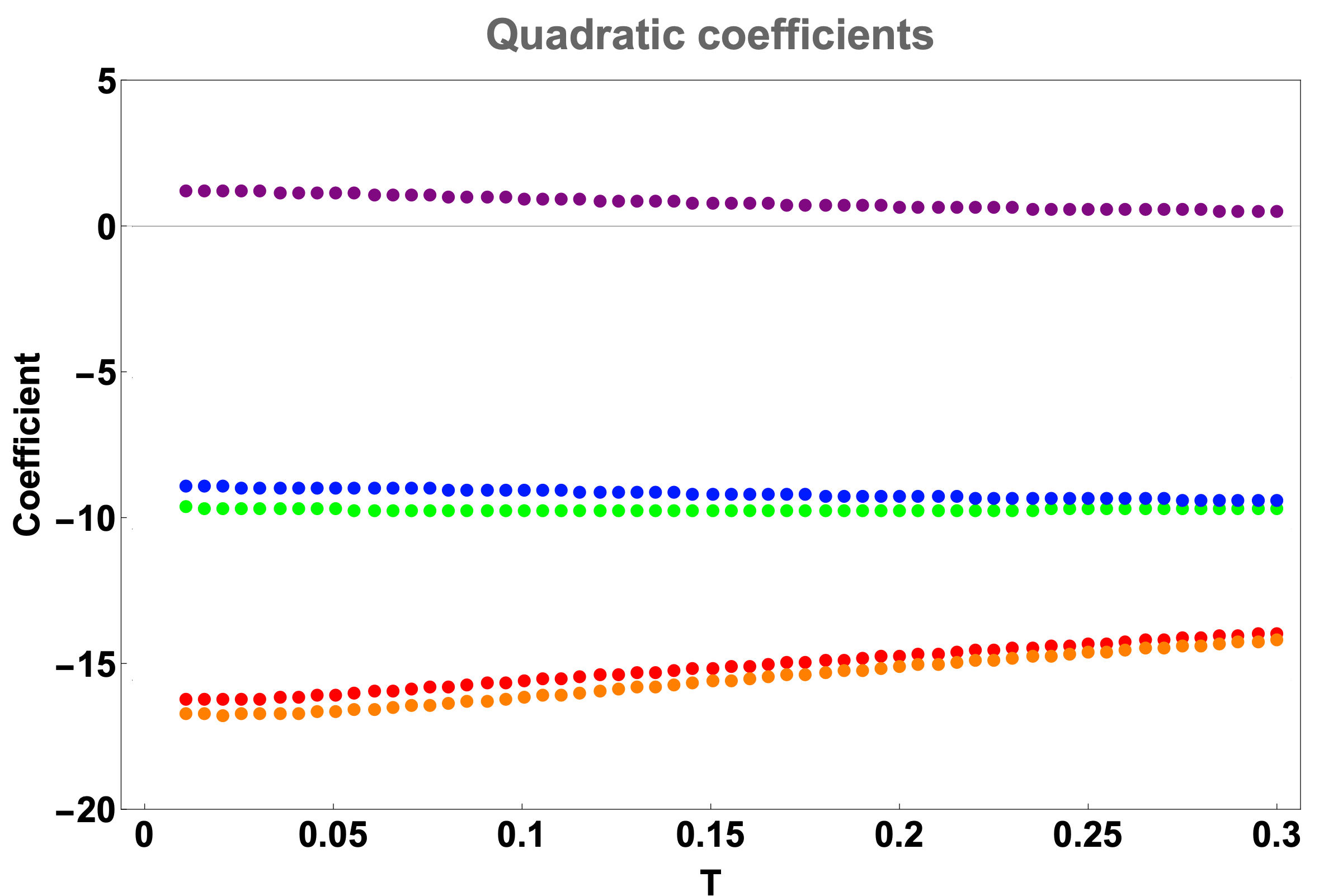}
    \includegraphics[height=4.8cm]{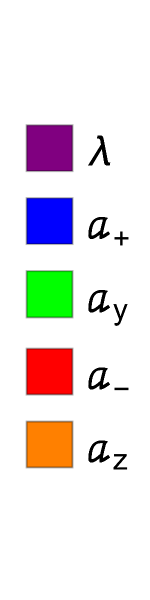}
    \caption{Quadratic coefficients $a_i-2/g$ in the Ginzburg-Landau theory as functions of temperature $T$ (in eV) at $Q=(2k_F, 2k_F)$, for $a_- \approx a_0$ (red), $a_z$ (orange), $a_+\approx a_x$ (blue), $a_y$ (green), and $\lambda$ (purple); $\tilde \lambda = 0$ exactly. Left: $t_d=0.16$eV, corresponding to the realistic value of $t_d$ quoted in Ref. ~\onlinecite{PhysRevB.74.245126}.  Right: $t_d=0.4$eV, much larger than the expected value.
    For both values of $t_d$, $a_-$ and $a_z$, as well as $a_+$ and $a_y$, are nearly degenerate and $|\lambda| \ll |a_i|$.}
    \label{fig:quadratic_coeff}
\end{figure*}

We now turn to the more general case where $t_d \neq 0$.
The GL theory contains all translationally invariant bilinears transforming in the trivial representation of $D_{2h}$ in Table \ref{tab:bilinears_irreps_d2h}.
To quadratic order:
\begin{eqnarray}
&&S_{\operatorname{eff}}^{(2)}=
\begin{pmatrix}
    \bar\Delta_{\bm{Q}_0}^0 &
    \bar\Delta_{\bm{Q}_0}^x 
\end{pmatrix}
\begin{pmatrix}
    a_0 & \lambda \\
    \lambda & a_x
\end{pmatrix}
\begin{pmatrix}
    \Delta_{\bm{Q}_0}^0 \\
    \Delta_{\bm{Q}_0}^x 
\end{pmatrix} \nonumber \\
&&+ \begin{pmatrix}
    \bar\Delta_{\bm{Q}_0}^y &
    \bar\Delta_{\bm{Q}_0}^z 
\end{pmatrix}
\begin{pmatrix}
    a_y & \tilde \lambda \\
    \tilde \lambda & a_z 
\end{pmatrix}
\begin{pmatrix}
    \Delta_{\bm{Q}_0}^y \\
    \Delta_{\bm{Q}_0}^z
\end{pmatrix},
\label{eq:gl_d2h}
\end{eqnarray}
where the coefficients may be written in terms of the susceptibilities (\ref{eq:susceptibility_components}) (see Appendix \ref{sec:deriving_gl} for the derivation):
\begin{eqnarray}
&&a_{0/z} = \frac{2}{g} - \Big[ 
\left(
\chi_{11}^{11} + \chi_{22}^{22}
\right)
\pm
\left(
\chi_{11}^{22}+\chi_{22}^{11}
\right) \Big], \nonumber \\
&& a_{x/y} = \frac{2}{g} - \Big[ 
\left(
\chi_{12}^{12} + \chi_{21}^{21}
\right)
\pm
\left(
\chi_{12}^{21}+\chi_{21}^{12}
\right)\Big].
\label{eq:coeff_susceptibilities_diag}
\end{eqnarray}
The generic presence of the off-diagonal terms with coefficients $\lambda$ and $\tilde{\lambda}$ is expected, since, according to Table \ref{tab:order_param_irreps_d2h}, the CDW order parameters $\Delta_{\bm{Q}_0}^0$ and $\Delta_{\bm{Q}_0}^x$  transform as the same irrep; similarly,  $\Delta_{\bm{Q}_0}^z$ and $\Delta_{\bm{Q}_0}^y$ also transform as the same irrep. Computing these off-diagonal coefficients (see Eq.~(\ref{eq:quadratic_coeffs_greens_fn}) in the Appendix), we find that the off-diagonal coefficient $\tilde\lambda$ vanishes for this simplified band dispersion, while the coefficient $\lambda$ is given by:
\begin{eqnarray}
    \lambda = - \left( \chi_{12}^{11}+
    \chi_{21}^{22}+
    \chi_{21}^{11}+
    \chi_{12}^{22}\right).
\label{eq:coeff_susceptibilities_offdiag}
\end{eqnarray}
Recall that susceptibility matrix elements of the form $\chi^{\beta \gamma}_{\alpha \delta}$ with $\alpha \neq \beta$ or $\delta \neq \gamma$ vanish when $t_d =0 $. Thus, since $t_d$ is small compared to the other couplings, $a_0$ and $a_z$ (as well as $a_x$ and $a_y$) are similar in magnitude.
Specifically, $a_z - a_0 = 2\left( \chi_{11}^{22}+\chi_{22}^{11}\right)$ and $a_y - a_x = 2\left(\chi_{12}^{21}+\chi_{21}^{12}\right)$, 
which are the sub-dominant components of the susceptibility, as discussed below Eq.~(\ref{eq:susceptibility_components}).
For the same reason, the off-diagonal coefficient $\lambda$ is much smaller than any of the diagonal coefficients $a_i$.

To obtain the phase diagram, we first diagonalize the effective quadratic action (\ref{eq:gl_d2h}):
\begin{eqnarray}
&&S_{\operatorname{eff}}^{(2)}=
\begin{pmatrix}
    \bar\Delta_{\bm{Q}_0}^+ &
    \bar\Delta_{\bm{Q}_0}^- 
\end{pmatrix}
\begin{pmatrix}
    a_+ & 0 \\
    0 & a_-
\end{pmatrix}
\begin{pmatrix}
    \Delta_{\bm{Q}_0}^+ \\
    \Delta_{\bm{Q}_0}^- 
\end{pmatrix} \nonumber \\
&&+ \begin{pmatrix}
    \bar\Delta_{\bm{Q}_0}^y &
    \bar\Delta_{\bm{Q}_0}^z 
\end{pmatrix}
\begin{pmatrix}
    a_y & 0 \\
    0 & a_z 
\end{pmatrix}
\begin{pmatrix}
    \Delta_{\bm{Q}_0}^y \\
    \Delta_{\bm{Q}_0}^z
\end{pmatrix},
\label{eq:quad_action_diag}
\end{eqnarray}
where $a_{\pm} = \big(a_0 + a_x \pm \sqrt{(a_0-a_x)^2 + 4 \lambda^2}\big)/2$ are eigenvalues in the $\Delta^{0/x}_{\bm{Q}_0}$-sector and $\bar\Delta_{\bm{Q}_0}^\pm$ are the corresponding linear combinations of $\Delta^{0}_{\bm{Q}_0}$ and $\Delta^{x}_{\bm{Q}_0}$. 
Let us also assume here that $a_0 < a_x$ and $\lambda \ll a_x - a_0$ so that $a_- \approx a_0$ and $a_+ \approx a_x$ for small values of $t_d$. These assumptions are supported by numerical evaluation of the coefficients in Eq.~(\ref{eq:quad_action_diag}), as illustrated in Fig. \ref{fig:quadratic_coeff}. Note that in this figure we plot $(a_i - 2/g)$, which implies that the most negative $(a_i - 2/g)$ corresponds to the leading instability of the system, which takes place when the condition $a_i=0$ is met.
Since $a_0 \approx a_z$, it follows that $a_- \approx a_z$ as long as $t_d$ is small.
This implies similar critical temperatures for $\Delta^-_{\bm{Q}_0}$ and $\Delta^z_{\bm{Q}_0}$, which become coincident in the $t_d \rightarrow 0$ limit. (In this limit, $\Delta^-_{\bm{Q}_0} \to \Delta^0_{\bm{Q}_0}$ and $\Delta^+_{\bm{Q}_0} \to \Delta^x_{\bm{Q}_0}$.) 

To analyze which (if any) coexistence state between the  $\Delta^-$  and  $\Delta^z$  CDWs emerges in this model, we proceed to analyze the action to quartic order. The irreducible representations of the $\Delta$-bilinears in Table \ref{tab:bilinears_irreps_d2h_detailed_pm_basis} yield the allowed quartic terms in the GL theory:

\begin{eqnarray}
&&S_{\operatorname{eff}}^{(4)} =
b_-
|\Delta^-_{\bm{Q}_0}|^4 + b_z|\Delta^z_{\bm{Q}_0}|^4 + 
c'\left(
 \bar\Delta^-_{\bm{Q}_0} \Delta^z_{\bm{Q}_0} +
 \bar\Delta^z_{\bm{Q}_0} \Delta^-_{\bm{Q}_0} \right)^2
 \nonumber \\
&&- c''\left(
 \bar\Delta^-_{\bm{Q}_0} \Delta^z_{\bm{Q}_0} -
 \bar\Delta^z_{\bm{Q}_0} \Delta^-_{\bm{Q}_0} \right)^2.
 \label{eq:quartic_gl_d2h}
\end{eqnarray}
This action is written such that each term is the square of an irreducible representation of the group $D_{2h}$. Note that in the action (\ref{eq:d2h_gl_td_zero}) for the $t_d=0$ case, $b_- = b_z$, $c' = 3b_z/2$, $c'' = b_z/2$ (see Appendix \ref{sec:deriving_gl}). To formulate the criterion for the double-$\Delta$ phase, it is more convenient to rewrite this action as:
\begin{eqnarray}
&&S_{\operatorname{eff}}^{(4)} =  
b_- |\Delta^-_{\bm{Q}_0}|^4 + b_z |\Delta^z_{\bm{Q}_0}|^4  \nonumber \\
&&+2\left[c' + c'' + (c'-c'')\cos 2\delta \alpha\right] \, |\Delta^-_{\bm{Q}_0}|^2 |\Delta^z_{\bm{Q}_0}|^2, \nonumber \\
&&
\label{eq:eff_action_cross_term}
\end{eqnarray}
where we allow for complex order parameters $\Delta^-_{\bm{Q}_0} = |\Delta^-_{\bm{Q}_0}| e^{i\alpha_-}$, $\Delta^z_{\bm{Q}_0} = |\Delta^z_{\bm{Q}_0}| e^{i\alpha_z}$ and $\delta \alpha = \alpha_z - \alpha_-$ is the relative phase. 
A double-$\Delta$ solution is energetically favorable when the cross-term coefficient is sufficiently small (assuming $b_i > 0$):
\begin{equation}
\left[(c' + c'') - |c'-c''|\right]^2  \le b_- b_z.
\label{eq:double_delta_criterion}
\end{equation}
If the equality holds, single-$\Delta$ and double-$\Delta$ solutions have the same free energy to this order. The type of coexistence double-$\Delta$ state, i.e. the value of the relative phase $\delta \alpha$, depends on the sign of $c'-c''$':
\begin{eqnarray}
&&\delta \alpha = \pm \frac{\pi}{2} \quad \text{if}\quad c' > c'' \nonumber, \\
&&\delta \alpha = 0, \, \pi \quad \text{if}\quad c' < c''.
 \label{eq:ccondition}
\end{eqnarray}
In the first case , the bilinears $\bar\Delta^-_{\bm{Q}_0} \Delta^z_{\bm{Q}_0} -\bar\Delta^z_{\bm{Q}_0} \Delta^-_{\bm{Q}_0} \propto \sin 2\alpha \ne 0$ and $\bar\Delta^-_{\bm{Q}_0} \Delta^z_{\bm{Q}_0} +\bar\Delta^z_{\bm{Q}_0} \Delta^-_{\bm{Q}_0} \propto \cos 2\alpha = 0$, while in the second case the situation is reversed.
As explained in the previous section, these bilinears characterize the type of symmetry breaking in the double-$\Delta$ phase, with the first bilinear transforming as $B_{2u}$ and the second as $B_{1g}$.

\begin{figure}
\includegraphics[height=5.5cm]
{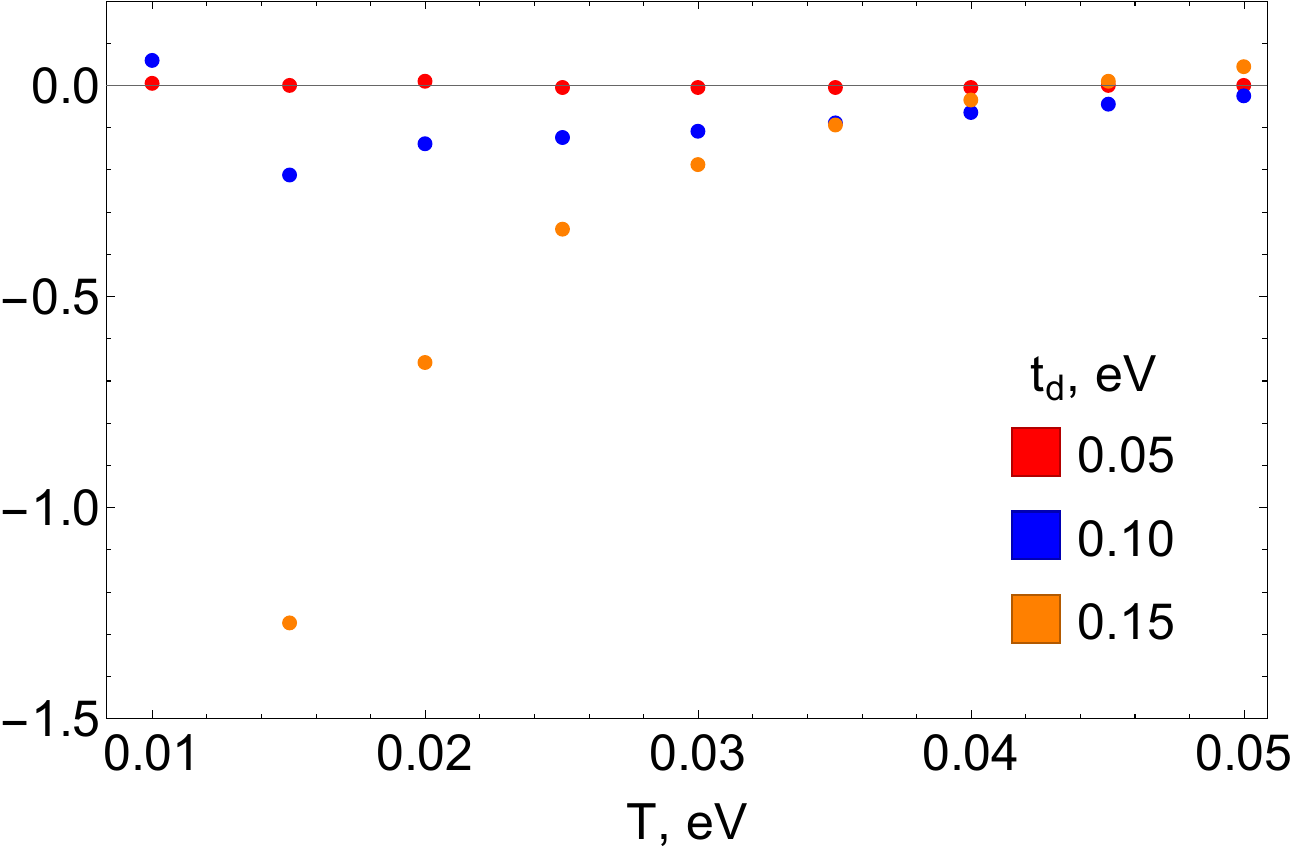}
\includegraphics[height=5.5cm]
{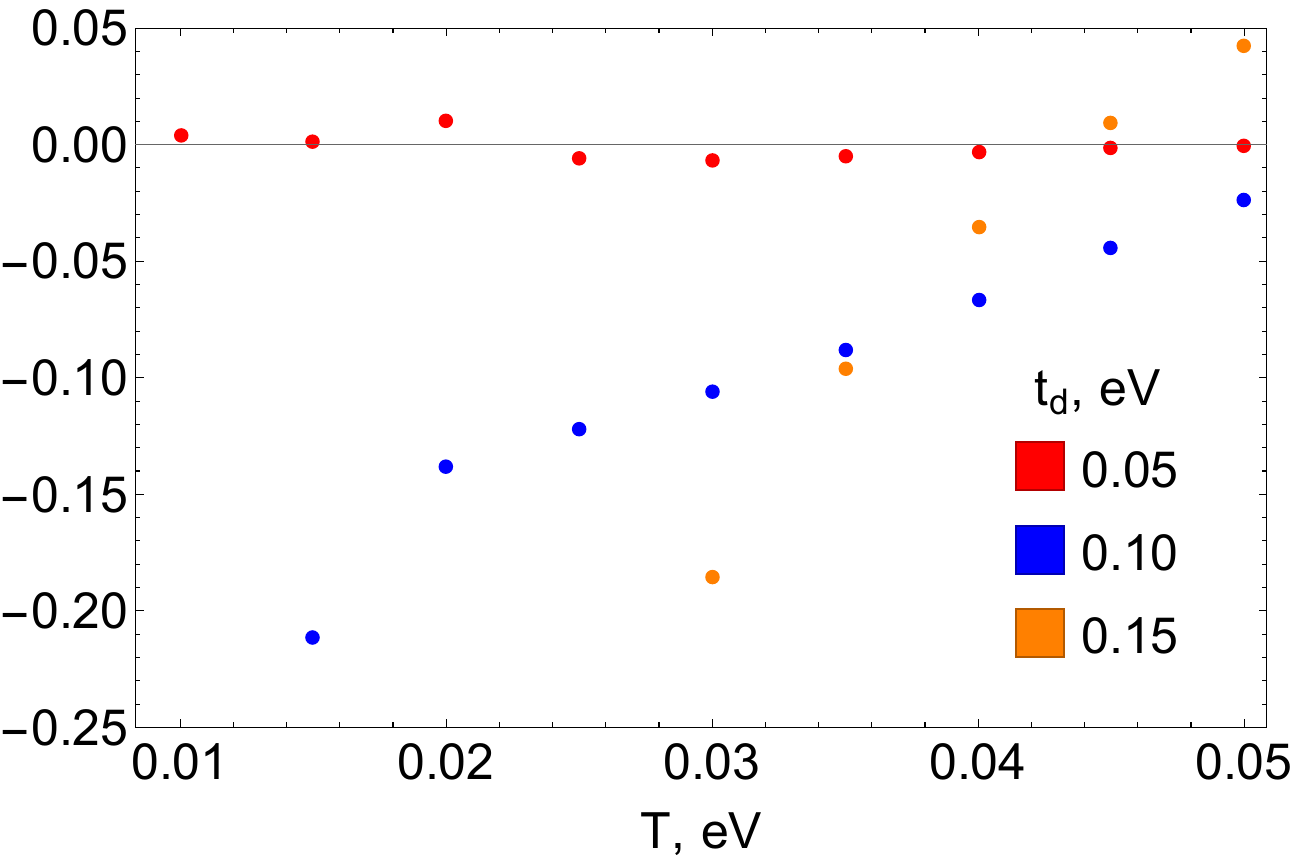}
    \caption{$(4c''^2 - b_-b_z)/4c''^2$ as a function of temperature $T$ (in eV) for $t_d = 0.05$eV (red), $t_d=0.10$eV (blue), $t_d=0.15$eV (orange). 
    When this quantity is negative, the double-$\Delta$ phase with $\delta\alpha = \pi/2$ is favored over the single-$\Delta$ phase. The lower plot displays a zoomed-in region ranging from -0.25 to 0.05. The coefficients $b_-$ and $b_z$ are positive in all cases.
} 
    \label{fig:DD_crit}
\end{figure}

Numerical evaluation of $c'$ and $c''$ reveals that $c' > c''$ and $c' + c'' > 0$ for small $t_d$ and most temperature values (see Fig. \ref{fig:c_sum_diff} in Appendix \ref{sec:deriving_gl}), 
which, from Eq~(\ref{eq:ccondition}), implies $\delta\alpha = \pm \pi/2$.
To verify that the coexistence condition in (\ref{eq:double_delta_criterion}) is simultaneously satisfied requires checking whether $4c''^2 <b_-b_z$, since $c' > c''$. Fig. \ref{fig:DD_crit} shows that this condition is satisfied for small $t_d$ across a wide range of temperatures, up to approximately $0.05\text{eV}$ $\approx 580$ K. 
For very small $t_d$, the condition is not satisfied at very low temperatures, but more realistic values of $t_d$, such as $t_d =0.15$eV, do not show this low-temperature behavior.
This result implies that the double-$\Delta$ phase is characterized by a non-zero $B_{2u}$-bilinear
$\bar\Delta^-_{\bm{Q}_0} \Delta^z_{\bm{Q}_0} - \bar\Delta^z_{\bm{Q}_0} \Delta^-_{\bm{Q}_0}$.

However, as discussed in the previous section for the $t_d=0$ case, these coefficients are derived assuming a purely weak-coupling electronic instability of the Fermi surface.
Introducing a phonon-CDW coupling changes the coefficients in the theory, altering the relation between $c'$ and $c''$ in (\ref{eq:quartic_gl_d2h}) by reducing $c'$ in a way analogous to Eq. \ref{eq:d2h_gl_td_zero_phonon}.
Such a contribution can then drive a double-$\Delta$ solution with $\delta \alpha = 0$, resulting in a coexistence phase characterized by the non-zero $B_{1g}$-bilinear $\bar\Delta^-_{\bm{Q}_0}\Delta^z_{\bm{Q}_0} + \bar\Delta^z_{\bm{Q}_0} \Delta^-_{\bm{Q}_0}$.
 
Let us now briefly return to the remaining order parameters $\Delta^y_{\bm{Q}_0} $ and $\Delta^+_{\bm{Q}_0}$, the latter being the other linear combination of $\Delta^{0,x}_{\bm{Q}_0}$, defined below Eq.~(\ref{eq:quad_action_diag}). 
Fig.~\ref{fig:quadratic_coeff} shows that for realistic parameters,  $\Delta^y_{\bm{Q}_0} $ and $\Delta^+_{\bm{Q}_0}$ do not acquire non-zero values immediately below the transition temperature; whether a second phase transition at an even lower temperature could take place is beyond the scope of our GL theory.
We emphasize that the transition temperature for a given order parameter is defined by the intersection of the corresponding quadratic coefficient curve in Fig. \ref{fig:quadratic_coeff} with the straight horizontal line  $-2/g$ set by the interaction.
In the cases of LaTe$_3$ and GdTe$_3$, 
the transition temperatures are given by
$T_{\text{CDW}}^{\text{Gd}} = 380$K and $T_{\text{CDW}}^{\text{La}} = 650$K,
which roughly correspond to $T = 0.03$eV and $T = 0.06$eV in Fig.~\ref{fig:quadratic_coeff}.
For the corresponding value of $g$, $a_{+,y}-2/g$ are always positive; hence, 
$\Delta^y_{\bm{Q}_0} $ and $\Delta^+_{\bm{Q}_0}$ remain zero.
Consequently, we may safely disregard these two order parameters, assuming they are zero in the region of the phase diagram under consideration.
Finally, since the tight-binding model in Eq.~(\ref{eq:non-int_h}) has $D_{4h}$ symmetry, for completeness, in Appendix \ref{sec:D4h_irreps} we derive the GL theory up to quadratic order for the point group $D_{4h}$ and show how to obtain the solutions with the same symmetry-breaking patterns described in this section in the broader $D_{4h}$ framework.

\section{\label{sec:experiment}Experimental Manifestations}

We now briefly discuss two experimental signatures of the coexistence CDW states with non-trivial orbital texture obtained in our model. Recent Raman experiments on GdTe$_3$, LaTe$_3$, and ErTe$_3$ show evidence that the collective CDW mode has axial symmetry \cite{wang2022axial,singh20204ferro}, whereas SHG measurements on LaTe$_3$ observe the lack of vertical mirrors in the CDW state. While in the parent $D_{4h}$ group this implies the existence of an order parameter with $A_{2g}$ symmetry, in the lower $D_{2h}$ group it corresponds to an order parameter with $B_{1g}$ symmetry. Indeed, $B_{1g}$ has the same transformation properties as rotations with respect to the $z$-axis, such that a $B_{1g}$ order parameter must be axial and break all vertical mirrors.
We now argue that this experimental result is only consistent with the double-$\Delta$, and not the single-$\Delta$ $B_{1g}$ phase.
Specifically, in the single-$\Delta$ $B_{1g}$ phase, the amplitude of the CDW is odd under mirror symmetry, but since the light intensity measures amplitude squared, the Raman measurement remains mirror-symmetric. In other words, a secondary zone-center mode with $B_{1g}$ symmetry is not induced in the single-$\Delta$ phase, and the system remains orthorhombic and non-axial.
In contrast, in the double-$\Delta$ phase, where both the $A_g$ and $B_{1g}$ CDW order parameters are present simultaneously, the squared amplitude contains interference terms that break mirror symmetry, allowing for the scattered light intensity to reflect the mirror symmetry breaking of the CDW phase. In other words, the admixture between $\Delta^0_{\bm{Q}_0}$ and $\Delta^z_{\bm{Q}_0}$ leads to a secondary zone-center mode with $B_{1g}$ symmetry, which is given precisely by the bilinear $\bar\Delta^-_{\bm{Q}_0} \Delta^z_{\bm{Q}_0} +\bar\Delta^z_{\bm{Q}_0} \Delta^-_{\bm{Q}_0} \propto \cos 2\alpha$, lowering the symmetry of the system to monoclinic.
Thus, the double-$\Delta$ $B_{1g}$ phase is the only phase in our theory consistent with the Raman, TEM, and SHG results reported in Refs.~\cite{wang2022axial,singh20204ferro}. 

\begin{figure}
\includegraphics[height=6.8cm]
{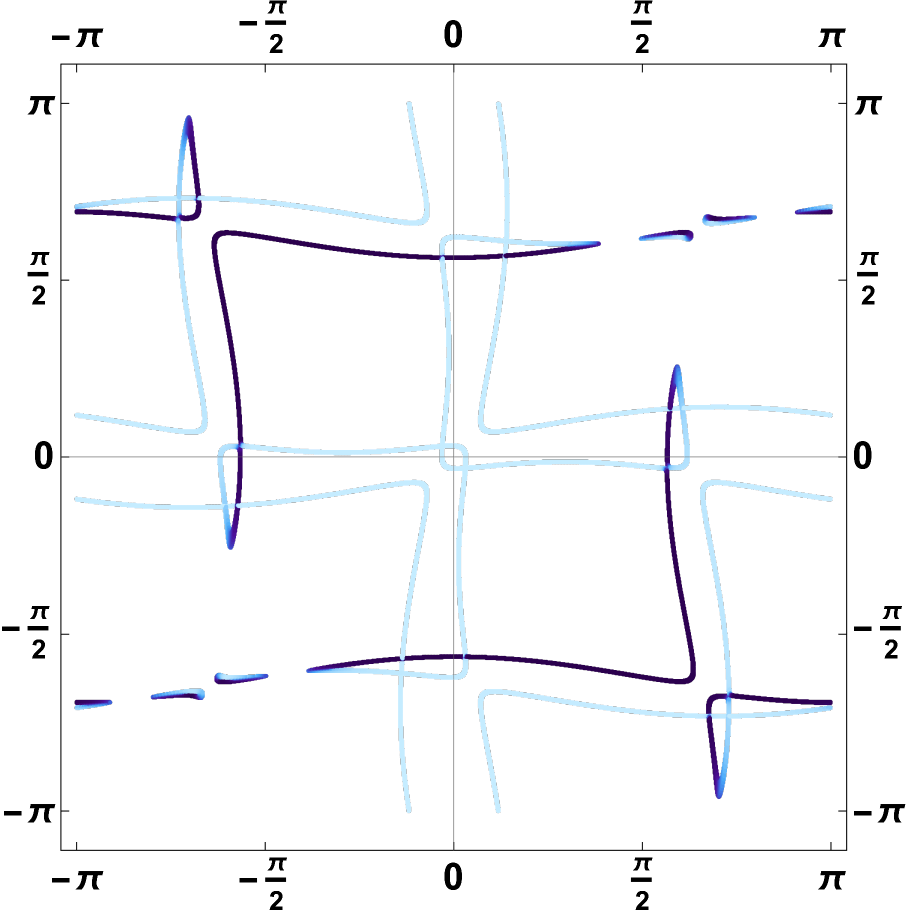}
\includegraphics[height=0.8cm]
{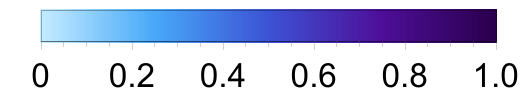}
    \caption{Fermi surface in the $B_{1g}$ double-$\Delta$ phase with $\Delta^0_{\bm{Q}_0}=0.1$ and $\Delta^z_{\bm{Q}_0}=0.05$, showing that both the diagonal mirrors are broken, while inversion is preserved. Here $t_d = 0.16$eV, $\mu= 1.53$eV. The color intensity indicates the spectral weight in the first Brillouin zone.
    } 
    \label{fig:fs_cdw}
\end{figure}

Angle-resolved photoemission spectroscopy,  by measuring the Fermi surface, provides another probe of the additional symmetry broken by the double-$\Delta$ phase.
Fig. \ref{fig:fs_cdw} demonstrates the distortion of the reconstructed Fermi surface in the $B_{1g}$ double-$\Delta$ phase, where both diagonal mirrors $\sigma_d$ are broken; the color code denotes the spectral weight intensity in the first Brillouin zone. 
In contrast, in the
$B_{1g}$ single-$\Delta$ phase, the Fermi surface remains inversion and mirror symmetric, similar to the $A_g$ single-$\Delta$ phase, see Fig. \ref{fig:fs_cdw_multiple} in Appendix \ref{sec:fs}. The Fermi surface also appears symmetric in the $B_{2u}$ double-$\Delta$ phase due to time-reversal symmetry, but would reveal inversion-breaking spin-splitting (i.e., a Rashba-like splitting) in the presence of spin-orbit coupling. Details of how the Fermi surface plots are obtained are provided in Appendix \ref{sec:fs}. 

\section{\label{sec:discussion} Discussion}

In this paper, we introduced a GL theory to describe CDWs with non-trivial orbital character in ``square net'' rare earth tritellurides.
Our theory shows that CDW ground states with non-trivial orbital textures are viable ground states for reasonable system parameters, owing to an emergent degeneracy of the microscopic model in the limit of decoupled $p_x$, $p_y$ orbitals.
These ground states are ``unconventional'' CDWs in the sense that they break crystal symmetries beyond those explicitly broken by the CDW $\bm{Q}$-vector. 

We identify four distinct phases, listed in Table~\ref{tab:solutions}, each characterized by a unique symmetry-breaking pattern. Two phases are characterized by the condensation of a single non-zero CDW order parameter transforming as the $A_{g}$ or $B_{1g}$ irreducible representations of $D_{2h}$ (i.e., either $\Delta^0_{\bm{Q}}$ or $\Delta^z_{\bm{Q}}$ is non-zero, respectively), where $\Delta^{p_x p_x}_{\bm{Q}}=\Delta^{p_y p_y}_{\bm{Q}}$ or $\Delta^{p_x p_x}_{\bm{Q}}=-\Delta^{p_y p_y}_{\bm{Q}}$. In a realistic system, domains of these two different states are likely to form, but in each domain, only one of the order parameters is non-zero, resulting in macroscopic phase separation.
In the other two phases, both $\Delta^0_{\bm{Q}}$ and $\Delta^z_{\bm{Q}}$ condense everywhere in the system, resulting in a microscopic coexistence state and promoting a non-trivial orbital texture that breaks additional symmetries in the CDW phase. In one case ($B_{1g}$ double-$\Delta$), the complex order parameters $\Delta^{p_x p_x}_{\bm{Q}}$ and $\Delta^{p_y p_y}_{\bm{Q}}$ have the same phase but different magnitudes, while in the second case ($B_{2u}$ double-$\Delta$) they have equal magnitude and phases that differ by $\pi/2$. 

The broken symmetries in each phase are as follows: both the $A_{g}$ and the $B_{1g}$ single-$\Delta$ phases share the same point group $D_{2h}$,  which is determined by the symmetries broken by the wave-vector $\bm{Q}$. The reason why the $B_{1g}$ single-$\Delta$ phase does not have a lower-symmetry point group with respect to the $A_{g}$ single-$\Delta$  phase is because the order parameter has finite momentum $\bm{Q}$ (recall that these are not irreps of the space group, but of the little group that leaves $\bm{Q}$ invariant or rotates $\bm{Q} \rightarrow - \bm{Q}$). This is analogous to the fact that a $d$-wave superconductor on a tetragonal lattice displays the same point group as an $s$-wave superconductor despite their order parameters transforming as different irreps. In contrast, the two double-$\Delta$ phases display point groups with lower symmetries as manifested by the fact that they support non-trivial zero-momentum bilinears $\bar\Delta^0_{\bm{Q}} \Delta^z_{\bm{Q}} \pm \bar\Delta^z_{\bm{Q}} \Delta^0_{\bm{Q}}$. This is analogous to the fact that a superconductor with coexisting $s$-wave and $d$-wave gaps will break tetragonal symmetry. In particular, the $B_{2u}$ double-$\Delta$ phase additionally breaks inversion and one of the diagonal mirrors $\sigma_d$, resulting in the orthorhombic point group $C_{2v}$. 
Conversely, the $B_{1g}$ double-$\Delta$ phase preserves inversion while breaking both diagonal mirrors, resulting in the monoclinic point group $C_{2h}$.

Which phase is favored depends on the coefficients of the GL theory. These coefficients, in turn, are defined by the underlying model and interactions considered.
We showed that a nesting-driven Fermi surface instability results in the inversion-breaking $B_{2u}$ double-$\Delta$ phase, while coupling to phonon modes can drive the system into the inversion-preserving ferroaxial $B_{1g}$ double-$\Delta$ phase. Note that, in our weak-coupling approach, the higher-order coefficients of the GL theory are determined solely by the band structure. However, interaction terms also contribute to these coefficients. Once interactions become stronger, these contributions may overcome those originating from the band structure and change the selected ground state, as shown for instance in the case of the spin density-wave instabilities of iron-pnictide superconductors \cite{Fernandes2015}.

Our theoretical framework provides insight into recent experiments showing evidence for unconventional CDW in $R$Te$_3$ ~\cite{wang2022axial, singh20204ferro}, particularly mirror symmetry-breaking.
Our work also motivates detailed first-principles studies to uncover the particular mechanism underlying the CDW in members of the $R$Te$_3$ family. It further lays the foundation for exploring the interplay between CDWs and magnetism in this family of materials \cite{pfuner2011incommensurate,lei2019charge,okuma2020fermionic,lei2021complex,raghavan2023atomic}.
More broadly, our work reveals a mechanism by which unconventional CDW states arise when order parameters with different orbital characters condense simultaneously. Interestingly, a somewhat related mechanism has been discussed in the context of the recently discovered kagome superconductors, attributing the breaking of time-reversal symmetry inside the CDW phase to the coexistence of different CDW-like order parameters \cite{Christensen2022,Xing2023}.

\begin{acknowledgments}
The authors are grateful for conversations and discussion of unpublished work with Kenneth Burch, Leslie Schoop, Birender Singh, and Ming Yi, as well as for conversations with J{\"o}rn Venderbos during the early stages of this work. 
This work was supported by the Air Force Office of Scientific Research under Grants No. FA9550-20-1-0260 (S.A.A.G., and J.C.) and FA9550-21-1-0423 (R.M.F.). J.C. is partially supported by the Alfred P. Sloan Foundation through a Sloan Research Fellowship. The Flatiron Institute is a division of the Simons Foundation. 
\end{acknowledgments}

\appendix
\section{\label{sec:deriving_gl}Derivation of Ginzburg-Landau theory}
To obtain the action in terms
of the order parameters $\Delta^i_{\bm{Q}}$, we start with the Hamiltonian $H = H_0 + V$ (see Eqs. (\ref{eq:non-int_h}) and (\ref{eq:coulomb_interaction})), write the partition function as the integral over Grassmann variables and then decouple the quartic term $V$ using the Hubbard-Stratonovich transformation, following, e.g., Ref.~\onlinecite{PhysRevB.85.024534}:
\begin{equation}
\label{eq:HS_transform}
Z \propto \int d \bar \Delta d \Delta\, \exp\left(
-\sum_{\bm{Q},\alpha\beta} \int d\tau \frac{\bar\Delta_{\bm{Q}}^{\alpha\beta} \Delta_{\bm{Q}}^{\alpha\beta}}{g}
\right)
\int d\bar\psi d\psi\, e^{-S}.
\end{equation}
Note that we do not consider higher CDW harmonics of the form $\Delta_{2\bm{Q}}$, and instead focus only on the dominant processes involving the nesting vector $\bm{Q}$. As discussed in Ref. \cite{PhysRevB.74.245126}, there are contributions from these higher harmonics to the quartic GL coefficients, but such contributions are strongly suppressed for nested bands. Here $S$ is the fermionic action modified by the presence of $\Delta_{\bm{Q}}$:
\begin{eqnarray}
    &&S = \int d\tau \bigg[
\sum_{\bm{k}}\psi^\dagger_{\bm{k}}\left(\partial_\tau + h_{\bm{k}} \right)\psi_{\bm{k}}+ 
\sum_{\bm{Q}, \alpha, \beta} \bar A_{\bm{Q}}^{\alpha\beta} \Delta_{\bm{Q}}^{\alpha\beta}  \nonumber \\
&& +\sum_{\bm{Q}, \alpha, \beta} \bar \Delta_{\bm{Q}}^{\alpha\beta} A_{\bm{Q}}^{\alpha\beta}\bigg],
\label{eq:mft_like}
\end{eqnarray}
where the sum over $\bm{Q}$ includes only $\bm{Q}_{0}$ for the point group $D_{2h}$ considered in the main text, and runs over $\bm{Q}_{0}$ and $\bm{Q}_{1}= C_4 \cdot \bm{Q}_0$ for the point group $D_{4h}$ considered in Appendix~\ref{sec:gl_d4h}; $A_{\bm{Q}}^{\alpha\beta}$ are the fermionic bilinears:
\begin{equation}
A_{\bm{Q}}^{\alpha\beta} = \sum_{\bm{k}}\bar\psi_{\bm{k-Q},\beta}\psi_{\bm{k},\alpha}.
\end{equation}
Note that before making the Hubbard-Stratonovich transformation, we restricted the summation over $\bm{Q}$ in the interaction term $V$ in Eq. (\ref{eq:coulomb_interaction}) only to the vectors $\pm\bm{Q}_0$ for $D_{2h}$ and to $\pm \bm{Q}_{0,1}$ for $D_{4h}$.
Integrating out the fermions in (\ref{eq:HS_transform}), one can obtain an effective action in terms of the order parameters $\Delta_{\bm{Q}}^{\alpha\beta}$.

The effective action for the order parameters takes the following form for the point group $D_{2h}$: 
\begin{equation}
    S_{\operatorname{eff}} = 
    \frac{1}{g}
\sum_{\alpha\beta}\bar\Delta_{\bm{Q}_0}^{\alpha\beta} \Delta_{\bm{Q}_0}^{\alpha\beta} - 
    \operatorname{Tr} \log \left(1- \mathcal{G} \mathcal{V}\right),
    \label{eq:eff_action_general_d2h}
\end{equation}
where $\mathcal{G}$ is a matrix of non-interacting Green's functions, $\mathcal{V}$ is the interaction matrix, and $\operatorname{Tr}$ indicates summation over momenta, Matsubara frequencies, and orbital indices:
\begin{equation}
    \operatorname{Tr} = \frac{1}{\beta} \sum_{i \omega_n} \int \frac{d \bm{k}}{(2 \pi)^2}\ \operatorname{tr}_{\alpha\beta}.
\end{equation}
The matrix of non-interacting Green's functions is defined as:
\begin{equation}
    \mathcal{G}_{\bm{k}}(i\omega_n) = \begin{pmatrix}
    G_{\bm{k}}(i\omega_n) & 0 \\
    0 & G_{\bm{k+Q}_0}(i\omega_n)
    \end{pmatrix},
\end{equation}
where $G_{\bm{k}}(i\omega_n)$, the Green's function of the original system (\ref{eq:non-int_h}), is a $2\times2$ matrix itself. The interaction matrix has the following form:
\begin{equation}
    \mathcal{V} = 
    \begin{pmatrix}
        0 & \Delta_{\bm{Q}_0}^\dagger \\
        \Delta_{\bm{Q}_0} & 0 
    \end{pmatrix}
\label{eq: int_matrix}
\end{equation}
with $ \Delta_{\bm{Q}_0} = \sum_{i=0}^{3} \Delta^i_{\bm{Q}_0} \sigma^i$ and $ \Delta_{\bm{Q}_0}^\dagger = \sum_{i=0}^{3} \bar\Delta^i_{\bm{Q}_0} \sigma^i$. Thus, the most generic form of $S_{\operatorname{eff}}$ for point group $D_{2h}$ up to the fourth order, which is obtained by expanding (\ref{eq:eff_action_general_d2h}) in powers of $\Delta^i_{\bm{Q}_0}$, is
\begin{eqnarray}
    &&S_{\operatorname{eff}} =  \frac{2}{g} 
    \sum_{i=\overline{0,3}}
    | \Delta^i_{\bm{Q}_0} |^2 + \nonumber \\
    &&+\sum_{ij} \operatorname{Tr} \left[ G_{\bm{k}}
    \sigma^i\,
    G_{\bm{k+Q}_0}
    \sigma^j 
    \right] 
    \bar \Delta^i_{\bm{Q}_0} \Delta^j_{\bm{Q}_0}  + \nonumber \\
    && + \frac{1}{2} \sum_{ijkl} 
    \operatorname{Tr} \left[ G_{\bm{k}}
    \sigma^i\,
    G_{\bm{k+Q}_0}
    \sigma^j\,
    G_{\bm{k}}
    \sigma^k\,
    G_{\bm{k+Q}_0}
    \sigma^l 
    \right] \nonumber \\
    && \times
    \bar \Delta^i_{\bm{Q}_0} \Delta^j_{\bm{Q}_0} \bar \Delta^k_{\bm{Q}_0} \Delta^l_{\bm{Q}_0}.
    \label{eq:gl_generic_d2h}
\end{eqnarray}
\begin{table}[t]
\begin{ruledtabular}
\begin{tabular}{ l l }
Representation & Bilinears \rule{0pt}{3ex}\\[1ex]
\colrule
$A_g$ & $|\Delta^0|^2,\ 
|\Delta^x|^2,\ 
|\Delta^y|^2,\ 
|\Delta^z|^2,$  \rule{0pt}{3ex}\\[1ex]
\  & $\bar \Delta^0 \Delta^x + \bar \Delta^x \Delta^0,\  
    \bar \Delta^y \Delta^z + \bar \Delta^z \Delta^y$  \rule{0pt}{3ex}\\[1ex]
$B_{3u}$ & $\bar\Delta^0 \Delta^x - \bar \Delta^x \Delta^0,\ 
    \bar \Delta^y \Delta^z - \bar \Delta^z \Delta^y$  \rule{0pt}{4ex}\\[1ex]
$B_{1g}$ & $\bar\Delta^0 \Delta^y + \bar \Delta^y \Delta^0,\ 
    \bar\Delta^x \Delta^y + \bar \Delta^y \Delta^x,$
    \rule{0pt}{4ex}\\[1ex]
\ & $\bar\Delta^0 \Delta^z + \bar \Delta^z \Delta^0,\ 
    \bar\Delta^x \Delta^z + \bar \Delta^z \Delta^x$  
    \rule{0pt}{3ex}\\[1ex]
$B_{2u}$ & $\bar\Delta^0 \Delta^y - \bar \Delta^y \Delta^0,\ 
    \bar\Delta^x \Delta^y - \bar \Delta^y \Delta^x,$
    \rule{0pt}{4ex}\\[1ex]
\ & $ \bar\Delta^0 \Delta^z - \bar \Delta^z \Delta^0,\ 
    \bar\Delta^x \Delta^z - \bar \Delta^z \Delta^x$  
    \rule{0pt}{3ex}\\[1ex]
\end{tabular}
\end{ruledtabular}
\caption{\label{tab:bilinears_irreps_d2h_detailed}
Irreducible representations of translationally invariant $\Delta$-bilinears in the basis $\{\Delta^0, \Delta^x, \Delta^y, \Delta^z \}$ for point group $D_{2h}$. Here the subscript $\bm{Q}_0$ is omitted for notational simplicity.
}
\end{table}
Comparing this expression to the quadratic order action (\ref{eq:gl_d2h}), obtained from symmetry considerations, we find that
\begin{eqnarray}
    &&a_0 = \frac{2}{g} + \operatorname{Tr} \left[ G_{\bm{k}}
    G_{\bm{k+Q}_0}
    \right], \nonumber \\
    &&a_x = \frac{2}{g} + \operatorname{Tr} \left[ G_{\bm{k}}
    \sigma^x\,
    G_{\bm{k+Q}_0}
    \sigma^x 
    \right], \nonumber \\ 
    &&a_y = \frac{2}{g} + \operatorname{Tr} \left[ G_{\bm{k}}
    \sigma^y\,
    G_{\bm{k+Q}_0}
    \sigma^y 
    \right], \nonumber \\ 
    &&a_z = \frac{2}{g} + \operatorname{Tr} \left[ G_{\bm{k}}
    \sigma^z\,
    G_{\bm{k+Q}_0}
    \sigma^z 
    \right], \nonumber \\ 
    && \lambda = \frac{1}{2}\big(
    \operatorname{Tr} \left[ G_{\bm{k}}
    \sigma^x\,
    G_{\bm{k+Q}_0}
    \right] + 
    \operatorname{Tr} \left[ G_{\bm{k}}
    G_{\bm{k+Q}_0}
    \sigma^x
    \right]\big), \nonumber \\ 
    && \tilde\lambda = \frac{1}{2}\big(
    \operatorname{Tr} \left[ G_{\bm{k}}
    \sigma^y\,
    G_{\bm{k+Q}_0}
    \sigma^z
    \right] + 
    \operatorname{Tr} \left[ G_{\bm{k}}
    \sigma^z\,
    G_{\bm{k+Q}_0}
    \sigma^y
    \right]\big). \nonumber \\ 
    \label{eq:quadratic_coeffs_greens_fn}
\end{eqnarray}
Taking the trace over orbital indices and comparing the obtained expressions to the susceptibility components (\ref{eq:susc_greens_function}), we arrive at the formulas (\ref{eq:coeff_susceptibilities_diag}) and (\ref{eq:coeff_susceptibilities_offdiag}). To show that $\tilde\lambda = 0$, one can use the symmetry property of the Green's function $G_{\bm{k}}^{\alpha\beta} = G_{\bm{k}}^{\beta\alpha}$, following from (\ref{eq:greens_function}) given that the matrix $a_m^\alpha(\bm{k})$ is real.

\begin{figure*}
\includegraphics[height=5.5cm]
{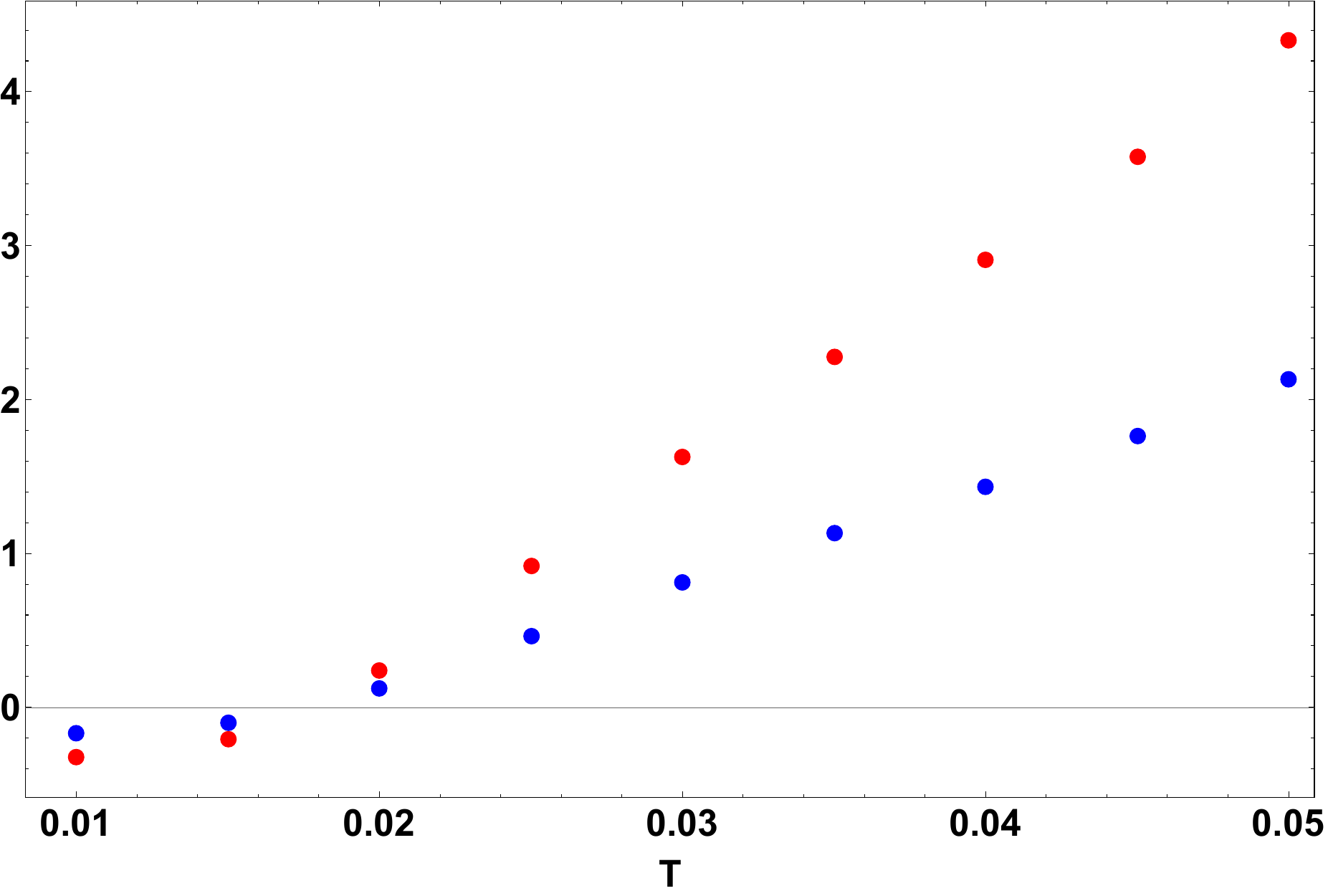}
\includegraphics[height=5.5cm]
{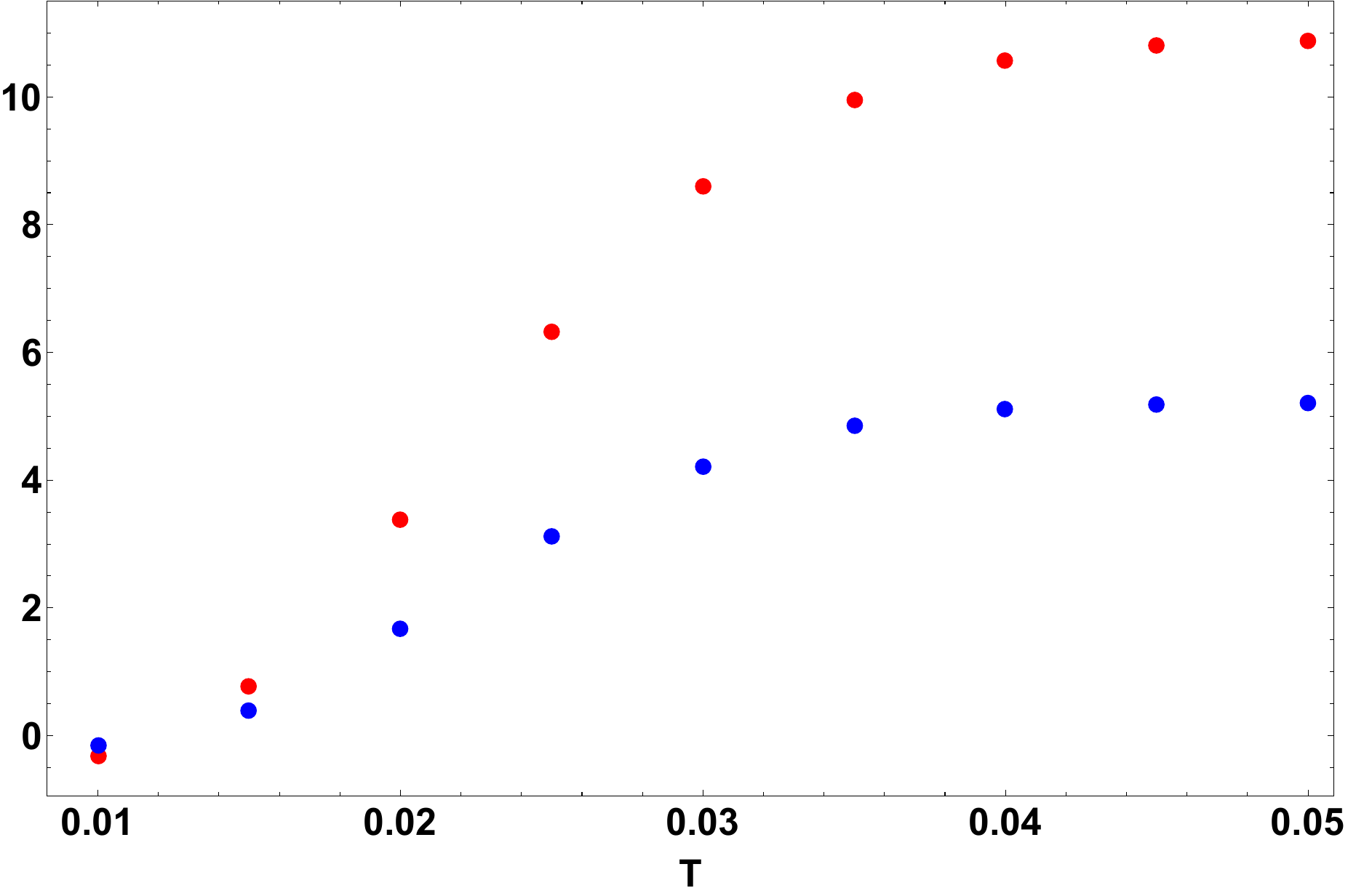}
\includegraphics[height=5.5cm]
{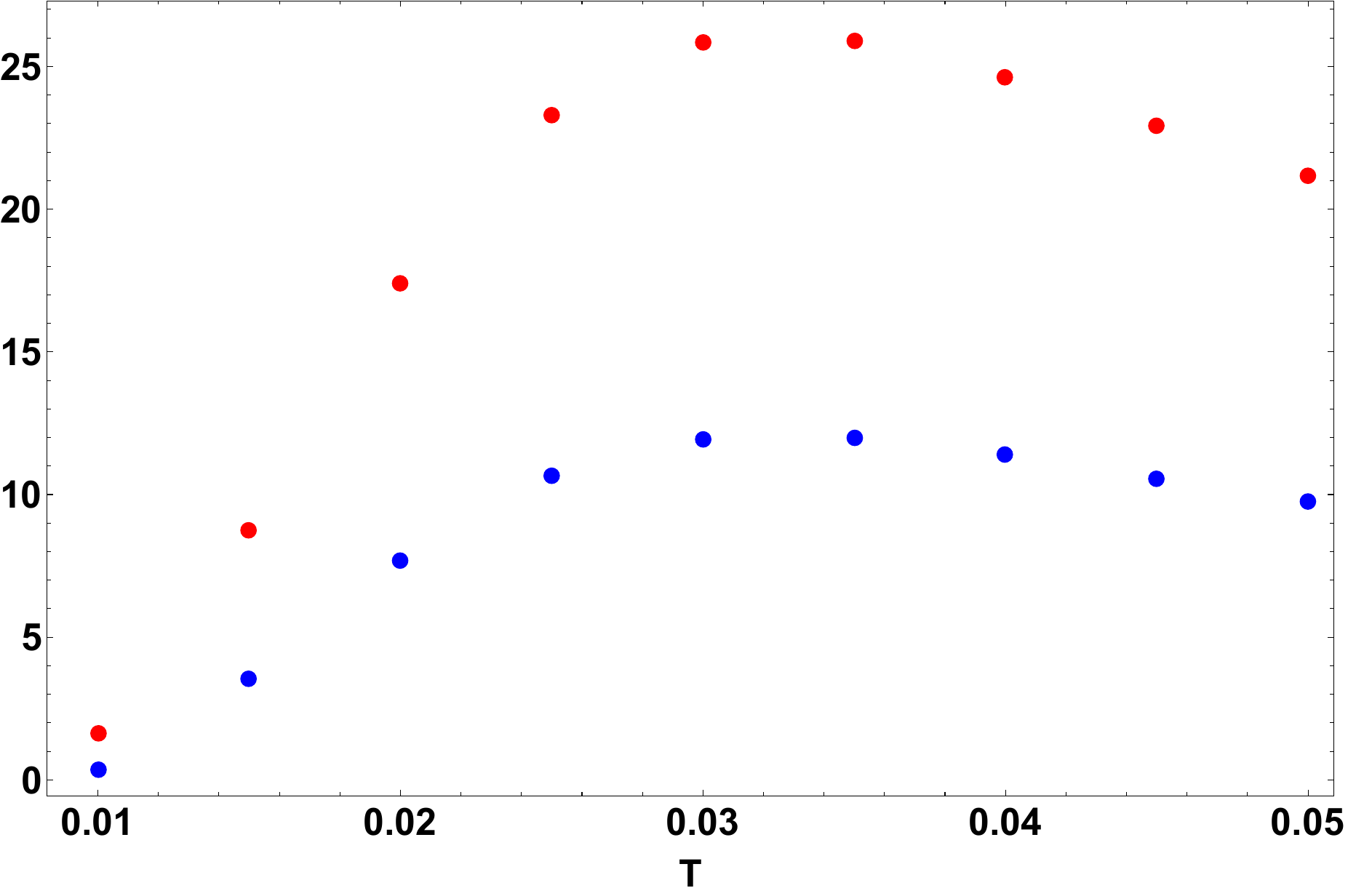}
\includegraphics[height=5.5cm]
{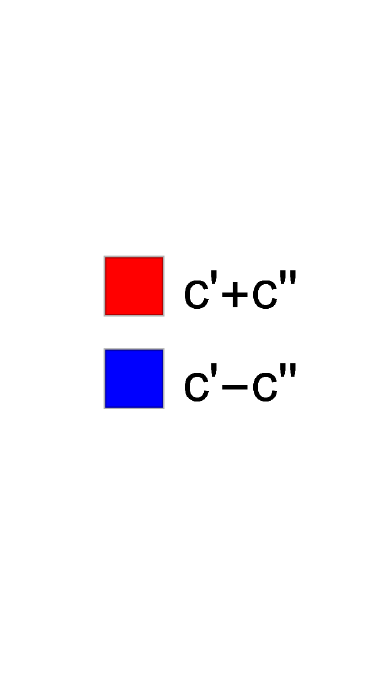}
    \caption{$c'-c''$ (blue) and $c' + c''$ (red) as functions of temperature $T$ (in eV). From left to right: $t_d = 0.05$eV, $t_d = 0.10$eV, $t_d = 0.15$eV. As  $c'-c''$ is mostly positive, the left-hand-side of (\ref{eq:double_delta_criterion}) is minimized for $\delta \alpha = \pm \pi/2$. 
}
    \label{fig:c_sum_diff}
\end{figure*}

To obtain the quartic order coefficients for $t_d \ne 0$ in Eq. (\ref{eq:quartic_gl_d2h}), we work in the basis formed by $\Delta^-_{\bm{Q}_0}$, $\Delta^+_{\bm{Q}_0}$, $\Delta^y_{\bm{Q}_0}$, and $\Delta^z_{\bm{Q}_0}$. Note that as both $\Delta^0_{\bm{Q}_0}$ and $\Delta^x_{\bm{Q}_0}$ belong to the same representation of $D_{2h}$, i.e. $A_{g}$, their linear combinations $\Delta^\pm_{\bm{Q}_0}$ are also in $A_{g}$. Expanding (\ref{eq:eff_action_general_d2h}) in powers of $ \Delta_{\bm{Q}_0} = \Delta^-_{\bm{Q}_0} \sigma^- + \Delta^+_{\bm{Q}_0} \sigma^+ + \Delta^y_{\bm{Q}_0} \sigma^y + \Delta^z_{\bm{Q}_0} \sigma^z$, we arrive at the expression of the same form as 
Eq. (\ref{eq:gl_generic_d2h}), where the summation is now over indices $\pm, y,z$. Explicit expressions for $\sigma^\pm$, which are not the usual ``raising'' and ``lowering'' forms of the Pauli matrices, can be found by equating $\Delta^-_{\bm{Q}_0} \sigma^- + \Delta^+_{\bm{Q}_0} \sigma^+$ and $\Delta^0_{\bm{Q}_0} \sigma^0 + \Delta^x_{\bm{Q}_0} \sigma^x$. Let us denote the fourth-order coefficients in Eq. (\ref{eq:gl_generic_d2h}) as
\begin{equation}
    A^{ijkl} = \frac{1}{2} 
    \operatorname{Tr} \left[ G_{\bm{k}}
    \sigma^i\,
    G_{\bm{k+Q}_0}
    \sigma^j\,
    G_{\bm{k}}
    \sigma^k\,
    G_{\bm{k+Q}_0}
    \sigma^l 
    \right],
    \label{eq:quartic_order_aux}
\end{equation}
where $ijkl$ take values in $\pm, y,z$ as previously mentioned. Comparing (\ref{eq:gl_generic_d2h}) to the effective action in fourth order (\ref{eq:quartic_gl_d2h}), obtained from symmetry considerations, we evaluate the fourth-order coefficients in Eq. (\ref{eq:quartic_gl_d2h}) as follows: 
\begin{eqnarray}
    &&b_- = A^{----} \nonumber, \\
    &&b_z = A^{zzzz} \nonumber, \\
    &&c' - c'' = A^{-z-z} = A^{z-z-} \nonumber, \\
    &&c' + c'' = \frac{1}{2}\big(A^{--zz}+A^{-zz-}+A^{zz--}+A^{z--z}\big). \nonumber \\
    \label{eq:quartic_coeffs_greens_fn}
\end{eqnarray}
As shown in Eq. (\ref{eq:ccondition}), the last two coefficients define the relative phase between $\Delta^-_{\bm{Q}_0}$ and $\Delta^z_{\bm{Q}_0}$. Fig. \ref{fig:c_sum_diff} shows that $0<(c'-c'')/(c'+c'')<1$, so the left-hand side in (\ref{eq:double_delta_criterion}) is minimized for $\delta \alpha = \pm \pi/2$.

\begin{table}[t]
\begin{ruledtabular}
\begin{tabular}{ l l }
Representation & Bilinears \rule{0pt}{3ex}\\[1ex]
\colrule
$A_g$ & $|\Delta^-|^2,\ 
|\Delta^+|^2,\ 
|\Delta^y|^2,\ 
|\Delta^z|^2,$  \rule{0pt}{3ex}\\[1ex]
\  & $\bar \Delta^- \Delta^+ + \bar \Delta^+ \Delta^-,\  
    \bar \Delta^y \Delta^z + \bar \Delta^z \Delta^y$  \rule{0pt}{3ex}\\[1ex]
$B_{3u}$ & $\bar\Delta^- \Delta^+ - \bar \Delta^+ \Delta^-,\ 
    \bar \Delta^y \Delta^z - \bar \Delta^z \Delta^y$  \rule{0pt}{4ex}\\[1ex]
$B_{1g}$ & $\bar\Delta^- \Delta^y + \bar \Delta^y \Delta^-,\ 
    \bar\Delta^+ \Delta^y + \bar \Delta^y \Delta^+,$
    \rule{0pt}{4ex}\\[1ex]
\ & $\bar\Delta^- \Delta^z + \bar \Delta^z \Delta^-,\ 
    \bar\Delta^+ \Delta^z + \bar \Delta^z \Delta^+$  
    \rule{0pt}{3ex}\\[1ex]
$B_{2u}$ & $\bar\Delta^- \Delta^y - \bar \Delta^y \Delta^-,\ 
    \bar\Delta^+ \Delta^y - \bar \Delta^y \Delta^+,$
    \rule{0pt}{4ex}\\[1ex]
\ & $ \bar\Delta^- \Delta^z - \bar \Delta^z \Delta^-,\ 
    \bar\Delta^+ \Delta^z - \bar \Delta^z \Delta^+$  
    \rule{0pt}{3ex}\\[1ex]
\end{tabular}
\end{ruledtabular}
\caption{\label{tab:bilinears_irreps_d2h_detailed_pm_basis}
Irreducible representations of translationally invariant $\Delta$-bilinears in the basis $\{\Delta^-, \Delta^+, \Delta^y, \Delta^z \}$ for the point group $D_{2h}$. Here the subscript $\bm{Q}_0$ is omitted for notational simplicity.
}
\end{table}

Finally, we examine the simplifications occurring for $t_d = 0$. In this case, the Green's function $G_{\bm{k}}^{\alpha\beta}$ is diagonal in orbital indices, so it follows from Eq. (\ref{eq:quadratic_coeffs_greens_fn}) that there is no mixing between $\Delta^0_{\bm{Q}_0}$ and $\Delta^x_{\bm{Q}_0}$ in the GL theory in quadratic order, i.e. $\lambda = \tilde\lambda = 0$. Moreover, it follows that $\Delta^0_{\bm{Q}_0}$ and $\Delta^z_{\bm{Q}_0}$ as well as $\Delta^x_{\bm{Q}_0}$ and $\Delta^y_{\bm{Q}_0}$ are exactly degenerate, i.e. $a_0 = a_z$ and $a_x = a_y$. Thus, $\Delta^-_{\bm{Q}_0} = \Delta^0_{\bm{Q}_0}$ and $\Delta^+_{\bm{Q}_0} = \Delta^x_{\bm{Q}_0}$ and one can prove the following set of relations for fourth-order coefficients, taking the trace over the orbital indices in Eqs. (\ref{eq:quartic_coeffs_greens_fn}) and (\ref{eq:quartic_order_aux}): $b_- = b_z = c'-c'' = (c'+c'')/2$.

For reference, we provide detailed Tables \ref{tab:bilinears_irreps_d2h_detailed} and \ref{tab:bilinears_irreps_d2h_detailed_pm_basis}, listing all the translationally invariant $\Delta$-bilinears and their corresponding representations in the two bases $\{\Delta^0_{\bm{Q}_0}, \Delta^x_{\bm{Q}_0}, \Delta^y_{\bm{Q}_0}, \Delta^z_{\bm{Q}_0} \}$ and $\{\Delta^-_{\bm{Q}_0}, \Delta^+_{\bm{Q}_0}, \Delta^y_{\bm{Q}_0}, \Delta^z_{\bm{Q}_0} \}$  respectively. Table \ref{tab:bilinears_irreps_d2h_detailed} is merely an expanded version of Table \ref{tab:bilinears_irreps_d2h} in the main text. Table \ref{tab:bilinears_irreps_d2h_detailed_pm_basis} is analogous to Table \ref{tab:bilinears_irreps_d2h_detailed}, and they become coincident in the $t_d \to 0$ limit as $\Delta^-_{\bm{Q}_0} \to \Delta^0_{\bm{Q}_0}$ and $\Delta^+_{\bm{Q}_0} \to \Delta^x_{\bm{Q}_0}$.

\section{\label{sec:fs}Fermi surfaces}
\begin{figure*}
\includegraphics[height=18.0cm]{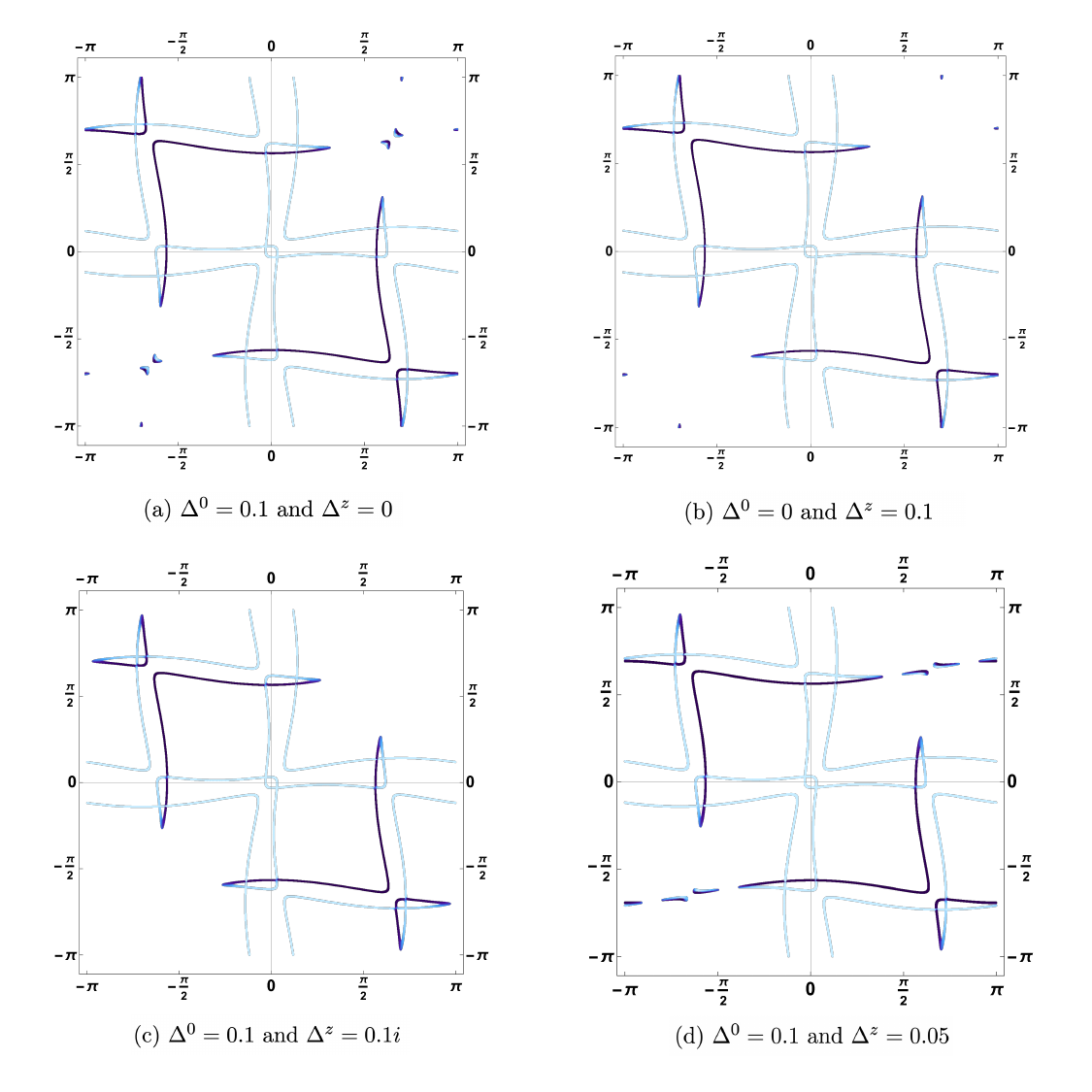}
\includegraphics[height=1.0cm]{graphics/FS/colorbar_horizontal.pdf}
    \caption{Fermi surfaces in the presence of CDW: (a) single-$\Delta$ $A_g$, (b) single-$\Delta$ $B_{1g}$, (c) double-$\Delta$ $B_{2u}$, (d) double-$\Delta$ $B_{1g}$. Here $t_d = 0.16$eV, $\mu= 1.53$eV. The color intensity indicates the spectral weight of the first Brillouin zone (see Eq. \ref{eq:ham_cdw}).
    Here the subscript $\bm{Q}_0$ is omitted for notational simplicity.
    }
    \label{fig:fs_cdw_multiple}
\end{figure*}

From Eq. (\ref{eq:mft_like}), the corresponding mean-field Hamiltonian in the presence of the CDW is derived as follows:
\begin{equation}
H_{\operatorname{MFT}} = 
\sum_{\bm{k}}\left[\psi^\dagger_{\bm{k}}h_{\bm{k}} \psi_{\bm{k}} + \psi^\dagger_{\bm{k}+\bm{Q}_0}\,
\Delta_{\bm{Q}_0} \psi_{\bm{k}} + 
\psi^\dagger_{\bm{k}-\bm{Q}_0}
\Delta_{\bm{Q}_0}^\dagger \psi_{\bm{k}}\right]
\end{equation}
To plot the Fermi surfaces for this Hamiltonian, we rewrite it in the following form and proceed with its diagonalization:
\begin{equation}
\mathcal{H}_{\bm{k}} = \begin{pmatrix}
h_{\bm{k}-\bm{Q}_0} & 0 & 0 \\
 0 &   h_{\bm{k}} & 0 \\
 0 &  0 & h_{\bm{k}+\bm{Q}_0}
\end{pmatrix}
+
\begin{pmatrix}
0 & \Delta_{\bm{Q}_0}^\dagger & 0 \\
\Delta_{\bm{Q}_0} & 0 & \Delta_{\bm{Q}_0}^\dagger \\
0 &  \Delta_{\bm{Q}_0} & 0
\end{pmatrix}.
\label{eq:ham_cdw}
\end{equation}
In Fig. \ref{fig:fs_cdw_multiple}, we depict the Fermi surfaces in four scenarios: when either $\Delta^0_{\bm{Q}_0}$ or $\Delta^z_{\bm{Q}_0}$ is non-zero while the other is zero, and when both are non-zero with relative phases of $0$ and $\pi/2$. We observe that $\Delta_{\bm{Q}_0}^z$ alone does not break the $\sigma_d$-symmetry of the Fermi surface. Combining $\Delta_{\bm{Q}_0}^0$ and $\Delta_{\bm{Q}_0}^z$ with the same phase results in the breaking of this symmetry in the Fermi surface. However, when their phase difference is $\pi/2$, mirror symmetry breaking is again not observed in the Fermi surface.

To provide some insight into why there is no additional mirror symmetry breaking when the relative phase is $\pi/2$, let us diagonalize the Hamiltonian (\ref{eq:ham_cdw}) perturbatively in the case $t_d = 0$. Assuming that the original Hamiltonian $h_{\bm{k}}$ takes the form
\begin{equation}
    h_{\bm{k}} = 
    \begin{pmatrix}
        \varepsilon_{\bm{k}, p_x}^{(0)} & 0 \\
        0 & \varepsilon_{\bm{k}, p_y}^{(0)}
    \end{pmatrix},
\end{equation}
we obtain the following corrections to the eigenvalues up to the second order in perturbation theory:
\begin{eqnarray}
    &&\varepsilon_{\bm{k}, p_x} = \varepsilon_{\bm{k}, p_x}^{(0)}+ 
    \frac{|\Delta^0_{\bm{Q}_0} + \Delta^z_{\bm{Q}_0}|^2}{\varepsilon_{\bm{k}-\bm{Q}, p_x}^{(0)} - 
    \varepsilon_{\bm{k}, p_x}^{(0)}} + 
    \frac{|\Delta^0_{\bm{Q}_0} + \Delta^z_{\bm{Q}_0}|^2}{\varepsilon_{\bm{k}+\bm{Q}, p_x}^{(0)} - 
    \varepsilon_{\bm{k}, p_x}^{(0)}} \nonumber \\
    &&\varepsilon_{\bm{k}, p_y} = \varepsilon_{\bm{k}, p_y}^{(0)}+ 
    \frac{|\Delta^0_{\bm{Q}_0} - \Delta^z_{\bm{Q}_0}|^2}{\varepsilon_{\bm{k}-\bm{Q}, p_y}^{(0)} - 
    \varepsilon_{\bm{k}, p_y}^{(0)}} + 
    \frac{|\Delta^0_{\bm{Q}_0} - \Delta^z_{\bm{Q}_0}|^2}{\varepsilon_{\bm{k}+\bm{Q}, p_y}^{(0)} - 
    \varepsilon_{\bm{k}, p_y}^{(0)}} \nonumber \\
    &&
\end{eqnarray}
Note that 
\begin{eqnarray}
    && |\Delta^0_{\bm{Q}_0} \pm \Delta^z_{\bm{Q}_0}|^2 = |\Delta^0_{\bm{Q}_0}|^2 + |\Delta^z_{\bm{Q}_0}|^2 \pm \nonumber \\ 
    && \pm \left(\bar\Delta^0_{\bm{Q}_0} \Delta^z_{\bm{Q}_0} +
 \bar\Delta^z_{\bm{Q}_0} \Delta^0_{\bm{Q}_0}\right), 
\end{eqnarray}
implying that the asymmetry between $\varepsilon_{\bm{k}, p_x}$ and $\varepsilon_{\bm{k}, p_y}$ is determined by the expression $\bar\Delta^0_{\bm{Q}_0} \Delta^z_{\bm{Q}_0} + \bar\Delta^z_{\bm{Q}_0} \Delta^0_{\bm{Q}_0}$, which equals $0$ when the relative phase is $\pi/2$.

\section{\label{sec:gl_d4h}CDW in the point group $D_{4h}$}
\subsection{\label{sec:D4h_irreps} Irreducible representations of order parameters}
Although the $R$Te$_3$ materials have an orthorhombic space group, they are often modeled in terms of an approximate tetragonal lattice. Here, we generalize our results to this case. The point group $D_{4h}$ additionally includes two four-fold rotations about the principal ($z$) axis ($2C_4$), two two-fold rotations about the axes $x$ and $y$ ($2C_2'$), two reflections through one of vertical mirror planes $x=0$ or $y=0$ ($2\sigma_v$), and two improper rotations ($2S_4$, where $S_4 = \sigma_h C_4$). Its irreducible representations are summarized in Table \ref{tab:char_tab_d4h}. For this point group, the space of order parameters is spanned by $\Delta^i_{\bm{Q}_n},\, n=\overline{0,3}$ with $\bm{Q}_0 = (2k_F,2k_F)$ and $\bm{Q}_n = C_4^n \cdot \bm{Q}_0$. 
This space is decomposed into irreducible representations of $D_{4h}$ in Table \ref{tab:order_param_irreps_d4h}.
\begin{table*}[t]
\begin{ruledtabular}
\begin{tabular}{c r r r r r r r r r r}
$D_{4h}$ & $E$ & $2C_4$ & $C_2$ & $2C_2'$ & $2C_2''$&
$I$ & $2S_4$ & $\sigma_h$ & $2\sigma_v$ & $2\sigma_d$  \rule{0pt}{3ex}\\[1ex]
\colrule
$A_{1g}$ &	$1$ & $1$ & $1$ & $1$ &	$1$ & $1$ & $1$ & $1$ & $1$ & $1$	\rule{0pt}{3ex}\\[1ex]
$A_{2g}$ &	$1$ & $1$ & $1$ & $-1$ & $-1$ & $1$ & $1$ & $1$ & $-1$ & $-1$	\rule{0pt}{3ex}\\[1ex]
$B_{1g}$ &	$1$ & $-1$ & $1$ & $1$ & $-1$ & $1$ & $-1$ & $1$ & $1$ & $-1$	\rule{0pt}{3ex}\\[1ex]
$B_{2g}$ &	$1$ & $-1$ & $1$ & $-1$ & $1$ & $1$ & $-1$ & $1$ & $-1$ & $1$	\rule{0pt}{3ex}\\[1ex]
$E_g$ &	$2$ & $0$ & $-2$ & $0$ & $0$ & $2$ & $0$ & $-2$ & $0$ & $0$	\rule{0pt}{3ex}\\[1ex]
$A_{1u}$ &	$1$ & $1$ & $1$ & $1$ &	$1$ & $-1$ & $-1$ & $-1$ & $-1$ & $-1$	\rule{0pt}{3ex}\\[1ex]
$A_{2u}$ &	$1$ & $1$ & $1$ & $-1$ & $-1$ & $-1$ & $-1$ & $-1$ & $1$ & $1$	\rule{0pt}{3ex}\\[1ex]
$B_{1u}$ &	$1$ & $-1$ & $1$ & $1$ & $-1$ & $-1$ & $1$ & $-1$ & $-1$ & $1$	\rule{0pt}{3ex}\\[1ex]
$B_{2u}$ &	$1$ & $-1$ & $1$ & $-1$ & $1$ & $-1$ & $1$ & $-1$ & $1$ & $-1$	\rule{0pt}{3ex}\\[1ex]
$E_u$ &	$2$ & $0$ & $-2$ & $0$ & $0$ & $-2$ & $0$ & $2$ & $0$ & $0$	\rule{0pt}{3ex}\\[1ex]
\end{tabular}
\end{ruledtabular}
\caption{\label{tab:char_tab_d4h}
Character table for point group $D_{4h}$.
}
\end{table*}

\begin{table}[t]
\begin{ruledtabular}
\begin{tabular}{@{\hspace{5em}} c c @{\hspace{5em}}}
Basis &\ Representation  \rule{0pt}{3ex}\\[1ex]
 \colrule
$\Re \left(\Delta^i_{\bm{Q}_0} + \Delta^i_{\bm{Q}_1} \right)$ & $A_{1g} \otimes \rho_i$ \rule{0pt}{3ex}\\[1ex]

$\Re 
\left(
\Delta^i_{\bm{Q}_0} - \Delta^i_{\bm{Q}_1}
\right)$ & $B_{2g} \otimes \rho_i$ \rule{0pt}{3ex}\\[1ex]

$ \left\{
\Im \Delta^i_{\bm{Q}_0},
\Im \Delta^i_{\bm{Q}_1}
\right \}$ & $E_u \otimes \rho_i$ \rule{0pt}{3ex}\\[1ex]
\end{tabular}
\end{ruledtabular}
\caption{\label{tab:order_param_irreps_d4h}
Irreducible representations of the CDW order parameters for point group $D_{4h}$. The index $i$ takes values in the set $\{0,x,y,z\}$ and $\rho_0 = A_{1g}$,   $\rho_x = B_{2g}$, $\rho_y = A_{2g}$, $\rho_z = B_{1g}$.
}
\end{table}

In the table, $\rho_i$ is the one-dimensional irreducible representation of $D_{4h}$ under which the Pauli matrix $\sigma^i$ transforms: $\rho_0 = A_{1g}$,   $\rho_x = B_{2g}$, $\rho_y = A_{2g}$, $\rho_z = B_{1g}$.

The only translationally invariant bilinears in order parameters are those where both $\bm{Q}_n$ and $-\bm{Q}_n$ are present, i.e. 
    $\Lambda^{ij}_n = \Delta^i_{\bm{Q}_n} \Delta^j_{-\bm{Q}_n}$.
This space of bilinears is decomposed into irreducible representations of $D_{4h}$ in a similar fashion as in Table \ref{tab:bilinears_irreps_d4h}. 

\begin{table}[t]
\begin{ruledtabular}
\begin{tabular}{@{\hspace{5em}} c c @{\hspace{5em}}}
Basis & Representation \rule{0pt}{3ex}\\[1ex]
 \colrule
$\Re\left(\Lambda_0^{ij} + \Lambda_1^{ij}\right)$ & $A_{1g} \otimes \rho_i \otimes \rho_j$ \rule{0pt}{3ex}\\[1ex]

$\Re\left(\Lambda_0^{ij} - \Lambda_1^{ij}\right)$ & $B_{2g} \otimes \rho_i \otimes \rho_j$ \rule{0pt}{3ex}\\[1ex]

$ \left
\{\Im\Lambda_0^{ij},\ \Im\Lambda_1^{ij}
\right\} $ & $E_u \otimes \rho_i \otimes \rho_j$ \rule{0pt}{3ex}\\[1ex]
\end{tabular}
\end{ruledtabular}
\caption{\label{tab:bilinears_irreps_d4h}
Irreducible representations of the translationally invariant $\Delta$-bilinears $\Lambda^{ij}_{0,1} = \Delta^i_{\bm{Q}_{0,1}} \Delta^j_{-\bm{Q}_{0,1}}$ for the point group $D_{4h}$. The index $i$ takes values in the set $\{0,x,y,z\}$ and $\rho_0 = A_{1g}$,   $\rho_x = B_{2g}$, $\rho_y = A_{2g}$, $\rho_z = B_{1g}$.
}
\end{table}

\subsection{Ginzburg-Landau theory}
The resulting action for point group $D_{4h}$ is similar to that in Eq.~(\ref{eq:eff_action_general_d2h}):
\begin{equation}
    S_{\operatorname{eff}} = 
    \frac{1}{g}
\sum_{\substack{n=0,1\\ \alpha\beta}}\bar\Delta_{\bm{Q}_n}^{\alpha\beta} \Delta_{\bm{Q}_n}^{\alpha\beta} - 
    \operatorname{Tr} \log \left(1- \mathcal{G} \mathcal{V}\right),
\end{equation}
with the only difference being that it also includes $\bm{Q}_1$, i.e. the $90^\circ$ rotated $\bm{Q}_0$. The matrix of the non-interacting Green's function $\mathcal{G}_{\bm{k}}$ and the interaction matrix $\mathcal{V}$ are modified as follows:
\begin{eqnarray}
    &&\mathcal{G}_{\bm{k}} = \operatorname{diag}
    \left(
    G_{\bm{k}},\,
    G_{\bm{k+Q}_0},\,
    G_{\bm{k+Q_1}}
    \right), \\
&&\mathcal{V} = \begin{pmatrix}
        0 & \Delta_{\bm{Q}_0}^\dagger & \Delta_{\bm{Q}_1}^\dagger \\
        \Delta_{\bm{Q}_0} & 0 & 0  \\
        \Delta_{\bm{Q}_1} & 0 & 0 
    \end{pmatrix}.
\end{eqnarray}

The action contains only the translationally invariant bilinears transforming trivially under $D_{4h}$, giving, to quadratic order:
\begin{eqnarray}
S_{\operatorname{eff}}^{(2)} &&=
\begin{pmatrix}
    \bar\Delta_{\bm{Q}_0}^0 &
    \bar\Delta_{\bm{Q}_0}^x 
\end{pmatrix}
\begin{pmatrix}
    a_0 & \lambda \\
    \lambda & a_x
\end{pmatrix}
\begin{pmatrix}
    \Delta_{\bm{Q}_0}^0 \\
    \Delta_{\bm{Q}_0}^x 
\end{pmatrix} + \nonumber \\
&& +\begin{pmatrix}
    \bar\Delta_{\bm{Q}_1}^0 & 
    \bar\Delta_{\bm{Q}_1}^x
\end{pmatrix}
\begin{pmatrix}
    a_0 & -\lambda\\
    -\lambda & a_x
\end{pmatrix}
\begin{pmatrix}
    \Delta_{\bm{Q}_1}^0 \\
    \Delta_{\bm{Q}_1}^x
\end{pmatrix} + \nonumber  \\ 
&&+ \begin{pmatrix}
    \bar\Delta_{\bm{Q}_0}^y &
    \bar\Delta_{\bm{Q}_0}^z 
\end{pmatrix}
\begin{pmatrix}
    a_y & \tilde \lambda \\
    \tilde \lambda & a_z 
\end{pmatrix}
\begin{pmatrix}
    \Delta_{\bm{Q}_0}^y \\
    \Delta_{\bm{Q}_0}^z
\end{pmatrix} + \nonumber \\
&&+ \begin{pmatrix}
    \bar\Delta_{\bm{Q}_1}^y &
    \bar\Delta_{\bm{Q}_1}^z 
\end{pmatrix}
\begin{pmatrix}
    a_y & -\tilde \lambda \\
    -\tilde \lambda & a_z
\end{pmatrix}
\begin{pmatrix}
    \Delta_{\bm{Q}_1}^y \\
    \Delta_{\bm{Q}_1}^z
\end{pmatrix},
\label{eq:d4h_gl_quadratic}
\end{eqnarray}
with the coefficients given by the same expressions as in (\ref{eq:coeff_susceptibilities_diag}), (\ref{eq:coeff_susceptibilities_offdiag}), with $\bm{Q}_0$ replaced by $\bm{Q}_1$ as appropriate; in particular, $\tilde \lambda = 0$. 
Note that the eigenvalues in the first two blocks are exactly equal to each other, as are the eigenvalues in the last two blocks. Thus, in addition to the near-degeneracy between $a_0$ and $a_z$ and between $a_x$ and $a_y$, there is an exact degeneracy between $\Delta^i_{\bm{Q}_0}$ and $\Delta^i_{\bm{Q}_1}$ in the $D_{4h}$ case. As in the $D_{2h}$ case, when $t_d = 0$, the degeneracy between $a_0$ and $a_z$ and between $a_x$ and $a_y$ is exact and $\lambda = 0$.

\subsection{Results}
\subsubsection{$t_d = 0$}
From Eq. (\ref{eq:d4h_gl_quadratic}), one obtains the following non-zero irreducible representations below the critical temperature for $\Delta^0$ and $\Delta^z$ assuming no relative phase between them:
\begin{eqnarray}
    && A_{1g}: \Re \left(\Delta_{\bm{Q}_0}^0 + \Delta_{\bm{Q}_1}^0 \right) 
    \quad\quad
    B_{1g}: \Re \left(\Delta_{\bm{Q}_0}^z + \Delta_{\bm{Q}_1}^z \right) \nonumber\\
    && B_{2g}: \Re \left(\Delta_{\bm{Q}_0}^0 - \Delta_{\bm{Q}_1}^0 \right) 
    \quad\quad
    A_{2g}: \Re \left(\Delta_{\bm{Q}_0}^z - \Delta_{\bm{Q}_1}^z \right). \nonumber \\
    &&
\label{eq:non_zero_irreps_d4h}
\end{eqnarray}
Alternatively, when the phase difference is $\pm \pi/2$, one obtains:
\begin{eqnarray}
    && A_{1g}: \Re \left(\Delta_{\bm{Q}_0}^0 + \Delta_{\bm{Q}_1}^0 \right) 
    \quad\quad
    E_u: \left\{
\Im \Delta^z_{\bm{Q}_0},\ 
\Im \Delta^z_{\bm{Q}_1}
\right \} \nonumber\\
    && B_{2g}: \Re \left(\Delta_{\bm{Q}_0}^0 - \Delta_{\bm{Q}_1}^0 \right).
\label{eq:non_zero_irreps_d4h_phase}
\end{eqnarray}

The fate of the system depends on whether the ground state is a single-$\bm{Q}$ solution, in which case either one of the sets of CDW order parameters $\{\Delta_{\bm{Q}_0}^i\}$  or $\{\Delta_{\bm{Q}_1}^i\}$ is zero, or a double-$\bm{Q}$ solution,  in which case the magnitudes are expected to be the same $|\Delta_{\bm{Q}_0}^i| = |\Delta_{\bm{Q}_1}^i|$. In each case, one can in principle still find double-$\Delta$  and single-$\Delta$ solutions, corresponding to the coexistence or not of the components $\Delta_{\bm{Q}_n}^0$ and $\Delta_{\bm{Q}_n}^z$. These combinations further extend the possible broken symmetries in the CDW ground state.

In the single-$\bm{Q}$ family of solutions, we can readily identify the single-$\Delta$ and double-$\Delta$ phases that we found in the $D_{2h}$ case directly from the translationally invariant bilinears listed Table \ref{tab:bilinears_irreps_d4h}. The breaking of tetragonal symmetry by selecting $\bm{Q}_0$ over  $\bm{Q}_1$ (or vice versa) is reflected by the bilinear $\Re\left(\Lambda_0^{ii} - \Lambda_1^{ii}\right)$, which transforms as the $B_{2g}$ irrep for any $i=0,\,z$ and is always non-zero in the single-$\bm{Q}$ case. As a result, the point group symmetry is lowered from $D_{4h}$ to $D_{2h}$. In the case of double-$\Delta$ solutions, there are other non-zero bilinears that indicate additional broken symmetries in the coexistence state. For instance, if the relative phase between  $\Delta_{\bm{Q}_n}^0$ and $\Delta_{\bm{Q}_n}^z$ is $0,\,\pi$ ,  the bilinear $\Re\left(\Lambda_0^{0z} - \Lambda_1^{0z}\right)$ is non-zero. Because it transforms as the $A_{2g} = B_{2g}\otimes A_{1g} \otimes B_{1g}$ irrep of $D_{4h}$, it corresponds to a ferroaxial moment. Due to the fact that the tetragonal symmetry is broken by the bilinear $\Re\left(\Lambda_0^{ii} - \Lambda_1^{ii}\right)$,  the off-diagonal  $A_{2g}$ bilinear can be equivalently labeled as the $B_{1g}$ irrep of the new point group $D_{2h}$.  Conversely, when the relative phase between  $\Delta_{\bm{Q}_n}^0$ and $\Delta_{\bm{Q}_n}^z$ is $\pm \pi/2$, it is  the bilinear $\Im\Lambda_0^{0z}$ that is non-zero, corresponding to an in-plane electric field. It transforms as one of the components of the $E_u$ irrep of $D_{4h}$ which, in terms of the lower point-group $D_{2h}$, corresponds to the $B_{2u}$ irrep.

\subsubsection{$t_d \ne 0$}
When $t_d$ is finite, due to mixing between $\Delta^0_{\bm{Q}_0
}$ and $\Delta^x_{\bm{Q}_0}$, one obtains the following non-zero order parameters in addition to those from Eq. (\ref{eq:non_zero_irreps_d4h}): 
\begin{eqnarray}
&&B_{2g}: \Re \left(\Delta_{\bm{Q}_0}^x + \Delta_{\bm{Q}_1}^x \right) \nonumber \\
&&A_{1g}: \Re \left(\Delta_{\bm{Q}_0}^x - \Delta_{\bm{Q}_1}^x \right). 
\end{eqnarray}
They transform in the same irreps as $\Re \left(\Delta_{\bm{Q}_0}^0 + \Delta_{\bm{Q}_1}^0 \right)$ and $\Re \left(\Delta_{\bm{Q}_0}^0 - \Delta_{\bm{Q}_1}^0 \right)$ and thus do not result in additional symmetry breaking compared to the $t_d = 0$ case.

\nocite{*}

\bibliography{apssamp}

\providecommand{\noopsort}[1]{}\providecommand{\singleletter}[1]{#1}%
\begin{thebibliography}{45}%
\makeatletter
\providecommand \@ifxundefined [1]{%
 \@ifx{#1\undefined}
}%
\providecommand \@ifnum [1]{%
 \ifnum #1\expandafter \@firstoftwo
 \else \expandafter \@secondoftwo
 \fi
}%
\providecommand \@ifx [1]{%
 \ifx #1\expandafter \@firstoftwo
 \else \expandafter \@secondoftwo
 \fi
}%
\providecommand \natexlab [1]{#1}%
\providecommand \enquote  [1]{``#1''}%
\providecommand \bibnamefont  [1]{#1}%
\providecommand \bibfnamefont [1]{#1}%
\providecommand \citenamefont [1]{#1}%
\providecommand \href@noop [0]{\@secondoftwo}%
\providecommand \href [0]{\begingroup \@sanitize@url \@href}%
\providecommand \@href[1]{\@@startlink{#1}\@@href}%
\providecommand \@@href[1]{\endgroup#1\@@endlink}%
\providecommand \@sanitize@url [0]{\catcode `\\12\catcode `\$12\catcode
  `\&12\catcode `\#12\catcode `\^12\catcode `\_12\catcode `\%12\relax}%
\providecommand \@@startlink[1]{}%
\providecommand \@@endlink[0]{}%
\providecommand \url  [0]{\begingroup\@sanitize@url \@url }%
\providecommand \@url [1]{\endgroup\@href {#1}{\urlprefix }}%
\providecommand \urlprefix  [0]{URL }%
\providecommand \Eprint [0]{\href }%
\providecommand \doibase [0]{https://doi.org/}%
\providecommand \selectlanguage [0]{\@gobble}%
\providecommand \bibinfo  [0]{\@secondoftwo}%
\providecommand \bibfield  [0]{\@secondoftwo}%
\providecommand \translation [1]{[#1]}%
\providecommand \BibitemOpen [0]{}%
\providecommand \bibitemStop [0]{}%
\providecommand \bibitemNoStop [0]{.\EOS\space}%
\providecommand \EOS [0]{\spacefactor3000\relax}%
\providecommand \BibitemShut  [1]{\csname bibitem#1\endcsname}%
\let\auto@bib@innerbib\@empty
\bibitem [{\citenamefont {DiMasi}\ \emph {et~al.}(1995)\citenamefont {DiMasi},
  \citenamefont {Aronson}, \citenamefont {Mansfield}, \citenamefont {Foran},\
  and\ \citenamefont {Lee}}]{dimasi1995chemical}%
  \BibitemOpen
  \bibfield  {author} {\bibinfo {author} {\bibfnamefont {E.}~\bibnamefont
  {DiMasi}}, \bibinfo {author} {\bibfnamefont {M.}~\bibnamefont {Aronson}},
  \bibinfo {author} {\bibfnamefont {J.}~\bibnamefont {Mansfield}}, \bibinfo
  {author} {\bibfnamefont {B.}~\bibnamefont {Foran}},\ and\ \bibinfo {author}
  {\bibfnamefont {S.}~\bibnamefont {Lee}},\ }\bibfield  {title} {\bibinfo
  {title} {Chemical pressure and charge-density waves in rare-earth
  tritellurides},\ }\href@noop {} {\bibfield  {journal} {\bibinfo  {journal}
  {Physical Review B}\ }\textbf {\bibinfo {volume} {52}},\ \bibinfo {pages}
  {14516} (\bibinfo {year} {1995})}\BibitemShut {NoStop}%
\bibitem [{\citenamefont {DiMasi}\ \emph {et~al.}(1996)\citenamefont {DiMasi},
  \citenamefont {Foran}, \citenamefont {Aronson},\ and\ \citenamefont
  {Lee}}]{dimasi1996stability}%
  \BibitemOpen
  \bibfield  {author} {\bibinfo {author} {\bibfnamefont {E.}~\bibnamefont
  {DiMasi}}, \bibinfo {author} {\bibfnamefont {B.}~\bibnamefont {Foran}},
  \bibinfo {author} {\bibfnamefont {M.}~\bibnamefont {Aronson}},\ and\ \bibinfo
  {author} {\bibfnamefont {S.}~\bibnamefont {Lee}},\ }\bibfield  {title}
  {\bibinfo {title} {Stability of charge-density waves under continuous
  variation of band filling in {LaTe$_{2- x}$Sb$_x$ $(0 \leq x \leq 1)$}},\
  }\href@noop {} {\bibfield  {journal} {\bibinfo  {journal} {Physical Review
  B}\ }\textbf {\bibinfo {volume} {54}},\ \bibinfo {pages} {13587} (\bibinfo
  {year} {1996})}\BibitemShut {NoStop}%
\bibitem [{\citenamefont {Brouet}\ \emph {et~al.}(2004)\citenamefont {Brouet},
  \citenamefont {Yang}, \citenamefont {Zhou}, \citenamefont {Hussain},
  \citenamefont {Ru}, \citenamefont {Shin}, \citenamefont {Fisher},\ and\
  \citenamefont {Shen}}]{brouet2004Fermi}%
  \BibitemOpen
  \bibfield  {author} {\bibinfo {author} {\bibfnamefont {V.}~\bibnamefont
  {Brouet}}, \bibinfo {author} {\bibfnamefont {W.~L.}\ \bibnamefont {Yang}},
  \bibinfo {author} {\bibfnamefont {X.~J.}\ \bibnamefont {Zhou}}, \bibinfo
  {author} {\bibfnamefont {Z.}~\bibnamefont {Hussain}}, \bibinfo {author}
  {\bibfnamefont {N.}~\bibnamefont {Ru}}, \bibinfo {author} {\bibfnamefont
  {K.~Y.}\ \bibnamefont {Shin}}, \bibinfo {author} {\bibfnamefont {I.~R.}\
  \bibnamefont {Fisher}},\ and\ \bibinfo {author} {\bibfnamefont {Z.~X.}\
  \bibnamefont {Shen}},\ }\bibfield  {title} {\bibinfo {title} {Fermi surface
  reconstruction in the {CDW} state of {CeTe$_3$} observed by photoemission},\
  }\href {https://doi.org/10.1103/PhysRevLett.93.126405} {\bibfield  {journal}
  {\bibinfo  {journal} {Phys. Rev. Lett.}\ }\textbf {\bibinfo {volume} {93}},\
  \bibinfo {pages} {126405} (\bibinfo {year} {2004})}\BibitemShut {NoStop}%
\bibitem [{\citenamefont {Komoda}\ \emph {et~al.}(2004)\citenamefont {Komoda},
  \citenamefont {Sato}, \citenamefont {Souma}, \citenamefont {Takahashi},
  \citenamefont {Ito},\ and\ \citenamefont {Suzuki}}]{komoda2004high}%
  \BibitemOpen
  \bibfield  {author} {\bibinfo {author} {\bibfnamefont {H.}~\bibnamefont
  {Komoda}}, \bibinfo {author} {\bibfnamefont {T.}~\bibnamefont {Sato}},
  \bibinfo {author} {\bibfnamefont {S.}~\bibnamefont {Souma}}, \bibinfo
  {author} {\bibfnamefont {T.}~\bibnamefont {Takahashi}}, \bibinfo {author}
  {\bibfnamefont {Y.}~\bibnamefont {Ito}},\ and\ \bibinfo {author}
  {\bibfnamefont {K.}~\bibnamefont {Suzuki}},\ }\bibfield  {title} {\bibinfo
  {title} {High-resolution angle-resolved photoemission study of incommensurate
  charge-density-wave compound {CeTe$_3$}},\ }\href
  {https://doi.org/10.1103/PhysRevB.70.195101} {\bibfield  {journal} {\bibinfo
  {journal} {Phys. Rev. B}\ }\textbf {\bibinfo {volume} {70}},\ \bibinfo
  {pages} {195101} (\bibinfo {year} {2004})}\BibitemShut {NoStop}%
\bibitem [{\citenamefont {Laverock}\ \emph {et~al.}(2005)\citenamefont
  {Laverock}, \citenamefont {Dugdale}, \citenamefont {Major}, \citenamefont
  {Alam}, \citenamefont {Ru}, \citenamefont {Fisher}, \citenamefont {Santi},\
  and\ \citenamefont {Bruno}}]{laverock2005Fermi}%
  \BibitemOpen
  \bibfield  {author} {\bibinfo {author} {\bibfnamefont {J.}~\bibnamefont
  {Laverock}}, \bibinfo {author} {\bibfnamefont {S.~B.}\ \bibnamefont
  {Dugdale}}, \bibinfo {author} {\bibfnamefont {Z.}~\bibnamefont {Major}},
  \bibinfo {author} {\bibfnamefont {M.~A.}\ \bibnamefont {Alam}}, \bibinfo
  {author} {\bibfnamefont {N.}~\bibnamefont {Ru}}, \bibinfo {author}
  {\bibfnamefont {I.~R.}\ \bibnamefont {Fisher}}, \bibinfo {author}
  {\bibfnamefont {G.}~\bibnamefont {Santi}},\ and\ \bibinfo {author}
  {\bibfnamefont {E.}~\bibnamefont {Bruno}},\ }\bibfield  {title} {\bibinfo
  {title} {Fermi surface nesting and charge-density wave formation in
  rare-earth tritellurides},\ }\href
  {https://doi.org/10.1103/PhysRevB.71.085114} {\bibfield  {journal} {\bibinfo
  {journal} {Phys. Rev. B}\ }\textbf {\bibinfo {volume} {71}},\ \bibinfo
  {pages} {085114} (\bibinfo {year} {2005})}\BibitemShut {NoStop}%
\bibitem [{\citenamefont {Kim}\ \emph {et~al.}(2006)\citenamefont {Kim},
  \citenamefont {Malliakas}, \citenamefont {Tomi\ifmmode~\acute{c}\else
  \'{c}\fi{}}, \citenamefont {Tessmer}, \citenamefont {Kanatzidis},\ and\
  \citenamefont {Billinge}}]{kim2006local}%
  \BibitemOpen
  \bibfield  {author} {\bibinfo {author} {\bibfnamefont {H.~J.}\ \bibnamefont
  {Kim}}, \bibinfo {author} {\bibfnamefont {C.~D.}\ \bibnamefont {Malliakas}},
  \bibinfo {author} {\bibfnamefont {A.~T.}\ \bibnamefont
  {Tomi\ifmmode~\acute{c}\else \'{c}\fi{}}}, \bibinfo {author} {\bibfnamefont
  {S.~H.}\ \bibnamefont {Tessmer}}, \bibinfo {author} {\bibfnamefont {M.~G.}\
  \bibnamefont {Kanatzidis}},\ and\ \bibinfo {author} {\bibfnamefont
  {S.~J.~L.}\ \bibnamefont {Billinge}},\ }\bibfield  {title} {\bibinfo {title}
  {Local atomic structure and discommensurations in the charge density wave of
  {CeTe$_3$}},\ }\href {https://doi.org/10.1103/PhysRevLett.96.226401}
  {\bibfield  {journal} {\bibinfo  {journal} {Phys. Rev. Lett.}\ }\textbf
  {\bibinfo {volume} {96}},\ \bibinfo {pages} {226401} (\bibinfo {year}
  {2006})}\BibitemShut {NoStop}%
\bibitem [{\citenamefont {Fang}\ \emph {et~al.}(2007)\citenamefont {Fang},
  \citenamefont {Ru}, \citenamefont {Fisher},\ and\ \citenamefont
  {Kapitulnik}}]{fang2007stm}%
  \BibitemOpen
  \bibfield  {author} {\bibinfo {author} {\bibfnamefont {A.}~\bibnamefont
  {Fang}}, \bibinfo {author} {\bibfnamefont {N.}~\bibnamefont {Ru}}, \bibinfo
  {author} {\bibfnamefont {I.~R.}\ \bibnamefont {Fisher}},\ and\ \bibinfo
  {author} {\bibfnamefont {A.}~\bibnamefont {Kapitulnik}},\ }\bibfield  {title}
  {\bibinfo {title} {{STM} studies of {TbTe$_3$}: Evidence for a fully
  incommensurate charge density wave},\ }\href
  {https://doi.org/10.1103/PhysRevLett.99.046401} {\bibfield  {journal}
  {\bibinfo  {journal} {Phys. Rev. Lett.}\ }\textbf {\bibinfo {volume} {99}},\
  \bibinfo {pages} {046401} (\bibinfo {year} {2007})}\BibitemShut {NoStop}%
\bibitem [{\citenamefont {Brouet}\ \emph {et~al.}(2008)\citenamefont {Brouet},
  \citenamefont {Yang}, \citenamefont {Zhou}, \citenamefont {Hussain},
  \citenamefont {Moore}, \citenamefont {He}, \citenamefont {Lu}, \citenamefont
  {Shen}, \citenamefont {Laverock}, \citenamefont {Dugdale}, \citenamefont
  {Ru},\ and\ \citenamefont {Fisher}}]{brouet2008angle}%
  \BibitemOpen
  \bibfield  {author} {\bibinfo {author} {\bibfnamefont {V.}~\bibnamefont
  {Brouet}}, \bibinfo {author} {\bibfnamefont {W.~L.}\ \bibnamefont {Yang}},
  \bibinfo {author} {\bibfnamefont {X.~J.}\ \bibnamefont {Zhou}}, \bibinfo
  {author} {\bibfnamefont {Z.}~\bibnamefont {Hussain}}, \bibinfo {author}
  {\bibfnamefont {R.~G.}\ \bibnamefont {Moore}}, \bibinfo {author}
  {\bibfnamefont {R.}~\bibnamefont {He}}, \bibinfo {author} {\bibfnamefont
  {D.~H.}\ \bibnamefont {Lu}}, \bibinfo {author} {\bibfnamefont {Z.~X.}\
  \bibnamefont {Shen}}, \bibinfo {author} {\bibfnamefont {J.}~\bibnamefont
  {Laverock}}, \bibinfo {author} {\bibfnamefont {S.~B.}\ \bibnamefont
  {Dugdale}}, \bibinfo {author} {\bibfnamefont {N.}~\bibnamefont {Ru}},\ and\
  \bibinfo {author} {\bibfnamefont {I.~R.}\ \bibnamefont {Fisher}},\ }\bibfield
   {title} {\bibinfo {title} {{Angle-resolved photoemission study of the
  evolution of band structure and charge density wave properties in {RTe$_3$}
  (R=Y, La, Ce, Sm, Gd, Tb, and Dy)}},\ }\href
  {https://doi.org/10.1103/PhysRevB.77.235104} {\bibfield  {journal} {\bibinfo
  {journal} {Phys. Rev. B}\ }\textbf {\bibinfo {volume} {77}},\ \bibinfo
  {pages} {235104} (\bibinfo {year} {2008})}\BibitemShut {NoStop}%
\bibitem [{\citenamefont {Ru}\ \emph {et~al.}(2008)\citenamefont {Ru},
  \citenamefont {Condron}, \citenamefont {Margulis}, \citenamefont {Shin},
  \citenamefont {Laverock}, \citenamefont {Dugdale}, \citenamefont {Toney},\
  and\ \citenamefont {Fisher}}]{ru2008effect}%
  \BibitemOpen
  \bibfield  {author} {\bibinfo {author} {\bibfnamefont {N.}~\bibnamefont
  {Ru}}, \bibinfo {author} {\bibfnamefont {C.~L.}\ \bibnamefont {Condron}},
  \bibinfo {author} {\bibfnamefont {G.~Y.}\ \bibnamefont {Margulis}}, \bibinfo
  {author} {\bibfnamefont {K.~Y.}\ \bibnamefont {Shin}}, \bibinfo {author}
  {\bibfnamefont {J.}~\bibnamefont {Laverock}}, \bibinfo {author}
  {\bibfnamefont {S.~B.}\ \bibnamefont {Dugdale}}, \bibinfo {author}
  {\bibfnamefont {M.~F.}\ \bibnamefont {Toney}},\ and\ \bibinfo {author}
  {\bibfnamefont {I.~R.}\ \bibnamefont {Fisher}},\ }\bibfield  {title}
  {\bibinfo {title} {Effect of chemical pressure on the charge density wave
  transition in rare-earth tritellurides {RTe$_3$}},\ }\href
  {https://doi.org/10.1103/PhysRevB.77.035114} {\bibfield  {journal} {\bibinfo
  {journal} {Phys. Rev. B}\ }\textbf {\bibinfo {volume} {77}},\ \bibinfo
  {pages} {035114} (\bibinfo {year} {2008})}\BibitemShut {NoStop}%
\bibitem [{\citenamefont {Sinchenko}\ \emph {et~al.}(2014)\citenamefont
  {Sinchenko}, \citenamefont {Lejay}, \citenamefont {Leynaud},\ and\
  \citenamefont {Monceau}}]{sinchenko2014unidirectional}%
  \BibitemOpen
  \bibfield  {author} {\bibinfo {author} {\bibfnamefont {A.}~\bibnamefont
  {Sinchenko}}, \bibinfo {author} {\bibfnamefont {P.}~\bibnamefont {Lejay}},
  \bibinfo {author} {\bibfnamefont {O.}~\bibnamefont {Leynaud}},\ and\ \bibinfo
  {author} {\bibfnamefont {P.}~\bibnamefont {Monceau}},\ }\bibfield  {title}
  {\bibinfo {title} {Unidirectional charge-density-wave sliding in
  two-dimensional rare-earth tritellurides},\ }\href@noop {} {\bibfield
  {journal} {\bibinfo  {journal} {Solid state communications}\ }\textbf
  {\bibinfo {volume} {188}},\ \bibinfo {pages} {67} (\bibinfo {year}
  {2014})}\BibitemShut {NoStop}%
\bibitem [{\citenamefont {Maschek}\ \emph {et~al.}(2015)\citenamefont
  {Maschek}, \citenamefont {Rosenkranz}, \citenamefont {Heid}, \citenamefont
  {Said}, \citenamefont {Giraldo-Gallo}, \citenamefont {Fisher},\ and\
  \citenamefont {Weber}}]{maschek2015wave}%
  \BibitemOpen
  \bibfield  {author} {\bibinfo {author} {\bibfnamefont {M.}~\bibnamefont
  {Maschek}}, \bibinfo {author} {\bibfnamefont {S.}~\bibnamefont {Rosenkranz}},
  \bibinfo {author} {\bibfnamefont {R.}~\bibnamefont {Heid}}, \bibinfo {author}
  {\bibfnamefont {A.~H.}\ \bibnamefont {Said}}, \bibinfo {author}
  {\bibfnamefont {P.}~\bibnamefont {Giraldo-Gallo}}, \bibinfo {author}
  {\bibfnamefont {I.~R.}\ \bibnamefont {Fisher}},\ and\ \bibinfo {author}
  {\bibfnamefont {F.}~\bibnamefont {Weber}},\ }\bibfield  {title} {\bibinfo
  {title} {Wave-vector-dependent electron-phonon coupling and the
  charge-density-wave transition in {TbTe$_3$}},\ }\href
  {https://doi.org/10.1103/PhysRevB.91.235146} {\bibfield  {journal} {\bibinfo
  {journal} {Phys. Rev. B}\ }\textbf {\bibinfo {volume} {91}},\ \bibinfo
  {pages} {235146} (\bibinfo {year} {2015})}\BibitemShut {NoStop}%
\bibitem [{\citenamefont {Kogar}\ \emph {et~al.}(2019)\citenamefont {Kogar},
  \citenamefont {Zong}, \citenamefont {Dolgirev}, \citenamefont {Shen},
  \citenamefont {Straquadine}, \citenamefont {Bie}, \citenamefont {Wang},
  \citenamefont {Rohwer}, \citenamefont {Tung}, \citenamefont {Yang},
  \citenamefont {Li}, \citenamefont {Yang}, \citenamefont {Weathersby},
  \citenamefont {Park}, \citenamefont {Kozina}, \citenamefont {Sie},
  \citenamefont {Wen}, \citenamefont {Jarillo-Herrero}, \citenamefont {Fisher},
  \citenamefont {Wang},\ and\ \citenamefont {Gedik}}]{kogar2019light}%
  \BibitemOpen
  \bibfield  {author} {\bibinfo {author} {\bibfnamefont {A.}~\bibnamefont
  {Kogar}}, \bibinfo {author} {\bibfnamefont {A.}~\bibnamefont {Zong}},
  \bibinfo {author} {\bibfnamefont {P.~E.}\ \bibnamefont {Dolgirev}}, \bibinfo
  {author} {\bibfnamefont {X.}~\bibnamefont {Shen}}, \bibinfo {author}
  {\bibfnamefont {J.}~\bibnamefont {Straquadine}}, \bibinfo {author}
  {\bibfnamefont {Y.-Q.}\ \bibnamefont {Bie}}, \bibinfo {author} {\bibfnamefont
  {X.}~\bibnamefont {Wang}}, \bibinfo {author} {\bibfnamefont {T.}~\bibnamefont
  {Rohwer}}, \bibinfo {author} {\bibfnamefont {I.-C.}\ \bibnamefont {Tung}},
  \bibinfo {author} {\bibfnamefont {Y.}~\bibnamefont {Yang}}, \bibinfo {author}
  {\bibfnamefont {R.}~\bibnamefont {Li}}, \bibinfo {author} {\bibfnamefont
  {J.}~\bibnamefont {Yang}}, \bibinfo {author} {\bibfnamefont {S.}~\bibnamefont
  {Weathersby}}, \bibinfo {author} {\bibfnamefont {S.}~\bibnamefont {Park}},
  \bibinfo {author} {\bibfnamefont {M.~E.}\ \bibnamefont {Kozina}}, \bibinfo
  {author} {\bibfnamefont {E.~J.}\ \bibnamefont {Sie}}, \bibinfo {author}
  {\bibfnamefont {H.}~\bibnamefont {Wen}}, \bibinfo {author} {\bibfnamefont
  {P.}~\bibnamefont {Jarillo-Herrero}}, \bibinfo {author} {\bibfnamefont
  {I.~R.}\ \bibnamefont {Fisher}}, \bibinfo {author} {\bibfnamefont
  {X.}~\bibnamefont {Wang}},\ and\ \bibinfo {author} {\bibfnamefont
  {N.}~\bibnamefont {Gedik}},\ }\bibfield  {title} {\bibinfo {title}
  {Light-induced charge density wave in {LaTe$_3$}},\ }\href
  {https://doi.org/10.1038/s41567-019-0705-3} {\bibfield  {journal} {\bibinfo
  {journal} {Nature Physics}\ }\textbf {\bibinfo {volume} {16}},\ \bibinfo
  {pages} {159–163} (\bibinfo {year} {2019})}\BibitemShut {NoStop}%
\bibitem [{\citenamefont {{Zong}}\ \emph {et~al.}(2019)\citenamefont {{Zong}},
  \citenamefont {{Kogar}}, \citenamefont {{Bie}}, \citenamefont {{Rohwer}},
  \citenamefont {{Lee}}, \citenamefont {{Baldini}}, \citenamefont
  {{Erge{\c{c}}en}}, \citenamefont {{Yilmaz}}, \citenamefont {{Freelon}},
  \citenamefont {{Sie}}, \citenamefont {{Zhou}}, \citenamefont {{Straquadine}},
  \citenamefont {{Walmsley}}, \citenamefont {{Dolgirev}}, \citenamefont
  {{Rozhkov}}, \citenamefont {{Fisher}}, \citenamefont {{Jarillo-Herrero}},
  \citenamefont {{Fine}},\ and\ \citenamefont {{Gedik}}}]{zong2019evidence}%
  \BibitemOpen
  \bibfield  {author} {\bibinfo {author} {\bibfnamefont {A.}~\bibnamefont
  {{Zong}}}, \bibinfo {author} {\bibfnamefont {A.}~\bibnamefont {{Kogar}}},
  \bibinfo {author} {\bibfnamefont {Y.-Q.}\ \bibnamefont {{Bie}}}, \bibinfo
  {author} {\bibfnamefont {T.}~\bibnamefont {{Rohwer}}}, \bibinfo {author}
  {\bibfnamefont {C.}~\bibnamefont {{Lee}}}, \bibinfo {author} {\bibfnamefont
  {E.}~\bibnamefont {{Baldini}}}, \bibinfo {author} {\bibfnamefont
  {E.}~\bibnamefont {{Erge{\c{c}}en}}}, \bibinfo {author} {\bibfnamefont
  {M.~B.}\ \bibnamefont {{Yilmaz}}}, \bibinfo {author} {\bibfnamefont
  {B.}~\bibnamefont {{Freelon}}}, \bibinfo {author} {\bibfnamefont {E.~J.}\
  \bibnamefont {{Sie}}}, \bibinfo {author} {\bibfnamefont {H.}~\bibnamefont
  {{Zhou}}}, \bibinfo {author} {\bibfnamefont {J.}~\bibnamefont
  {{Straquadine}}}, \bibinfo {author} {\bibfnamefont {P.}~\bibnamefont
  {{Walmsley}}}, \bibinfo {author} {\bibfnamefont {P.~E.}\ \bibnamefont
  {{Dolgirev}}}, \bibinfo {author} {\bibfnamefont {A.~V.}\ \bibnamefont
  {{Rozhkov}}}, \bibinfo {author} {\bibfnamefont {I.~R.}\ \bibnamefont
  {{Fisher}}}, \bibinfo {author} {\bibfnamefont {P.}~\bibnamefont
  {{Jarillo-Herrero}}}, \bibinfo {author} {\bibfnamefont {B.~V.}\ \bibnamefont
  {{Fine}}},\ and\ \bibinfo {author} {\bibfnamefont {N.}~\bibnamefont
  {{Gedik}}},\ }\bibfield  {title} {\bibinfo {title} {{Evidence for topological
  defects in a photoinduced phase transition}},\ }\href
  {https://doi.org/10.1038/s41567-018-0311-9} {\bibfield  {journal} {\bibinfo
  {journal} {Nature Physics}\ }\textbf {\bibinfo {volume} {15}},\ \bibinfo
  {pages} {27} (\bibinfo {year} {2019})},\ \Eprint
  {https://arxiv.org/abs/1806.02766} {arXiv:1806.02766 [cond-mat.mtrl-sci]}
  \BibitemShut {NoStop}%
\bibitem [{\citenamefont {Walmsley}\ \emph {et~al.}(2020)\citenamefont
  {Walmsley}, \citenamefont {Aeschlimann}, \citenamefont {Straquadine},
  \citenamefont {Giraldo-Gallo}, \citenamefont {Riggs}, \citenamefont {Chan},
  \citenamefont {McDonald},\ and\ \citenamefont
  {Fisher}}]{walmsley2020magnetic}%
  \BibitemOpen
  \bibfield  {author} {\bibinfo {author} {\bibfnamefont {P.}~\bibnamefont
  {Walmsley}}, \bibinfo {author} {\bibfnamefont {S.}~\bibnamefont
  {Aeschlimann}}, \bibinfo {author} {\bibfnamefont {J.~A.~W.}\ \bibnamefont
  {Straquadine}}, \bibinfo {author} {\bibfnamefont {P.}~\bibnamefont
  {Giraldo-Gallo}}, \bibinfo {author} {\bibfnamefont {S.~C.}\ \bibnamefont
  {Riggs}}, \bibinfo {author} {\bibfnamefont {M.~K.}\ \bibnamefont {Chan}},
  \bibinfo {author} {\bibfnamefont {R.~D.}\ \bibnamefont {McDonald}},\ and\
  \bibinfo {author} {\bibfnamefont {I.~R.}\ \bibnamefont {Fisher}},\ }\bibfield
   {title} {\bibinfo {title} {Magnetic breakdown and charge density wave
  formation: A quantum oscillation study of the rare-earth tritellurides},\
  }\href {https://doi.org/10.1103/PhysRevB.102.045150} {\bibfield  {journal}
  {\bibinfo  {journal} {Phys. Rev. B}\ }\textbf {\bibinfo {volume} {102}},\
  \bibinfo {pages} {045150} (\bibinfo {year} {2020})}\BibitemShut {NoStop}%
\bibitem [{\citenamefont {Sharma}\ \emph {et~al.}(2020)\citenamefont {Sharma},
  \citenamefont {Singh}, \citenamefont {Ahmed}, \citenamefont {Yu},
  \citenamefont {Walmsley}, \citenamefont {Fisher},\ and\ \citenamefont
  {Boyer}}]{sharma2020interplay}%
  \BibitemOpen
  \bibfield  {author} {\bibinfo {author} {\bibfnamefont {B.}~\bibnamefont
  {Sharma}}, \bibinfo {author} {\bibfnamefont {M.}~\bibnamefont {Singh}},
  \bibinfo {author} {\bibfnamefont {B.}~\bibnamefont {Ahmed}}, \bibinfo
  {author} {\bibfnamefont {B.}~\bibnamefont {Yu}}, \bibinfo {author}
  {\bibfnamefont {P.}~\bibnamefont {Walmsley}}, \bibinfo {author}
  {\bibfnamefont {I.~R.}\ \bibnamefont {Fisher}},\ and\ \bibinfo {author}
  {\bibfnamefont {M.~C.}\ \bibnamefont {Boyer}},\ }\bibfield  {title} {\bibinfo
  {title} {Interplay of charge density wave states and strain at the surface of
  {CeTe$_2$}},\ }\href {https://doi.org/10.1103/PhysRevB.101.245423} {\bibfield
   {journal} {\bibinfo  {journal} {Phys. Rev. B}\ }\textbf {\bibinfo {volume}
  {101}},\ \bibinfo {pages} {245423} (\bibinfo {year} {2020})}\BibitemShut
  {NoStop}%
\bibitem [{\citenamefont {Liu}\ \emph {et~al.}(2020)\citenamefont {Liu},
  \citenamefont {Huan}, \citenamefont {Liu}, \citenamefont {Liu}, \citenamefont
  {Liu}, \citenamefont {Lu}, \citenamefont {Huang}, \citenamefont {Jiang},
  \citenamefont {Wang}, \citenamefont {Yu}, \citenamefont {Zou}, \citenamefont
  {Guo},\ and\ \citenamefont {Shen}}]{liu2020electronic}%
  \BibitemOpen
  \bibfield  {author} {\bibinfo {author} {\bibfnamefont {J.~S.}\ \bibnamefont
  {Liu}}, \bibinfo {author} {\bibfnamefont {S.~C.}\ \bibnamefont {Huan}},
  \bibinfo {author} {\bibfnamefont {Z.~H.}\ \bibnamefont {Liu}}, \bibinfo
  {author} {\bibfnamefont {W.~L.}\ \bibnamefont {Liu}}, \bibinfo {author}
  {\bibfnamefont {Z.~T.}\ \bibnamefont {Liu}}, \bibinfo {author} {\bibfnamefont
  {X.~L.}\ \bibnamefont {Lu}}, \bibinfo {author} {\bibfnamefont
  {Z.}~\bibnamefont {Huang}}, \bibinfo {author} {\bibfnamefont {Z.~C.}\
  \bibnamefont {Jiang}}, \bibinfo {author} {\bibfnamefont {X.}~\bibnamefont
  {Wang}}, \bibinfo {author} {\bibfnamefont {N.}~\bibnamefont {Yu}}, \bibinfo
  {author} {\bibfnamefont {Z.~Q.}\ \bibnamefont {Zou}}, \bibinfo {author}
  {\bibfnamefont {Y.~F.}\ \bibnamefont {Guo}},\ and\ \bibinfo {author}
  {\bibfnamefont {D.~W.}\ \bibnamefont {Shen}},\ }\bibfield  {title} {\bibinfo
  {title} {Electronic structure of the high-mobility two-dimensional
  antiferromagnetic metal {GdTe$_3$}},\ }\href
  {https://doi.org/10.1103/PhysRevMaterials.4.114005} {\bibfield  {journal}
  {\bibinfo  {journal} {Phys. Rev. Mater.}\ }\textbf {\bibinfo {volume} {4}},\
  \bibinfo {pages} {114005} (\bibinfo {year} {2020})}\BibitemShut {NoStop}%
\bibitem [{\citenamefont {Lei}\ \emph {et~al.}(2021{\natexlab{a}})\citenamefont
  {Lei}, \citenamefont {Teicher}, \citenamefont {Topp}, \citenamefont {Cai},
  \citenamefont {Lin}, \citenamefont {Cheng}, \citenamefont {Salters},
  \citenamefont {Rodolakis}, \citenamefont {McChesney}, \citenamefont
  {Lapidus}, \citenamefont {Yao}, \citenamefont {Krivenkov}, \citenamefont
  {Marchenko}, \citenamefont {Varykhalov}, \citenamefont {Ast}, \citenamefont
  {Car}, \citenamefont {Cano}, \citenamefont {Vergniory}, \citenamefont {Ong},\
  and\ \citenamefont {Schoop}}]{lei2021band}%
  \BibitemOpen
  \bibfield  {author} {\bibinfo {author} {\bibfnamefont {S.}~\bibnamefont
  {Lei}}, \bibinfo {author} {\bibfnamefont {S.~M.~L.}\ \bibnamefont {Teicher}},
  \bibinfo {author} {\bibfnamefont {A.}~\bibnamefont {Topp}}, \bibinfo {author}
  {\bibfnamefont {K.}~\bibnamefont {Cai}}, \bibinfo {author} {\bibfnamefont
  {J.}~\bibnamefont {Lin}}, \bibinfo {author} {\bibfnamefont {G.}~\bibnamefont
  {Cheng}}, \bibinfo {author} {\bibfnamefont {T.~H.}\ \bibnamefont {Salters}},
  \bibinfo {author} {\bibfnamefont {F.}~\bibnamefont {Rodolakis}}, \bibinfo
  {author} {\bibfnamefont {J.~L.}\ \bibnamefont {McChesney}}, \bibinfo {author}
  {\bibfnamefont {S.}~\bibnamefont {Lapidus}}, \bibinfo {author} {\bibfnamefont
  {N.}~\bibnamefont {Yao}}, \bibinfo {author} {\bibfnamefont {M.}~\bibnamefont
  {Krivenkov}}, \bibinfo {author} {\bibfnamefont {D.}~\bibnamefont
  {Marchenko}}, \bibinfo {author} {\bibfnamefont {A.}~\bibnamefont
  {Varykhalov}}, \bibinfo {author} {\bibfnamefont {C.~R.}\ \bibnamefont {Ast}},
  \bibinfo {author} {\bibfnamefont {R.}~\bibnamefont {Car}}, \bibinfo {author}
  {\bibfnamefont {J.}~\bibnamefont {Cano}}, \bibinfo {author} {\bibfnamefont
  {M.~G.}\ \bibnamefont {Vergniory}}, \bibinfo {author} {\bibfnamefont {N.~P.}\
  \bibnamefont {Ong}},\ and\ \bibinfo {author} {\bibfnamefont {L.~M.}\
  \bibnamefont {Schoop}},\ }\bibfield  {title} {\bibinfo {title} {Band
  engineering of {Dirac} semimetals using charge density waves},\ }\bibfield
  {journal} {\bibinfo  {journal} {Advanced Materials}\ }\textbf {\bibinfo
  {volume} {33}},\ \href {https://doi.org/10.1002/adma.202101591}
  {10.1002/adma.202101591} (\bibinfo {year} {2021}{\natexlab{a}})\BibitemShut
  {NoStop}%
\bibitem [{\citenamefont {Zhou}\ \emph {et~al.}(2021)\citenamefont {Zhou},
  \citenamefont {Williams}, \citenamefont {Sun}, \citenamefont {Malliakas},
  \citenamefont {Kanatzidis}, \citenamefont {Kemper},\ and\ \citenamefont
  {Ruan}}]{zhou2021nonequilibrium}%
  \BibitemOpen
  \bibfield  {author} {\bibinfo {author} {\bibfnamefont {F.}~\bibnamefont
  {Zhou}}, \bibinfo {author} {\bibfnamefont {J.}~\bibnamefont {Williams}},
  \bibinfo {author} {\bibfnamefont {S.}~\bibnamefont {Sun}}, \bibinfo {author}
  {\bibfnamefont {C.~D.}\ \bibnamefont {Malliakas}}, \bibinfo {author}
  {\bibfnamefont {M.~G.}\ \bibnamefont {Kanatzidis}}, \bibinfo {author}
  {\bibfnamefont {A.~F.}\ \bibnamefont {Kemper}},\ and\ \bibinfo {author}
  {\bibfnamefont {C.-Y.}\ \bibnamefont {Ruan}},\ }\bibfield  {title} {\bibinfo
  {title} {Nonequilibrium dynamics of spontaneous symmetry breaking into a
  hidden state of charge-density wave},\ }\href@noop {} {\bibfield  {journal}
  {\bibinfo  {journal} {Nature communications}\ }\textbf {\bibinfo {volume}
  {12}},\ \bibinfo {pages} {566} (\bibinfo {year} {2021})}\BibitemShut
  {NoStop}%
\bibitem [{\citenamefont {Gonz{\'a}lez-Vallejo}\ \emph
  {et~al.}(2022)\citenamefont {Gonz{\'a}lez-Vallejo}, \citenamefont {Jacques},
  \citenamefont {Boschetto}, \citenamefont {Rizza}, \citenamefont {Hadj-Azzem},
  \citenamefont {Faure}, \citenamefont {Bolloc'h} \emph
  {et~al.}}]{gonzalez2022time}%
  \BibitemOpen
  \bibfield  {author} {\bibinfo {author} {\bibfnamefont {I.}~\bibnamefont
  {Gonz{\'a}lez-Vallejo}}, \bibinfo {author} {\bibfnamefont {V.}~\bibnamefont
  {Jacques}}, \bibinfo {author} {\bibfnamefont {D.}~\bibnamefont {Boschetto}},
  \bibinfo {author} {\bibfnamefont {G.}~\bibnamefont {Rizza}}, \bibinfo
  {author} {\bibfnamefont {A.}~\bibnamefont {Hadj-Azzem}}, \bibinfo {author}
  {\bibfnamefont {J.}~\bibnamefont {Faure}}, \bibinfo {author} {\bibfnamefont
  {L.}~\bibnamefont {Bolloc'h}}, \emph {et~al.},\ }\bibfield  {title} {\bibinfo
  {title} {Time-resolved structural dynamics of the out-of-equilibrium charge
  density wave phase transition in {GdTe$_3$}},\ }\href@noop {} {\bibfield
  {journal} {\bibinfo  {journal} {Structural Dynamics}\ }\textbf {\bibinfo
  {volume} {9}} (\bibinfo {year} {2022})}\BibitemShut {NoStop}%
\bibitem [{\citenamefont {Wang}\ \emph {et~al.}(2022)\citenamefont {Wang},
  \citenamefont {Petrides}, \citenamefont {McNamara}, \citenamefont {Hosen},
  \citenamefont {Lei}, \citenamefont {Wu}, \citenamefont {Hart}, \citenamefont
  {Lv}, \citenamefont {Yan}, \citenamefont {Xiao}, \citenamefont {Cha},
  \citenamefont {Narang}, \citenamefont {Schoop},\ and\ \citenamefont
  {Burch}}]{wang2022axial}%
  \BibitemOpen
  \bibfield  {author} {\bibinfo {author} {\bibfnamefont {Y.}~\bibnamefont
  {Wang}}, \bibinfo {author} {\bibfnamefont {I.}~\bibnamefont {Petrides}},
  \bibinfo {author} {\bibfnamefont {G.}~\bibnamefont {McNamara}}, \bibinfo
  {author} {\bibfnamefont {M.~M.}\ \bibnamefont {Hosen}}, \bibinfo {author}
  {\bibfnamefont {S.}~\bibnamefont {Lei}}, \bibinfo {author} {\bibfnamefont
  {Y.-C.}\ \bibnamefont {Wu}}, \bibinfo {author} {\bibfnamefont {J.~L.}\
  \bibnamefont {Hart}}, \bibinfo {author} {\bibfnamefont {H.}~\bibnamefont
  {Lv}}, \bibinfo {author} {\bibfnamefont {J.}~\bibnamefont {Yan}}, \bibinfo
  {author} {\bibfnamefont {D.}~\bibnamefont {Xiao}}, \bibinfo {author}
  {\bibfnamefont {J.~J.}\ \bibnamefont {Cha}}, \bibinfo {author} {\bibfnamefont
  {P.}~\bibnamefont {Narang}}, \bibinfo {author} {\bibfnamefont {L.~M.}\
  \bibnamefont {Schoop}},\ and\ \bibinfo {author} {\bibfnamefont {K.~S.}\
  \bibnamefont {Burch}},\ }\bibfield  {title} {\bibinfo {title} {{Axial Higgs
  mode detected by quantum pathway interference in RTe3}},\ }\href
  {https://doi.org/10.1038/s41586-022-04746-6} {\bibfield  {journal} {\bibinfo
  {journal} {Nature}\ }\textbf {\bibinfo {volume} {606}},\ \bibinfo {pages}
  {896} (\bibinfo {year} {2022})}\BibitemShut {NoStop}%
\bibitem [{\citenamefont {Straquadine}\ \emph {et~al.}(2022)\citenamefont
  {Straquadine}, \citenamefont {Ikeda},\ and\ \citenamefont
  {Fisher}}]{straquadine2022evidence}%
  \BibitemOpen
  \bibfield  {author} {\bibinfo {author} {\bibfnamefont {J.~A.~W.}\
  \bibnamefont {Straquadine}}, \bibinfo {author} {\bibfnamefont {M.~S.}\
  \bibnamefont {Ikeda}},\ and\ \bibinfo {author} {\bibfnamefont {I.~R.}\
  \bibnamefont {Fisher}},\ }\bibfield  {title} {\bibinfo {title} {Evidence for
  realignment of the charge density wave state in ${\mathrm{erte}}_{3}$ and
  ${\mathrm{tmte}}_{3}$ under uniaxial stress via elastocaloric and
  elastoresistivity measurements},\ }\href
  {https://doi.org/10.1103/PhysRevX.12.021046} {\bibfield  {journal} {\bibinfo
  {journal} {Phys. Rev. X}\ }\textbf {\bibinfo {volume} {12}},\ \bibinfo
  {pages} {021046} (\bibinfo {year} {2022})}\BibitemShut {NoStop}%
\bibitem [{\citenamefont {Singh}\ \emph {et~al.}(2023)\citenamefont {Singh},
  \citenamefont {Bachmann}, \citenamefont {Sanchez}, \citenamefont {Pandey},
  \citenamefont {Kapitulnik}, \citenamefont {Kim}, \citenamefont {Ryan},
  \citenamefont {Kivelson},\ and\ \citenamefont {Fisher}}]{Singh2023}%
  \BibitemOpen
  \bibfield  {author} {\bibinfo {author} {\bibfnamefont {A.~G.}\ \bibnamefont
  {Singh}}, \bibinfo {author} {\bibfnamefont {M.~D.}\ \bibnamefont {Bachmann}},
  \bibinfo {author} {\bibfnamefont {J.~J.}\ \bibnamefont {Sanchez}}, \bibinfo
  {author} {\bibfnamefont {A.}~\bibnamefont {Pandey}}, \bibinfo {author}
  {\bibfnamefont {A.}~\bibnamefont {Kapitulnik}}, \bibinfo {author}
  {\bibfnamefont {J.~W.}\ \bibnamefont {Kim}}, \bibinfo {author} {\bibfnamefont
  {P.~J.}\ \bibnamefont {Ryan}}, \bibinfo {author} {\bibfnamefont {S.~A.}\
  \bibnamefont {Kivelson}},\ and\ \bibinfo {author} {\bibfnamefont {I.~R.}\
  \bibnamefont {Fisher}},\ }\bibfield  {title} {\bibinfo {title} {Emergent
  tetragonality in a fundamentally orthorhombic material},\ }\href@noop {}
  {\bibfield  {journal} {\bibinfo  {journal} {arXiv:2306.14755}\ } (\bibinfo
  {year} {2023})}\BibitemShut {NoStop}%
\bibitem [{\citenamefont {Kivelson}\ \emph {et~al.}(2023)\citenamefont
  {Kivelson}, \citenamefont {Pandey}, \citenamefont {Singh}, \citenamefont
  {Kapitulnik},\ and\ \citenamefont {Fisher}}]{Kivelson2023}%
  \BibitemOpen
  \bibfield  {author} {\bibinfo {author} {\bibfnamefont {S.~A.}\ \bibnamefont
  {Kivelson}}, \bibinfo {author} {\bibfnamefont {A.}~\bibnamefont {Pandey}},
  \bibinfo {author} {\bibfnamefont {A.~G.}\ \bibnamefont {Singh}}, \bibinfo
  {author} {\bibfnamefont {A.}~\bibnamefont {Kapitulnik}},\ and\ \bibinfo
  {author} {\bibfnamefont {I.~R.}\ \bibnamefont {Fisher}},\ }\bibfield  {title}
  {\bibinfo {title} {Emergent {${\mathbb{Z}}_{2}$} symmetry near a charge
  density wave multicritical point},\ }\href
  {https://doi.org/10.1103/PhysRevB.108.205141} {\bibfield  {journal} {\bibinfo
   {journal} {Phys. Rev. B}\ }\textbf {\bibinfo {volume} {108}},\ \bibinfo
  {pages} {205141} (\bibinfo {year} {2023})}\BibitemShut {NoStop}%
\bibitem [{\citenamefont {Chikina}\ \emph {et~al.}(2023)\citenamefont
  {Chikina}, \citenamefont {Lund}, \citenamefont {Bianchi}, \citenamefont
  {Curcio}, \citenamefont {Dalgaard}, \citenamefont {Bremholm}, \citenamefont
  {Lei}, \citenamefont {Singha}, \citenamefont {Schoop},\ and\ \citenamefont
  {Hofmann}}]{chikina2023charge}%
  \BibitemOpen
  \bibfield  {author} {\bibinfo {author} {\bibfnamefont {A.}~\bibnamefont
  {Chikina}}, \bibinfo {author} {\bibfnamefont {H.}~\bibnamefont {Lund}},
  \bibinfo {author} {\bibfnamefont {M.}~\bibnamefont {Bianchi}}, \bibinfo
  {author} {\bibfnamefont {D.}~\bibnamefont {Curcio}}, \bibinfo {author}
  {\bibfnamefont {K.~J.}\ \bibnamefont {Dalgaard}}, \bibinfo {author}
  {\bibfnamefont {M.}~\bibnamefont {Bremholm}}, \bibinfo {author}
  {\bibfnamefont {S.}~\bibnamefont {Lei}}, \bibinfo {author} {\bibfnamefont
  {R.}~\bibnamefont {Singha}}, \bibinfo {author} {\bibfnamefont {L.~M.}\
  \bibnamefont {Schoop}},\ and\ \bibinfo {author} {\bibfnamefont
  {P.}~\bibnamefont {Hofmann}},\ }\bibfield  {title} {\bibinfo {title} {Charge
  density wave generated fermi surfaces in {NdTe$_3$}},\ }\href
  {https://doi.org/10.1103/PhysRevB.107.L161103} {\bibfield  {journal}
  {\bibinfo  {journal} {Phys. Rev. B}\ }\textbf {\bibinfo {volume} {107}},\
  \bibinfo {pages} {L161103} (\bibinfo {year} {2023})}\BibitemShut {NoStop}%
\bibitem [{\citenamefont {Raghavan}\ \emph {et~al.}(2023)\citenamefont
  {Raghavan}, \citenamefont {Romanelli}, \citenamefont {May-Mann},
  \citenamefont {Aishwarya}, \citenamefont {Aggarwal}, \citenamefont {Singh},
  \citenamefont {Bachmann}, \citenamefont {Schoop}, \citenamefont {Fradkin},
  \citenamefont {Fisher},\ and\ \citenamefont {Madhavan}}]{raghavan2023atomic}%
  \BibitemOpen
  \bibfield  {author} {\bibinfo {author} {\bibfnamefont {A.}~\bibnamefont
  {Raghavan}}, \bibinfo {author} {\bibfnamefont {M.}~\bibnamefont {Romanelli}},
  \bibinfo {author} {\bibfnamefont {J.}~\bibnamefont {May-Mann}}, \bibinfo
  {author} {\bibfnamefont {A.}~\bibnamefont {Aishwarya}}, \bibinfo {author}
  {\bibfnamefont {L.}~\bibnamefont {Aggarwal}}, \bibinfo {author}
  {\bibfnamefont {A.~G.}\ \bibnamefont {Singh}}, \bibinfo {author}
  {\bibfnamefont {M.~D.}\ \bibnamefont {Bachmann}}, \bibinfo {author}
  {\bibfnamefont {L.~M.}\ \bibnamefont {Schoop}}, \bibinfo {author}
  {\bibfnamefont {E.}~\bibnamefont {Fradkin}}, \bibinfo {author} {\bibfnamefont
  {I.~R.}\ \bibnamefont {Fisher}},\ and\ \bibinfo {author} {\bibfnamefont
  {V.}~\bibnamefont {Madhavan}},\ }\bibfield  {title} {\bibinfo {title}
  {Atomic-scale visualization of a cascade of magnetic orders in the layered
  antiferromagnet {GdTe$_3$}},\ }\href@noop {} {\bibfield  {journal} {\bibinfo
  {journal} {arXiv:2308.15691}\ } (\bibinfo {year} {2023})}\BibitemShut
  {NoStop}%
\bibitem [{\citenamefont {Kim}\ \emph {et~al.}(2024)\citenamefont {Kim},
  \citenamefont {Orenstein}, \citenamefont {Singh}, \citenamefont {Fisher},
  \citenamefont {Reis},\ and\ \citenamefont {Trigo}}]{Kim2024}%
  \BibitemOpen
  \bibfield  {author} {\bibinfo {author} {\bibfnamefont {S.}~\bibnamefont
  {Kim}}, \bibinfo {author} {\bibfnamefont {G.}~\bibnamefont {Orenstein}},
  \bibinfo {author} {\bibfnamefont {A.~G.}\ \bibnamefont {Singh}}, \bibinfo
  {author} {\bibfnamefont {I.~R.}\ \bibnamefont {Fisher}}, \bibinfo {author}
  {\bibfnamefont {D.~A.}\ \bibnamefont {Reis}},\ and\ \bibinfo {author}
  {\bibfnamefont {M.}~\bibnamefont {Trigo}},\ }\bibfield  {title} {\bibinfo
  {title} {Ultrafast measurements under anisotropic strain reveal near
  equivalence of competing charge orders in {TbTe$_3$}},\ }\href@noop {}
  {\bibfield  {journal} {\bibinfo  {journal} {arXiv:2401.17437}\ } (\bibinfo
  {year} {2024})}\BibitemShut {NoStop}%
\bibitem [{\citenamefont {Singh}\ \emph {et~al.}(2024)\citenamefont {Singh},
  \citenamefont {McNamara}, \citenamefont {Kim}, \citenamefont {Siddique},
  \citenamefont {Funni}, \citenamefont {Zhang}, \citenamefont {Luo},
  \citenamefont {Sakrikar}, \citenamefont {Kenny}, \citenamefont {Singha},
  \citenamefont {Alekseev}, \citenamefont {Ghorashi}, \citenamefont {Hicken},
  \citenamefont {Baine}, \citenamefont {Luetkens}, \citenamefont {Wang},
  \citenamefont {Plisson}, \citenamefont {annd Connor A.~Occhialini},
  \citenamefont {Comin}, \citenamefont {Graf}, \citenamefont {Zhao},
  \citenamefont {Cano}, \citenamefont {Fernandes}, \citenamefont {Cha},
  \citenamefont {Schoop},\ and\ \citenamefont {Burch}}]{singh20204ferro}%
  \BibitemOpen
  \bibfield  {author} {\bibinfo {author} {\bibfnamefont {B.}~\bibnamefont
  {Singh}}, \bibinfo {author} {\bibfnamefont {G.}~\bibnamefont {McNamara}},
  \bibinfo {author} {\bibfnamefont {K.-M.}\ \bibnamefont {Kim}}, \bibinfo
  {author} {\bibfnamefont {S.}~\bibnamefont {Siddique}}, \bibinfo {author}
  {\bibfnamefont {S.~D.}\ \bibnamefont {Funni}}, \bibinfo {author}
  {\bibfnamefont {W.}~\bibnamefont {Zhang}}, \bibinfo {author} {\bibfnamefont
  {X.}~\bibnamefont {Luo}}, \bibinfo {author} {\bibfnamefont {P.}~\bibnamefont
  {Sakrikar}}, \bibinfo {author} {\bibfnamefont {E.~M.}\ \bibnamefont {Kenny}},
  \bibinfo {author} {\bibfnamefont {R.}~\bibnamefont {Singha}}, \bibinfo
  {author} {\bibfnamefont {S.}~\bibnamefont {Alekseev}}, \bibinfo {author}
  {\bibfnamefont {S.~A.~A.}\ \bibnamefont {Ghorashi}}, \bibinfo {author}
  {\bibfnamefont {T.}~\bibnamefont {Hicken}}, \bibinfo {author} {\bibfnamefont
  {C.}~\bibnamefont {Baine}}, \bibinfo {author} {\bibfnamefont
  {H.}~\bibnamefont {Luetkens}}, \bibinfo {author} {\bibfnamefont
  {Y.}~\bibnamefont {Wang}}, \bibinfo {author} {\bibfnamefont {V.}~\bibnamefont
  {Plisson}}, \bibinfo {author} {\bibfnamefont {M.~G.}\ \bibnamefont {annd
  Connor A.~Occhialini}}, \bibinfo {author} {\bibfnamefont {R.}~\bibnamefont
  {Comin}}, \bibinfo {author} {\bibfnamefont {M.~J.}\ \bibnamefont {Graf}},
  \bibinfo {author} {\bibfnamefont {L.}~\bibnamefont {Zhao}}, \bibinfo {author}
  {\bibfnamefont {J.}~\bibnamefont {Cano}}, \bibinfo {author} {\bibfnamefont
  {R.~M.}\ \bibnamefont {Fernandes}}, \bibinfo {author} {\bibfnamefont {J.~J.}\
  \bibnamefont {Cha}}, \bibinfo {author} {\bibfnamefont {L.~M.}\ \bibnamefont
  {Schoop}},\ and\ \bibinfo {author} {\bibfnamefont {K.~S.}\ \bibnamefont
  {Burch}},\ }\bibfield  {title} {\bibinfo {title} {Ferro-rotational charge
  density wave order as the origin of the axial {Higgs} mode in {RTe$_3$}},\
  }\href@noop {} {\bibfield  {journal} {\bibinfo  {journal}
  {\textit{submitted}}\ } (\bibinfo {year} {2024})}\BibitemShut {NoStop}%
\bibitem [{\citenamefont {Akatsuka}\ \emph {et~al.}(2024)\citenamefont
  {Akatsuka}, \citenamefont {Esser}, \citenamefont {Okumura}, \citenamefont
  {Yambe}, \citenamefont {Yamada}, \citenamefont {Hirschmann}, \citenamefont
  {Aji}, \citenamefont {White}, \citenamefont {Gao}, \citenamefont {Onuki},
  \citenamefont {Arima}, \citenamefont {Nakajima},\ and\ \citenamefont
  {Hirschberger}}]{akatsuka2024noncoplanar}%
  \BibitemOpen
  \bibfield  {author} {\bibinfo {author} {\bibfnamefont {S.}~\bibnamefont
  {Akatsuka}}, \bibinfo {author} {\bibfnamefont {S.}~\bibnamefont {Esser}},
  \bibinfo {author} {\bibfnamefont {S.}~\bibnamefont {Okumura}}, \bibinfo
  {author} {\bibfnamefont {R.}~\bibnamefont {Yambe}}, \bibinfo {author}
  {\bibfnamefont {R.}~\bibnamefont {Yamada}}, \bibinfo {author} {\bibfnamefont
  {M.~M.}\ \bibnamefont {Hirschmann}}, \bibinfo {author} {\bibfnamefont
  {S.}~\bibnamefont {Aji}}, \bibinfo {author} {\bibfnamefont {J.~S.}\
  \bibnamefont {White}}, \bibinfo {author} {\bibfnamefont {S.}~\bibnamefont
  {Gao}}, \bibinfo {author} {\bibfnamefont {Y.}~\bibnamefont {Onuki}}, \bibinfo
  {author} {\bibfnamefont {T.-H.}\ \bibnamefont {Arima}}, \bibinfo {author}
  {\bibfnamefont {T.}~\bibnamefont {Nakajima}},\ and\ \bibinfo {author}
  {\bibfnamefont {M.}~\bibnamefont {Hirschberger}},\ }\href@noop {} {\bibinfo
  {title} {Non-coplanar helimagnetism in the layered van-der-{Waals} metal
  {DyTe$_3$}}} (\bibinfo {year} {2024}),\ \Eprint
  {https://arxiv.org/abs/2306.04854} {arXiv:2306.04854 [cond-mat.mtrl-sci]}
  \BibitemShut {NoStop}%
\bibitem [{\citenamefont {Yao}\ \emph {et~al.}(2006)\citenamefont {Yao},
  \citenamefont {Robertson}, \citenamefont {Kim},\ and\ \citenamefont
  {Kivelson}}]{PhysRevB.74.245126}%
  \BibitemOpen
  \bibfield  {author} {\bibinfo {author} {\bibfnamefont {H.}~\bibnamefont
  {Yao}}, \bibinfo {author} {\bibfnamefont {J.~A.}\ \bibnamefont {Robertson}},
  \bibinfo {author} {\bibfnamefont {E.-A.}\ \bibnamefont {Kim}},\ and\ \bibinfo
  {author} {\bibfnamefont {S.~A.}\ \bibnamefont {Kivelson}},\ }\bibfield
  {title} {\bibinfo {title} {Theory of stripes in quasi-two-dimensional
  rare-earth tellurides},\ }\href {https://doi.org/10.1103/PhysRevB.74.245126}
  {\bibfield  {journal} {\bibinfo  {journal} {Phys. Rev. B}\ }\textbf {\bibinfo
  {volume} {74}},\ \bibinfo {pages} {245126} (\bibinfo {year}
  {2006})}\BibitemShut {NoStop}%
\bibitem [{\citenamefont {Johannes}\ and\ \citenamefont
  {Mazin}(2008)}]{johannes2008fermi}%
  \BibitemOpen
  \bibfield  {author} {\bibinfo {author} {\bibfnamefont {M.}~\bibnamefont
  {Johannes}}\ and\ \bibinfo {author} {\bibfnamefont {I.}~\bibnamefont
  {Mazin}},\ }\bibfield  {title} {\bibinfo {title} {Fermi surface nesting and
  the origin of charge density waves in metals},\ }\href@noop {} {\bibfield
  {journal} {\bibinfo  {journal} {Physical Review B}\ }\textbf {\bibinfo
  {volume} {77}},\ \bibinfo {pages} {165135} (\bibinfo {year}
  {2008})}\BibitemShut {NoStop}%
\bibitem [{\citenamefont {Klemenz}\ \emph {et~al.}(2020)\citenamefont
  {Klemenz}, \citenamefont {Schoop},\ and\ \citenamefont
  {Cano}}]{PhysRevB.101.165121}%
  \BibitemOpen
  \bibfield  {author} {\bibinfo {author} {\bibfnamefont {S.}~\bibnamefont
  {Klemenz}}, \bibinfo {author} {\bibfnamefont {L.}~\bibnamefont {Schoop}},\
  and\ \bibinfo {author} {\bibfnamefont {J.}~\bibnamefont {Cano}},\ }\bibfield
  {title} {\bibinfo {title} {Systematic study of stacked square nets: From
  {Dirac} fermions to material realizations},\ }\href
  {https://doi.org/10.1103/PhysRevB.101.165121} {\bibfield  {journal} {\bibinfo
   {journal} {Phys. Rev. B}\ }\textbf {\bibinfo {volume} {101}},\ \bibinfo
  {pages} {165121} (\bibinfo {year} {2020})}\BibitemShut {NoStop}%
\bibitem [{\citenamefont {Graser}\ \emph {et~al.}(2009)\citenamefont {Graser},
  \citenamefont {Maier}, \citenamefont {Hirschfeld},\ and\ \citenamefont
  {Scalapino}}]{Graser_2009}%
  \BibitemOpen
  \bibfield  {author} {\bibinfo {author} {\bibfnamefont {S.}~\bibnamefont
  {Graser}}, \bibinfo {author} {\bibfnamefont {T.~A.}\ \bibnamefont {Maier}},
  \bibinfo {author} {\bibfnamefont {P.~J.}\ \bibnamefont {Hirschfeld}},\ and\
  \bibinfo {author} {\bibfnamefont {D.~J.}\ \bibnamefont {Scalapino}},\
  }\bibfield  {title} {\bibinfo {title} {Near-degeneracy of several pairing
  channels in multiorbital models for the {Fe} pnictides},\ }\href
  {https://doi.org/10.1088/1367-2630/11/2/025016} {\bibfield  {journal}
  {\bibinfo  {journal} {New Journal of Physics}\ }\textbf {\bibinfo {volume}
  {11}},\ \bibinfo {pages} {025016} (\bibinfo {year} {2009})}\BibitemShut
  {NoStop}%
\bibitem [{\citenamefont {Chubukov}(2009)}]{chubukov2009renormalization}%
  \BibitemOpen
  \bibfield  {author} {\bibinfo {author} {\bibfnamefont {A.}~\bibnamefont
  {Chubukov}},\ }\bibfield  {title} {\bibinfo {title} {Renormalization group
  analysis of competing orders and the pairing symmetry in {Fe}-based
  superconductors},\ }\href@noop {} {\bibfield  {journal} {\bibinfo  {journal}
  {Physica C: Superconductivity}\ }\textbf {\bibinfo {volume} {469}},\ \bibinfo
  {pages} {640} (\bibinfo {year} {2009})}\BibitemShut {NoStop}%
\bibitem [{\citenamefont {Fernandes}\ and\ \citenamefont
  {Schmalian}(2010)}]{Fernandes2010}%
  \BibitemOpen
  \bibfield  {author} {\bibinfo {author} {\bibfnamefont {R.~M.}\ \bibnamefont
  {Fernandes}}\ and\ \bibinfo {author} {\bibfnamefont {J.}~\bibnamefont
  {Schmalian}},\ }\bibfield  {title} {\bibinfo {title} {Competing order and
  nature of the pairing state in the iron pnictides},\ }\href
  {https://doi.org/10.1103/PhysRevB.82.014521} {\bibfield  {journal} {\bibinfo
  {journal} {Phys. Rev. B}\ }\textbf {\bibinfo {volume} {82}},\ \bibinfo
  {pages} {014521} (\bibinfo {year} {2010})}\BibitemShut {NoStop}%
\bibitem [{\citenamefont {Chen}\ \emph {et~al.}(2019)\citenamefont {Chen},
  \citenamefont {Wang}, \citenamefont {Wu}, \citenamefont {Ma}, \citenamefont
  {Wen}, \citenamefont {Wu}, \citenamefont {Li}, \citenamefont {Zhao},
  \citenamefont {Wang}, \citenamefont {Zhang}, \citenamefont {Huang},
  \citenamefont {Li},\ and\ \citenamefont {Huang}}]{chen2019raman}%
  \BibitemOpen
  \bibfield  {author} {\bibinfo {author} {\bibfnamefont {Y.}~\bibnamefont
  {Chen}}, \bibinfo {author} {\bibfnamefont {P.}~\bibnamefont {Wang}}, \bibinfo
  {author} {\bibfnamefont {M.}~\bibnamefont {Wu}}, \bibinfo {author}
  {\bibfnamefont {J.}~\bibnamefont {Ma}}, \bibinfo {author} {\bibfnamefont
  {S.}~\bibnamefont {Wen}}, \bibinfo {author} {\bibfnamefont {X.}~\bibnamefont
  {Wu}}, \bibinfo {author} {\bibfnamefont {G.}~\bibnamefont {Li}}, \bibinfo
  {author} {\bibfnamefont {Y.}~\bibnamefont {Zhao}}, \bibinfo {author}
  {\bibfnamefont {K.}~\bibnamefont {Wang}}, \bibinfo {author} {\bibfnamefont
  {L.}~\bibnamefont {Zhang}}, \bibinfo {author} {\bibfnamefont
  {L.}~\bibnamefont {Huang}}, \bibinfo {author} {\bibfnamefont
  {W.}~\bibnamefont {Li}},\ and\ \bibinfo {author} {\bibfnamefont
  {M.}~\bibnamefont {Huang}},\ }\bibfield  {title} {\bibinfo {title} {{Raman
  spectra and dimensional effect on the charge density wave transition in
  {GdTe$_3$}}},\ }\href {https://doi.org/10.1063/1.5118870} {\bibfield
  {journal} {\bibinfo  {journal} {Applied Physics Letters}\ }\textbf {\bibinfo
  {volume} {115}},\ \bibinfo {pages} {151905} (\bibinfo {year}
  {2019})}\BibitemShut {NoStop}%
\bibitem [{\citenamefont {Cheong}\ \emph {et~al.}(2018)\citenamefont {Cheong},
  \citenamefont {Talbayev}, \citenamefont {Kiryukhin},\ and\ \citenamefont
  {Saxena}}]{Cheong2018}%
  \BibitemOpen
  \bibfield  {author} {\bibinfo {author} {\bibfnamefont {S.-W.}\ \bibnamefont
  {Cheong}}, \bibinfo {author} {\bibfnamefont {D.}~\bibnamefont {Talbayev}},
  \bibinfo {author} {\bibfnamefont {V.}~\bibnamefont {Kiryukhin}},\ and\
  \bibinfo {author} {\bibfnamefont {A.}~\bibnamefont {Saxena}},\ }\bibfield
  {title} {\bibinfo {title} {Broken symmetries, non-reciprocity, and
  multiferroicity},\ }\href@noop {} {\bibfield  {journal} {\bibinfo  {journal}
  {npj Quantum Materials}\ }\textbf {\bibinfo {volume} {3}},\ \bibinfo {pages}
  {19} (\bibinfo {year} {2018})}\BibitemShut {NoStop}%
\bibitem [{\citenamefont {Hayami}\ \emph {et~al.}(2022)\citenamefont {Hayami},
  \citenamefont {Oiwa},\ and\ \citenamefont {Kusunose}}]{Hayami2022}%
  \BibitemOpen
  \bibfield  {author} {\bibinfo {author} {\bibfnamefont {S.}~\bibnamefont
  {Hayami}}, \bibinfo {author} {\bibfnamefont {R.}~\bibnamefont {Oiwa}},\ and\
  \bibinfo {author} {\bibfnamefont {H.}~\bibnamefont {Kusunose}},\ }\bibfield
  {title} {\bibinfo {title} {Electric ferro-axial moment as nanometric rotator
  and source of longitudinal spin current},\ }\href@noop {} {\bibfield
  {journal} {\bibinfo  {journal} {Journal of the Physical Society of Japan}\
  }\textbf {\bibinfo {volume} {91}},\ \bibinfo {pages} {113702} (\bibinfo
  {year} {2022})}\BibitemShut {NoStop}%
\bibitem [{\citenamefont {Wang}\ \emph {et~al.}(2015)\citenamefont {Wang},
  \citenamefont {Kang},\ and\ \citenamefont {Fernandes}}]{Fernandes2015}%
  \BibitemOpen
  \bibfield  {author} {\bibinfo {author} {\bibfnamefont {X.}~\bibnamefont
  {Wang}}, \bibinfo {author} {\bibfnamefont {J.}~\bibnamefont {Kang}},\ and\
  \bibinfo {author} {\bibfnamefont {R.~M.}\ \bibnamefont {Fernandes}},\
  }\bibfield  {title} {\bibinfo {title} {Magnetic order without
  tetragonal-symmetry-breaking in iron arsenides: Microscopic mechanism and
  spin-wave spectrum},\ }\href {https://doi.org/10.1103/PhysRevB.91.024401}
  {\bibfield  {journal} {\bibinfo  {journal} {Phys. Rev. B}\ }\textbf {\bibinfo
  {volume} {91}},\ \bibinfo {pages} {024401} (\bibinfo {year}
  {2015})}\BibitemShut {NoStop}%
\bibitem [{\citenamefont {Pfuner}\ \emph {et~al.}(2011)\citenamefont {Pfuner},
  \citenamefont {Gvasaliya}, \citenamefont {Zaharko}, \citenamefont {Keller},
  \citenamefont {Mesot}, \citenamefont {Pomjakushin}, \citenamefont {Chu},
  \citenamefont {Fisher},\ and\ \citenamefont
  {Degiorgi}}]{pfuner2011incommensurate}%
  \BibitemOpen
  \bibfield  {author} {\bibinfo {author} {\bibfnamefont {F.}~\bibnamefont
  {Pfuner}}, \bibinfo {author} {\bibfnamefont {S.}~\bibnamefont {Gvasaliya}},
  \bibinfo {author} {\bibfnamefont {O.}~\bibnamefont {Zaharko}}, \bibinfo
  {author} {\bibfnamefont {L.}~\bibnamefont {Keller}}, \bibinfo {author}
  {\bibfnamefont {J.}~\bibnamefont {Mesot}}, \bibinfo {author} {\bibfnamefont
  {V.}~\bibnamefont {Pomjakushin}}, \bibinfo {author} {\bibfnamefont
  {J.}~\bibnamefont {Chu}}, \bibinfo {author} {\bibfnamefont {I.}~\bibnamefont
  {Fisher}},\ and\ \bibinfo {author} {\bibfnamefont {L.}~\bibnamefont
  {Degiorgi}},\ }\bibfield  {title} {\bibinfo {title} {Incommensurate magnetic
  order in {TbTe$_3$}},\ }\href@noop {} {\bibfield  {journal} {\bibinfo
  {journal} {Journal of Physics: Condensed Matter}\ }\textbf {\bibinfo {volume}
  {24}},\ \bibinfo {pages} {036001} (\bibinfo {year} {2011})}\BibitemShut
  {NoStop}%
\bibitem [{\citenamefont {Lei}\ \emph {et~al.}(2019)\citenamefont {Lei},
  \citenamefont {Duppel}, \citenamefont {Lippmann}, \citenamefont {Nuss},
  \citenamefont {Lotsch},\ and\ \citenamefont {Schoop}}]{lei2019charge}%
  \BibitemOpen
  \bibfield  {author} {\bibinfo {author} {\bibfnamefont {S.}~\bibnamefont
  {Lei}}, \bibinfo {author} {\bibfnamefont {V.}~\bibnamefont {Duppel}},
  \bibinfo {author} {\bibfnamefont {J.~M.}\ \bibnamefont {Lippmann}}, \bibinfo
  {author} {\bibfnamefont {J.}~\bibnamefont {Nuss}}, \bibinfo {author}
  {\bibfnamefont {B.~V.}\ \bibnamefont {Lotsch}},\ and\ \bibinfo {author}
  {\bibfnamefont {L.~M.}\ \bibnamefont {Schoop}},\ }\bibfield  {title}
  {\bibinfo {title} {Charge density waves and magnetism in topological
  semimetal candidates {GdSb$_x$Te$_{2-x-\delta}$}},\ }\href@noop {} {\bibfield
   {journal} {\bibinfo  {journal} {Advanced Quantum Technologies}\ }\textbf
  {\bibinfo {volume} {2}},\ \bibinfo {pages} {1900045} (\bibinfo {year}
  {2019})}\BibitemShut {NoStop}%
\bibitem [{\citenamefont {Okuma}\ \emph {et~al.}(2020)\citenamefont {Okuma},
  \citenamefont {Ueta}, \citenamefont {Kuniyoshi}, \citenamefont {Fujisawa},
  \citenamefont {Smith}, \citenamefont {Hsu}, \citenamefont {Inagaki},
  \citenamefont {Si}, \citenamefont {Kawae}, \citenamefont {Lin}, \citenamefont
  {Chuang}, \citenamefont {Masuda}, \citenamefont {Kobayashi},\ and\
  \citenamefont {Okada}}]{okuma2020fermionic}%
  \BibitemOpen
  \bibfield  {author} {\bibinfo {author} {\bibfnamefont {R.}~\bibnamefont
  {Okuma}}, \bibinfo {author} {\bibfnamefont {D.}~\bibnamefont {Ueta}},
  \bibinfo {author} {\bibfnamefont {S.}~\bibnamefont {Kuniyoshi}}, \bibinfo
  {author} {\bibfnamefont {Y.}~\bibnamefont {Fujisawa}}, \bibinfo {author}
  {\bibfnamefont {B.}~\bibnamefont {Smith}}, \bibinfo {author} {\bibfnamefont
  {C.}~\bibnamefont {Hsu}}, \bibinfo {author} {\bibfnamefont {Y.}~\bibnamefont
  {Inagaki}}, \bibinfo {author} {\bibfnamefont {W.}~\bibnamefont {Si}},
  \bibinfo {author} {\bibfnamefont {T.}~\bibnamefont {Kawae}}, \bibinfo
  {author} {\bibfnamefont {H.}~\bibnamefont {Lin}}, \bibinfo {author}
  {\bibfnamefont {F.}~\bibnamefont {Chuang}}, \bibinfo {author} {\bibfnamefont
  {T.}~\bibnamefont {Masuda}}, \bibinfo {author} {\bibfnamefont
  {R.}~\bibnamefont {Kobayashi}},\ and\ \bibinfo {author} {\bibfnamefont
  {Y.}~\bibnamefont {Okada}},\ }\bibfield  {title} {\bibinfo {title} {Fermionic
  order by disorder in a van der {Waals} antiferromagnet},\ }\href@noop {}
  {\bibfield  {journal} {\bibinfo  {journal} {Scientific reports}\ }\textbf
  {\bibinfo {volume} {10}},\ \bibinfo {pages} {15311} (\bibinfo {year}
  {2020})}\BibitemShut {NoStop}%
\bibitem [{\citenamefont {Lei}\ \emph {et~al.}(2021{\natexlab{b}})\citenamefont
  {Lei}, \citenamefont {Saltzman},\ and\ \citenamefont
  {Schoop}}]{lei2021complex}%
  \BibitemOpen
  \bibfield  {author} {\bibinfo {author} {\bibfnamefont {S.}~\bibnamefont
  {Lei}}, \bibinfo {author} {\bibfnamefont {A.}~\bibnamefont {Saltzman}},\ and\
  \bibinfo {author} {\bibfnamefont {L.~M.}\ \bibnamefont {Schoop}},\ }\bibfield
   {title} {\bibinfo {title} {Complex magnetic phases enriched by charge
  density waves in the topological semimetals {GdSb$_x$Te$_{2-x-\delta}$}},\
  }\href {https://doi.org/10.1103/PhysRevB.103.134418} {\bibfield  {journal}
  {\bibinfo  {journal} {Phys. Rev. B}\ }\textbf {\bibinfo {volume} {103}},\
  \bibinfo {pages} {134418} (\bibinfo {year} {2021}{\natexlab{b}})}\BibitemShut
  {NoStop}%
\bibitem [{\citenamefont {Christensen}\ \emph {et~al.}(2022)\citenamefont
  {Christensen}, \citenamefont {Birol}, \citenamefont {Andersen},\ and\
  \citenamefont {Fernandes}}]{Christensen2022}%
  \BibitemOpen
  \bibfield  {author} {\bibinfo {author} {\bibfnamefont {M.~H.}\ \bibnamefont
  {Christensen}}, \bibinfo {author} {\bibfnamefont {T.}~\bibnamefont {Birol}},
  \bibinfo {author} {\bibfnamefont {B.~M.}\ \bibnamefont {Andersen}},\ and\
  \bibinfo {author} {\bibfnamefont {R.~M.}\ \bibnamefont {Fernandes}},\
  }\bibfield  {title} {\bibinfo {title} {Loop currents in {$A$V$_3$Sb$_5$}
  kagome metals: Multipolar and toroidal magnetic orders},\ }\href
  {https://doi.org/10.1103/PhysRevB.106.144504} {\bibfield  {journal} {\bibinfo
   {journal} {Phys. Rev. B}\ }\textbf {\bibinfo {volume} {106}},\ \bibinfo
  {pages} {144504} (\bibinfo {year} {2022})}\BibitemShut {NoStop}%
\bibitem [{\citenamefont {Xing}\ \emph {et~al.}(2023)\citenamefont {Xing},
  \citenamefont {Bae}, \citenamefont {Ritz}, \citenamefont {Yang},
  \citenamefont {Birol}, \citenamefont {Salinas}, \citenamefont {Ortiz},
  \citenamefont {Wilson}, \citenamefont {Wang}, \citenamefont {Fernandes},\
  and\ \citenamefont {Madhavan}}]{Xing2023}%
  \BibitemOpen
  \bibfield  {author} {\bibinfo {author} {\bibfnamefont {Y.}~\bibnamefont
  {Xing}}, \bibinfo {author} {\bibfnamefont {S.}~\bibnamefont {Bae}}, \bibinfo
  {author} {\bibfnamefont {E.}~\bibnamefont {Ritz}}, \bibinfo {author}
  {\bibfnamefont {F.}~\bibnamefont {Yang}}, \bibinfo {author} {\bibfnamefont
  {T.}~\bibnamefont {Birol}}, \bibinfo {author} {\bibfnamefont {A.~N.}\
  \bibnamefont {Salinas}}, \bibinfo {author} {\bibfnamefont {B.~R.}\
  \bibnamefont {Ortiz}}, \bibinfo {author} {\bibfnamefont {S.~D.}\ \bibnamefont
  {Wilson}}, \bibinfo {author} {\bibfnamefont {Z.}~\bibnamefont {Wang}},
  \bibinfo {author} {\bibfnamefont {R.~M.}\ \bibnamefont {Fernandes}},\ and\
  \bibinfo {author} {\bibfnamefont {V.}~\bibnamefont {Madhavan}},\ }\bibfield
  {title} {\bibinfo {title} {Optical manipulation of the charge density wave
  state in {RbV$_3$Sb$_5$}},\ }\href@noop {} {\bibfield  {journal} {\bibinfo
  {journal} {arXiv:2308.04128}\ } (\bibinfo {year} {2023})}\BibitemShut
  {NoStop}%
\bibitem [{\citenamefont {Fernandes}\ \emph {et~al.}(2012)\citenamefont
  {Fernandes}, \citenamefont {Chubukov}, \citenamefont {Knolle}, \citenamefont
  {Eremin},\ and\ \citenamefont {Schmalian}}]{PhysRevB.85.024534}%
  \BibitemOpen
  \bibfield  {author} {\bibinfo {author} {\bibfnamefont {R.~M.}\ \bibnamefont
  {Fernandes}}, \bibinfo {author} {\bibfnamefont {A.~V.}\ \bibnamefont
  {Chubukov}}, \bibinfo {author} {\bibfnamefont {J.}~\bibnamefont {Knolle}},
  \bibinfo {author} {\bibfnamefont {I.}~\bibnamefont {Eremin}},\ and\ \bibinfo
  {author} {\bibfnamefont {J.}~\bibnamefont {Schmalian}},\ }\bibfield  {title}
  {\bibinfo {title} {Preemptive nematic order, pseudogap, and orbital order in
  the iron pnictides},\ }\href {https://doi.org/10.1103/PhysRevB.85.024534}
  {\bibfield  {journal} {\bibinfo  {journal} {Phys. Rev. B}\ }\textbf {\bibinfo
  {volume} {85}},\ \bibinfo {pages} {024534} (\bibinfo {year}
  {2012})}\BibitemShut {NoStop}%
\end{thebibliography}%

\end{document}